\documentclass[10pt]{article} 

\usepackage[utf8]{inputenc} 

\usepackage{geometry} 
\usepackage{graphicx} 

\usepackage{booktabs} 
\usepackage{array} 
\usepackage{paralist} 
\usepackage{verbatim} 

\usepackage{fancyhdr} 
\usepackage{sectsty}
\allsectionsfont{\sffamily\mdseries\upshape} 

\usepackage[nottoc,notlof,notlot]{tocbibind} 
\usepackage[titles,subfigure]{tocloft} 

\usepackage{cite} 
\usepackage{hyperref} 
\usepackage[retainorgcmds]{IEEEtrantools} 
\usepackage{amsmath} 
\usepackage{physics} 
\usepackage{qcircuit} 
\usepackage{xcolor}
\usepackage[utf8]{inputenc}
\usepackage[english]{babel}
\usepackage{graphicx}
\usepackage[skaknew]{chessboard,skak}
\usepackage{algpseudocode}
\usepackage{subcaption}
\usepackage{tikz}
\usepackage{tabu,multirow}
\usepackage{multicol}
\usetikzlibrary{tikzmark,decorations.pathmorphing}
\usepackage [autostyle, english = american]{csquotes}
\MakeOuterQuote{"}

\newtheorem{definition}{Definition}

\newtheorem{algorithm}{Algorithm}

\title{Quantum Chess: Developing a Mathematical Framework and Design Methodology for Creating Quantum Games}
\author{Christopher Cantwell\\
ccantwel@usc.edu\\}

\begin{document}
\maketitle

\section{Introduction}

Quantum phenomena have remained largely inaccessible to the general public. This can be attributed to the fact that we do not experience quantum mechanics on a tangible level in our daily lives. In order to explore how quantum systems work one generally needs access to a lab and must understand some mathematics that most people aren't ever exposed to. But to gain an intuitive understanding it might not be necessary to understand the math. Consider baseball as an analogy.  One doesn't need to understand classical mechanics to play baseball, and playing baseball will not teach one classical mechanics. However, playing baseball may help to more intuitively understand how things move. Games can provide an environment in which people can experience the strange behavior of the quantum world in a fun and mentally engaging way. 

With progress being made quantum computers one might pose the question, "Will there be quantum games played on quantum computers?" If one assumes the answer is yes, then the next question is, "What will those games be?" Games could offer an interesting test bed for near term quantum devices. Games can be tailored to support varying amounts of quantum behavior through simple rule changes, which can be useful when dealing with limited resources. AI algorithms that can play such games can be designed to tap into quantum computing resources to assist in evaluating next moves. We may see stronger AI players evolve as more advanced quantum hardware becomes available. Here we explore the design of Quantum Chess while keeping in mind the goal of allowing people the opportunity to experience quantum phenomena.

\section{Overview of Previous Work} 
There has been some investigation into a formal quantum game theory \cite{Prisoners:1999,Meyer:2000,QuantumGamesReview:2018,2017arXiv170102193D}. This paper will focus more on the concept of implementing playable quantum games. The body of previous work is somewhat expectedly small. There are a handful of games that have been proposed, but to our knowledge never implemented \cite{GordonGoren:2012,Neto:online,Nagy:2012,ranchin:2016}. The following games have indeed been implemented, and their properties will help direct the work to be done: Quantum Tic-Tic-Tac-Toe ~\cite{Goff:2006}, Quantum Minesweeper ~\cite{GordonGoren:2010}, QCraft \cite{qCraft:online}, and another version of Quantum Chess ~\cite{akl:2010}. More recently there have been citizen science type games like Decodoku \cite{Wooten:2017}, Quantum Moves \cite{Sorensen:2016}, and the Alice Challenge \cite{HeckE11231:2018} that have been used to help solve quantum problems, but we will not discuss these here as their goal is different than ours and their design did not influence the design of Quantum Chess. In examining the games mentioned we can find some gaps to fill when designing Quantum Chess.

\subsection{Quantum Tic-Tac-Toe}
Quantum Tic-Tac-Toe is a variant of Tic-Tac-Toe that allows players to place marks in superposition of being in two places at once.  This allows for the possibility of a square to be occupied by more than one mark. In this way squares can be seen to be entangled. Under the condition of cyclic entanglement collapse occurs. The player who did not create the cycle chooses the mark to collapse to, causing the entire cycle to collapse.

This game does illustrate superposition and entanglement.  However, it is deterministic in its measurement outcomes (collapse), where quantum mechanical measurements are specifically non-deterministic. There is also no representation of interference.

\subsection{Quantum Minesweeper}
Quantum Minesweeper is a variant of minesweeper that is played on a board that begins in a superposition of different mine configurations. The player then performs different types of measurements to try and guess the superposition. The measurements available are classical, interaction free, and entanglement measurements. In this way the player may gain some understanding of these quantum phenomena. However interference is not present, and the player is not actively generating quantum effects.

\subsection{QCraft}
QCraft is a mod that adds a quantum twist to the world of Minecraft. It adds certain quantum properties to Minecraft like observational dependency, teleportation, superposition, and entanglement. It is a great game for introducing people to select aspects of quantum mechanics in a fun macroscopic environment. However, qCraft's own makers state that isn't, nor is it meant to be, an accurate quantum simulation. It is just meant to be a fun introduction to quantum concepts.

\subsection{Selim Akl's Quantum Chess}
Quantum Chess by Selim Akl is a chess variant designed with the purpose of evening the playing field between computer and human players.  Pieces begin the game in superposition of being a number of different types of pieces at once until the player chooses which one to move.  At that point in time a measurement occurs and the piece is collapsed to a single type, after which it can be moved. The piece remains in a classical state until it lands on a black square, at which time it is placed back into a superposition of being multiple different types. The game illustrates superposition, but in the absence of other quantum effects it seems to only add an element of randomness to chess, at least as far as attempting to formulate a strategy for play goes.

\section{Quantum Chess Design Goals}\label{sec:DesignGoals}
An examination of the previous work highlighted one major gap that could be filled in the design of a Quantum Chess: interference.  None of these games exhibited quantum interference effects in any way. This realization, and other more subjective desires for a quantum game, led to the following list of design goals
\begin{enumerate}
\item \label{goal1}The game would include a trifecta of quantum evolutionary phenomena: superposition, entanglement, and interference.
\item \label{goal2}Said trifecta would be implemented in a way that could lead to the development of quantum strategies for play.  In other words, players should be able to create and use superposition, entanglement, and interference effects.\footnote{See \cite{Griffiths:2005} for more information on the concepts of superposition, entanglement, and interference.}
\item \label{goal3}Measurements must be non-deterministic.
\item \label{goal4}The game would not teach quantum mechanics. Instead it would provide an environment in which players could interact with quantum phenomena and, hopefully, develop a more intuitive understanding of how said phenomena work as a side effect of playing a fun game.
\end{enumerate}
The last goal stems from a desire to make the game accessible to the largest audience possible. With these design goals in mind we lay down the following criteria for how we will build Quantum Chess:

\begin{enumerate}
\item \label{req1}Movement is accomplished through unitary evolution of a quantum state. \footnote{Quantum state evolution is described by the application of unitary operators. See \cite{Griffiths:2005}}
\item \label{req2}Players have access to a set of moves through which they can create superposition.
\end{enumerate}

\section{Quantum Chess Algebraic Notation}
Classical chess has an established algebraic notation \cite{AlgebraicNotation:online,FIDE:online}. Here we define a simple Quantum Chess Algebraic Notation for describing different types of moves.
\begin{itemize}
	\item Non-quantum Move: $(source)(target)$, e.g. $a1a2$.
	\item Quantum Split Move: $(source)\wedge (target1)(target2)$, e.g. $a1\wedge a2a3$
	\item Quantum Merge Move: $(source1)(source2)\wedge (target)$, e.g. $a3a2\wedge a1$
	\item Pawn Promotion: $(source)(target)(piece)$, e.g. $a7a8Q$
\end{itemize}
All non-quantum moves involve a single source and single target, even castling can be described completely by the king source and king  target. Adding further characters to a move string to give a more detailed description, as is common in standard Chess, is a tricky prospect given the nature of superposition. Another consideration is that in Quantum Chess we will also have measurements (see section \ref{sec:NoDoubleOccupancyRule}). We will expand the notation to describe if a move requires a measurement, and the outcome of said measurement. For the purposes of this paper we will design all of our measurements to have two outcomes\footnote{It is possible to design more general measurements which could have three or more outcomes. See \cite{Griffiths:2005}\cite{Nielsen:2011}. This would require additional notation.}, thus we can append the following measurement notation to a move string to achieve:
\begin{itemize}
	\item Move with Measurement Outcome 0: $(move string).(measure string)0$, e.g. $a1a2.m0$
	\item Move with Measurement Outcome 1: $(move string).(measure string)1$, e.g. $a1a2.m1$
\end{itemize}
We can define a set of measurement strings to clarify the type of measurement being applied, but if we know the entire history it is enough to leave it generic. The type of measurement is defined by the type of move that corresponds to the move string acting on the current game state.


\section{State Description}
For a mathematical framework it will be useful to define a vector format that describes the state. We will use a hybrid classical / quantum state representation to describe the board. 

Our quantum state will be composed of 64 qubits to describe the "occupancy" of the board\footnote{We use standard Dirac, or Bra-ket,	 notation to describe the quantum state. See \cite{Griffiths:2005}\cite{Nielsen:2011}\cite{BraKet:online}}.
\begin{equation}\label{eq:OccupancySuperposition}
	\ket{\psi_\mathcal{B}} = \sum_{i} A_i \ket{q^{(i)}_0,...q^{(i)}_{63}} , q_j\in\{0=empty,1=occupied\}
\end{equation}
This state representation is similar to the bitboard representation used in classical chess\cite{FIDE:online} where our state is analogous to a superposition of the "all pieces" bitboard.

On top of this superposition of occupancy bitboards we will store classical type information for each square. This information consists of a single 64-element vector that describes what piece, if any, is occupying an given square. 
\begin{equation}\label{eq:ValueMap}
	\vec{v} = \{v_0,...v_{63}\} , v_i\in\{0,P,N,B,R,Q,K,p,n,b,r,q,k\}
\end{equation}
The values $v_i$ correspond to standard FEN values for chess pieces\cite{FEN:online} with lower case letters representing black pieces, upper case white pieces, and 0 empty. We say a square is occupied, or partially occupied, by a piece if there is a non-zero probability of finding that piece in that square. Note, this representation does not account for the possibility of superpositions of piece type, but we exclude that possibility with the "No Double Occupancy" rule described in section \ref{sec:NoDoubleOccupancyRule}. 

Finally we will need a set of classical information encoding the color of the current player, as well as any special flags for moves such as castling and en passant (e.p.). We define the following:
\begin{subequations}
\begin{IEEEeqnarray}{rCl}
	\mathcal{F} &=& \{F_c,F_K,F_Q,F_k,F_q,F_ep\} \label{eq:Flags}\\
	F_c &\in& \{w,b\} \text{, is the color of the current player}\\
	F_K &=& \text{True, if king side castling is legal for white player}\\
	F_Q &=& \text{True, if queen side castling is legal for white player}\\
	F_k &=& \text{True, if king side castling is legal for black player}\\
	F_q &=& \text{True, if queen side castling is legal for black player}\\
	F_{ep} &=& \begin{cases}
		file \in \{a,b,c,d,e,f,g,h\}, & \text{file of the capturable pawn if e.p. is legal}\\
		-, & \text{otherwise}\\
		\end{cases}
\end{IEEEeqnarray}
\end{subequations}
With our state representation complete we can then consider the action of a move. Each move will have the following parts:
\begin{enumerate}
	\item A unitary that acts to update the occupancy superposition $\ket{\psi_\mathcal{B}}$.
	\item An operation that updates the classical piece type information $\vec{v}$.
	\item An operation that updates the extra flags $\mathcal{F}$.
\end{enumerate}

\section{Movement Unitary Design}\label{sec:Movement}
In standard Chess a simple, non-capturing move from a source square to a target square can be accomplished by swapping the values of the pieces in the source square and target square. This realization leads to a natural unitary to accomplish movement on a quantum board: the quantum Swap gate\cite{Nielsen:2011}. We can also base a move that allows for the creation of superposition on the $\sqrt{Swap}$ unitary. For arbitrary reasons we choose the iSwap instead of Swap, but this does not change the conceptual effect of the move. Our basic unitaries are thus:

\begin{subequations}
\begin{IEEEeqnarray}{rCl}
	\label{eq:iSwap} U_{iSwap} &=& \begin{pmatrix}
		1 & 0 & 0 & 0\\
		0 & 0 & i & 0\\
		0 & i & 0 & 0\\
		0 & 0 & 0 & 1\\
	\end{pmatrix}\\
	\label{eq:iSwapRoot}U_{\sqrt{iSwap}} &=& \begin{pmatrix}
		1 & 0 & 0 & 0\\
		0 & \frac{1}{\sqrt{2}} & \frac{i}{\sqrt{2}} & 0\\
		0 & \frac{i}{\sqrt{2}} & \frac{1}{\sqrt{2}} & 0\\
		0 & 0 & 0 & 1\\
	\end{pmatrix}
\end{IEEEeqnarray}
\end{subequations}
However we can't just use these operators out of the box. We must also consider pathing and the piece types involved. Here we will detail the design of a set of movement unitaries, based on the idea of using Swap, that can be used to construct moves. The procedure for designing the unitary operators that act on the occupancy superposition will be as follows:
\begin{enumerate}
	\item Express the desired action of the move, acting on the relevant subspace of otherwise empty board, in the form of an operator being applied to a Hilbert space.
	\item Consider specific desired outcomes in the presence of superposition and modify the operator, keeping in mind unitarity, to reflect these outcomes. This may involve expanding the Hilbert space on which the operator acts.
	\item Use the unitary requirement to complete the operator.
\end{enumerate}
Once we have the unitary operators designed we can then use them to implement the various types of movement allowed in Quantum Chess
\subsection{Jump Unitary}\label{sec:JumpUnitary}
The Jump Unitary is the simplest of the movement unitaries. It will be useful for pieces that do not slide along some path, like the knight or pieces that take a single step. The operator acts on a subspace of the board involving only two squares: a source (s) and target (t). We can define the unitary operator as it acts on a reduced Hilbert space defined by the basis $\ket{t,s}$. This unitary then takes the simple form of the iSwap we introduced in equation \ref{eq:iSwap}.
\begin{equation}\label{eq:JumpUnitary}
U_{jump}= \begin{pmatrix}
				1 & 0 & 0 & 0\\
				0 & 0 & i & 0\\
				0 & i & 0 & 0\\
				0 & 0 & 0 & 1\\
			\end{pmatrix} 
\end{equation}

\subsection{Slide Unitary}\label{sec:SlideUnitary}
The Slide Unitary is the similar to the jump but we must add a control qubit. It will be useful for pieces that slide along some path, like the bishop, rook, and queen. We must expand the basis of the Hilbert space to be $\ket{p,t,s}$, where p is an ancilla qubit that we will set to be 1 if there is a piece occupying the path between s and t. We can set the state of p by applying a sequence of CNOT operators commonly seen in quantum computing circuits\cite{Nielsen:2011}. This unitary then takes the simple form of a controlled-iSwap, where it acts as identity when the path qubit is set to 1.
\begin{equation}\label{eq:SlideUnitary}
	U_{slide} = \begin{pmatrix}
		1 & 0 & 0 & 0 & 0 & 0 & 0 & 0\\
		0 & 0 & i & 0 & 0 & 0 & 0 & 0\\
		0 & i & 0 & 0 & 0 & 0 & 0 & 0\\
		0 & 0 & 0 & 1 & 0 & 0 & 0 & 0\\
		0 & 0 & 0 & 0 & 1 & 0 & 0 & 0\\
		0 & 0 & 0 & 0 & 0 & 1 & 0 & 0\\
		0 & 0 & 0 & 0 & 0 & 0 & 1 & 0\\
		0 & 0 & 0 & 0 & 0 & 0 & 0 & 1\\
	\end{pmatrix}\\
\end{equation}

\subsection{Split Jump Unitary}\label{sec:SplitJumpUnitary}
The Split Jump unitary is designed to allow a player to create superposition. This unitary will be based on the square root of an iSwap introduced in equation \ref{eq:iSwapRoot}. If $U_{\sqrt{iSwap}}$ is applied to the basis states of a two qubit Hilbert space we see
\begin{IEEEeqnarray*}{rCl}
	\ket{00} &\rightarrow& \ket{00} , \ket{11} \rightarrow \ket{11}\\
	\ket{01} &\rightarrow& \frac{1}{\sqrt{2}}(\ket{01} + i \ket{10})\\
	\ket{10} &\rightarrow& \frac{1}{\sqrt{2}}(\ket{10} + i \ket{01})\\
\end{IEEEeqnarray*}
This can be interpreted as a piece both moving and not moving. Not moving a piece might not be very desirable for a player. One option to counteract this is to allow the move to have twice the range so that on average a player accomplishes the same amount of work. This was the solution in a previous variant of Quantum Chess\cite{QuantumChess:online,Cantwell:online} but we have chosen to drop it for a number of reasons. Instead we choose to allow the player to "split" the piece to exist in superposition on two different squares. For this we need two targets: $t_1$ and $t_2$. We can accomplish the Split Jump by first performing our $\sqrt{iSwap}$ between $s$ and $t_1$, and then performing a full iSwap between $s$ and $t_2$. In the $\ket{t_2,t_1,s}$ basis these unitaries take the form:

\begin{equation*}\label{eq:TwoTargetUnitaries}
\begin{tabular}{l l}
\scalebox{0.85}{
$U_{\sqrt{iSwap}}(s,t_1) = 
	\begin{pmatrix}
		1 & 0 & 0 & 0 & 0 & 0 & 0 & 0\\
		0 & \frac{1}{\sqrt{2}}  & \frac{i}{\sqrt{2}}  & 0 & 0 & 0 & 0 & 0\\
		0 & \frac{i}{\sqrt{2}}  & \frac{1}{\sqrt{2}}  & 0 & 0 & 0 & 0 & 0\\
		0 & 0 & 0 & 1 & 0 & 0 & 0 & 0\\
		0 & 0 & 0 & 0 & 1 & 0 & 0 & 0\\
		0 & 0 & 0 & 0 & 0 & \frac{1}{\sqrt{2}}  & \frac{i}{\sqrt{2}}  & 0\\
		0 & 0 & 0 & 0 & 0 & \frac{i}{\sqrt{2}}  & \frac{1}{\sqrt{2}}  & 0\\
		0 & 0 & 0 & 0 & 0 & 0 & 0 & 1\\
	\end{pmatrix}$
	}
&
\scalebox{0.85}{
$U_{iSwap}(s,t_2) = \begin{pmatrix}
		1 & 0 & 0 & 0 & 0 & 0 & 0 & 0\\
		0 & 0 & 0 & 0 & i & 0 & 0 & 0\\
		0 & 0 & 1 & 0 & 0 & 0 & 0 & 0\\
		0 & 0 & 0 & 0 & 0 & 0 & i & 0\\
		0 & i & 0 & 0 & 1 & 0 & 0 & 0\\
		0 & 0 & 0 & 0 & 0 & 1 & 0 & 0\\
		0 & 0 & 0 & i & 0 & 0 & 0 & 0\\
		0 & 0 & 0 & 0 & 0 & 0 & 0 & 1\\
	\end{pmatrix}$
	}
\end{tabular}
\end{equation*}
which gives us a total Split Jump unitary:
\begin{equation}\label{eq:SplitJumpUnitary}
	U_{split} = \begin{pmatrix}
		1 & 0 & 0 & 0 & 0 & 0 & 0 & 0\\
		0 & 0 & 0 & 0 & i & 0 & 0 & 0\\
		0 & \frac{i}{\sqrt{2}} & \frac{1}{\sqrt{2}} & 0  & 0 & 0 & 0 & 0\\
		0 & 0 & 0 & 0 & 0 & \frac{-1}{\sqrt{2}} & \frac{i}{\sqrt{2}} & 0\\
		0 & \frac{i}{\sqrt{2}}& \frac{-1}{\sqrt{2}} & 0 & 0 & 0 & 0 & 0\\
		0 & 0 & 0 & 0 &  0 & \frac{i}{\sqrt{2}} & \frac{-1}{\sqrt{2}} & 0\\
		0 & 0 & 0 & i & 0 & 0 & 0 & 0\\
		0 & 0 & 0 & 0 & 0 & 0 & 0 & 1\\
	\end{pmatrix}
\end{equation}
 A quick examination of the non-trivial action of $U_{split}$ acting on the basis states of the $\ket{t_2,t_1,s}$ Hilbert space shows us that this unitary conserves piece number(occupancy number), which is an important consideration if we don't want pieces randomly appearing or disappearing in the game.
\begin{subequations}
\begin{multicols}{2}
\noindent
\begin{IEEEeqnarray}{l}
\scalebox{0.95}{$U_{split}\ket{001} = \frac{1}{\sqrt{2}}(i\ket{010} + i\ket{100})$}\\
\scalebox{0.95}{$U_{split}\ket{010} = \frac{1}{\sqrt{2}}(\ket{010} - \ket{100})$}\\
\scalebox{0.95}{$U_{split}\ket{100} = i\ket{001}$}
\end{IEEEeqnarray}
\begin{IEEEeqnarray}{l}
\scalebox{0.95}{$U_{split}\ket{011} = i\ket{110}$}\\
\scalebox{0.95}{$U_{split}\ket{101} = \frac{1}{\sqrt{2}}(-\ket{011} + i\ket{101})$}\\
\scalebox{0.95}{$U_{split}\ket{110} = \frac{1}{\sqrt{2}}(i\ket{011} - \ket{101})$}
\end{IEEEeqnarray}
\end{multicols}
\end{subequations}
The zero and three piece subspaces are trivially conserved by identity.

\subsection{Split Slide Unitary}\label{sec:SplitSlideUnitary}
The Split Slide unitary is a variation on the Split Jump unitary that takes into account the path a piece might take. When performing a Split Slide there are two paths to consider. We expand our basis to include two path control qubits: $p_1$ is the path from $s$ to $t_1$, and $p_2$ is the path from $s$ to $t_2$. We thus define the Split Slide unitary in the basis $\ket{p_2,p_1,t_2,t_1,s}$:
\begin{equation}\label{eq:SplitSlideUnitary}
		U_{split\_slide} = \begin{pmatrix}
		U_{split} & \underline{0} & \underline{0} & \underline{0}\\
		\underline{0} & U_{iSwap}(s,t_2) & \underline{0} & \underline{0}\\
		\underline{0} & \underline{0} & U_{iSwap}(s,t_1) & \underline{0}\\
		\underline{0} & \underline{0} & \underline{0} & I_{8x8}
	\end{pmatrix}_{32x32}
\end{equation}
We have chosen to have our move act as full iSwaps in the event that a single path is blocked. 

\subsection{Merge Jump Unitary}\label{sec:MergeJumpUnitary}
A natural extension of allowing players to split pieces is to allow them to merge pieces back together. With the Merge Jump unitary a player may undo a Split Jump. We define the Merge Jump operator to be the hermitian conjugate of the Split Jump operator. We must define our basis to be $\ket{s_1,s_2,t}$ if we want this unitary to act as the hermitian conjugate of a Split Jump on the basis $\ket{t_2,t_1,s}$ described above. Thus our Merge Jump unitary is:
\begin{equation}\label{eq:MergeJumpUnitary}
	U_{merge} = \begin{pmatrix}
		1 & 0 & 0 & 0 & 0 & 0 & 0 & 0\\
		0 & 0 & \frac{-i}{\sqrt{2}} & 0 & \frac{-i}{\sqrt{2}} & 0 & 0 & 0 \\
		0 & 0 & \frac{1}{\sqrt{2}} & 0 & \frac{-1}{\sqrt{2}} & 0 & 0 & 0 \\
		0 & 0 & 0 & 0 & 0 & 0 & -i & 0\\
		0 & -i & 0 & 0 & 0 & 0 & 0 & 0\\
		0 & 0 & 0 & \frac{-1}{\sqrt{2}} & 0 & \frac{-i}{\sqrt{2}} & 0 & 0\\
		0 & 0 & 0 & \frac{-i}{\sqrt{2}} & 0 & \frac{-1}{\sqrt{2}} & 0 & 0\\
		0 & 0 & 0 & 0 & 0 & 0 & 0 & 1\\
	\end{pmatrix}
\end{equation}
And again we see that the non-trivial effect on the basis states of the $\ket{s_1,s_2,t}$ Hilbert space conserves piece number.
\begin{subequations}
\begin{multicols}{2}
\noindent
\begin{IEEEeqnarray}{l}
\scalebox{0.95}{$U_{merge}\ket{001} = -i\ket{100}$}\\
\scalebox{0.95}{$U_{merge}\ket{010} = \frac{-1}{\sqrt{2}}(i\ket{001} - \ket{010})$}\\
\scalebox{0.95}{$U_{merge}\ket{100} = \frac{-1}{\sqrt{2}}(i\ket{001} + \ket{010})$}
\end{IEEEeqnarray}
\begin{IEEEeqnarray}{l}
\scalebox{0.95}{$U_{merge}\ket{011} = \frac{-1}{\sqrt{2}}(\ket{101} + i\ket{110})$}\\
\scalebox{0.95}{$U_{merge}\ket{101} = \frac{-1}{\sqrt{2}}(i\ket{101} +\ket{110})$}\\
\scalebox{0.95}{$U_{merge}\ket{110} = -i\ket{011}$}
\end{IEEEeqnarray}
\end{multicols}
\end{subequations}

\subsection{Merge Slide Unitary}\label{sec:MergeSlideUnitary}
As with the Split Jump, for some pieces we must consider the occupancy of the paths: $p_1$ from $s_1$ to $t$, and $p_2$ and from $s_2$ to $t$. In the basis $\ket{p_2,p_1,s_1,s_2,t}$ we define our Merge Slide Unitary:
\begin{equation}\label{eq:MergeSlideUnitary}
	U_{merge\_slide} = \begin{pmatrix}
		U_{mg} & \underline{0} & \underline{0} & \underline{0}\\
		\underline{0} & U^\dag_{iSwap}(s_2,t) & \underline{0} & \underline{0}\\
		\underline{0} & \underline{0} & U^\dag_{iSwap}(s_1,t) & \underline{0}\\
		\underline{0} & \underline{0} & \underline{0} & I_{8x8}
	\end{pmatrix}_{32x32}
\end{equation}

\section{No Double Occupancy Rule}\label{sec:NoDoubleOccupancyRule}
Before moving on to the design of the actual moves available to all of the pieces in the game, we need to discuss the No Double Occupancy Rule. Given that pieces can exist in superposition on a board it is possible that a move could result in a square being occupied by more than one type of piece.\footnote{More accurately, given our definition of state a move unitary could cause interference between different types of pieces, but then the classical piece type information is ambiguous. We may instead consider a Hilbert space that includes piece type, and then a move can be seen as placing two different pieces in the same square, i.e. Double Occupancy.} This leads to a number of complications, both in execution of the game and in visual representation of the board for a player. For these, and various other, reasons we have chosen to add the "No Double Occupancy" rule.

\begin{definition}{No Double Occupancy Rule:}
The no double occupancy rule states that at no point in the game do we allow a square to have a non-zero probability of being occupied by two or more pieces with different piece values.
\end{definition}

This language is chosen to allow for the interaction of like pieces. To enforce this rule we use a set of carefully designed projective measurements to ensure that the board is always in a state in which a move can't possibly lead to double occupancy. Whenever we perform a move we first look at the classical piece type information to determine whether double occupancy is a possible outcome. If it is, we perform a measurement that projects the state into a subspace where double occupancy will not occur. The general procedure for constructing measurement operators will be as follows:
\begin{enumerate}
	\item Consider the conditions that could lead to double occupancy after applying a move to a superposition of boards.
	\item Construct two sets of mutually exclusive basis states, $\mathcal{M}_0$ and $\mathcal{M}_1$ for the Hilbert space on which the move unitaries act. These sets should be constructed such that double occupancy will never occur for a move acting on any state that has elements in only one subset or the other. And $\mathcal{M}_0$ should hold basis states for which the move must do nothing, while $\mathcal{M}_1$ holds basis states for which the move may have some non-trivial effect (i.e. the piece attempting the move does in fact move).
	\item From $\mathcal{M}_0$ and $\mathcal{M}_1$ create two measurement operators, $M_0$ and $M_1$ that will be used to project the superposition into one of the two mutually exclusive subspaces.
	\item There will be some basis states left over that can never conflict with the No Double Occupancy rule (e.g. $\ket{000}$ for moves that act on 3 qubits). Complete $M_0$ and $M_1$ with the goal of being able to describe what the measurement does to a player in a single sentence.\footnote{This is a somewhat arbitrary requirement for completing the operators. Any choice of design goal will be equally valid. For example one may choose to try and maintain as much superposition as possible during measurement, or reduce the total quantum circuit area of the combined measurement and move.}
\end{enumerate}

\section{Quantum Chess Moves}\label{sec:QuantumChessMoves}
Now that our complete framework is in place we can detail the moves we allow in Quantum Chess. For each move (m) we will define its \text{possibility}. 
\begin{equation}
\mathcal{P}_m(\vec{v},\mathcal{F}) = \bigwedge\limits_{i}C_{i}(m,\vec{v},\mathcal{F})
\end{equation}
where $C_{m,i}(\vec{v},\mathcal{F})$ are a set of constraints on the classical information of a given state for move $m$ to be possible. We impose an additional condition to determine the \textit{legality} of a move.
\begin{definition}{Legal Move:}\label{def:LegalMove}
A move (m) is said to be legal if it satisfies all of its possibility constraints, i.e. $\mathcal{P}_m(\vec{v},\mathcal{F}) = true$, and it acts non-trivially on the occupancy superposition, i.e. $\ket{\psi_{\mathcal{B}}'} \neq \ket{\psi_{\mathcal{B}}}$ where $\ket{\psi_{\mathcal{B}}'}$ is the occupancy state after applying the move.
\end{definition}
We do not define a procedure for determining whether the superposition has changed. If executed on a quantum computer this would require some type of state tomography \cite{Tomography:1995}\cite{CompressedTomography:2010}\cite{ContinuousTomography:2009}. There are shortcuts one may take to make this determination when simulating on a classical computer. 

We can now outline the measuring and non-measuring movement procedures all moves will follow using pseudocode.\footnote{We use a mix of classical pseudocode and quantum pseudocode introduced in \cite{QuantumPseudocode:online}.} Our non-measuring move procedure is:
\begin{algorithm}{Execute Move:}\label{alg:ExecuteMove}
Consider a game with classical and quantum state $\{\ket{\psi_{\mathcal{B}}},\vec{v},\mathcal{F}\}$. Let $\underline{\psi_{\mathcal{B}}}$ be a quantum register initialized with state $\ket{\psi_{\mathcal{B}}}$. Let $U_m$ be the unitary(s) associated with a move (m). To apply $m$ to the game we use the following procedure.
\begin{algorithmic}
\Procedure{Apply}{$U_m,\underline{\psi_{\mathcal{B}}},\vec{v},\mathcal{F}$}
	\If{$\mathcal{P}_m(\vec{v},\mathcal{F}) = true$}
		\State $\underline{\psi_{\mathcal{B}}'} \gets U_m \underline{\psi_{\mathcal{B}}}$
		\If{$\underline{\psi_{\mathcal{B}}'} \neq \underline{\psi_{\mathcal{B}}}$}
			\State Update $\vec{v}$ to reflect the correct pieces.
			\State Update castling and e.p. flags in $\mathcal{F}$ appropriately.
			\State Update $\mathcal{F}_c$ to indicate change of player.
		\EndIf
	\EndIf
\EndProcedure
\end{algorithmic}
\end{algorithm}
This must be modified if we wish to include a measurement. The measurements presented here will always be two-outcome so it will be enough to supply $M_1$, as $M_0 = I - M_1$. Our algorithm thus becomes:
\begin{algorithm}{Execute Measuring Move:}\label{alg:ExecuteMeasuringMove}
Consider a game with state $\{\ket{\psi_{\mathcal{B}}},\vec{v},\mathcal{F}\}$. Let $\underline{\psi_{\mathcal{B}}}$ be a quantum register initialized with state $\ket{\psi_{\mathcal{B}}}$. Let $U_m$ and $M_1$ be the unitary(s) and measurement associated with a move (m). We define $Measure(M_1,\underline{a},\underline{\psi_{\mathcal{B}}})$ to be a subroutine that applies a quantum circuit to $\underline{\psi_{\mathcal{B}}}$ to encode the measurement $M_1$ into ancilla qubit $\underline{a}$, measures $\underline{a}$, and returns the result ($a$) as well as the newly projected state $\underline{\psi_{\mathcal{B}}'}$. To execute $m$ on the game we apply the following procedure.
\begin{algorithmic}
\Procedure{Apply}{$U_m,M_1,\underline{\psi_{\mathcal{B}}},\vec{v},\mathcal{F}$}
	\If{$\mathcal{P}_m(\vec{v},\mathcal{F}) = true$}
		\State $\underline{a} \gets \underline{0}$
		\State $(a,\underline{\psi_{\mathcal{B}}'}) \gets Measure(M_1,\underline{a},\underline{\psi_{\mathcal{B}}})$
		\If{a = 1}
			\State $\underline{\psi_{\mathcal{B}}'} \gets U_m \underline{\psi_{\mathcal{B}}'}$
		\EndIf
		\If{$\underline{\psi_{\mathcal{B}}'} \neq \underline{\psi_{\mathcal{B}}}$}
			\State Update $\vec{v}$ to reflect the correct pieces.
			\State Update castling and e.p. flags in $\mathcal{F}$ appropriately.
			\State Update $\mathcal{F}_c$ to indicate change of player.
		\EndIf
	\EndIf
\EndProcedure
\end{algorithmic}
\end{algorithm}
The following sections will detail the types of moves available to all pieces in the game. All pieces in Quantum Chess follow the same movement patterns as in standard Chess. For a full description of the patterns see Article 2 of the FIDE Laws of Chess\cite{FIDE:online}. Unlike in standard Chess, we do not define a notion of "check" for Quantum Chess. Therefore any rules that standard Chess applies to the legality of a King's moves when under check do not apply to Quantum Chess. It is perfectly legal to move one's King into a position where it is being attacked by an opponent's piece, or to castle through a position that is being attacked. 

Each section will define the possibility equation for the move being constructed from the perspective of the white player.\footnote{For the black player simply substitute capital FEN letters for lower case ones.} For this it will be helpful to define the notion of a valid move:
\begin{definition}{Valid:}
	A move is \textit{valid} if $s-t$ is a valid movement pattern for piece $v_s$ under the rules of standard Chess. In functional form:

	$valid(t,s,v_s) = True$, if $t$ is reachable from $s$ by piece $v_s$ under standard Chess rules.
\end{definition}
After defining a move's possibility equation the measurement operators $M_0$ and $M_1$ will be built if necessary. Finally the unitary procedure needed to update the occupancy superposition will be described.

\subsection{Standard Jump}\label{sec:StandardJump}
\begin{figure}[htbp]
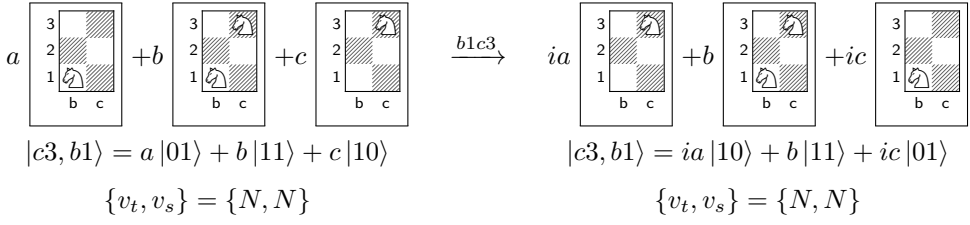

\captionsetup{width=.8\textwidth}
	\begin{centering}
	\def\arraystretch{1.5}
	\begin{tabular}{c c c}
		\hbox{
	    	$a$
		    \raisebox{-0.3\height}{\fbox{\chessboard[
			    setfen=8/8/8/8/8/8/8/1N6 w - - 0 1,
			    showmover=false,
			    printarea=b1-c3,
			    margintopwidth=0pt,
			    marginrightwidth=0pt,
			    marginleftwidth=8pt,
			    marginbottomwidth=10pt,
			    tinyboard
		    ]}} 
		    $+ b$
		    \raisebox{-0.3\height}{\fbox{\chessboard[
			    setfen=8/8/8/8/8/2N5/8/1N6 w - - 0 1,
			    showmover=false,
			    printarea=b1-c3,
			    margintopwidth=0pt,
			    marginrightwidth=0pt,
			    marginleftwidth=8pt,
			    marginbottomwidth=10pt,
			    tinyboard
		    ]}} 
		    $+ c$
		    \raisebox{-0.3\height}{\fbox{\chessboard[
			    setfen=8/8/8/8/8/2N5/8/8 w - - 0 1,
			    showmover=false,
			    printarea=b1-c3,
			    margintopwidth=0pt,
			    marginrightwidth=0pt,
			    marginleftwidth=8pt,
			    marginbottomwidth=10pt,
			    tinyboard
		    ]}} 
	    }
	    & $\xrightarrow{\text{$b1c3$}}$ &
	    \hbox{
	    	$ia$
		    \raisebox{-0.3\height}{\fbox{\chessboard[
			    setfen=8/8/8/8/8/2N5/8/8 w - - 0 1,
			    showmover=false,
			    printarea=b1-c3,
			    margintopwidth=0pt,
			    marginrightwidth=0pt,
			    marginleftwidth=8pt,
			    marginbottomwidth=10pt,
			    tinyboard
		    ]}} 
		    $+ b$
		    \raisebox{-0.3\height}{\fbox{\chessboard[
			    setfen=8/8/8/8/8/2N5/8/1N6 w - - 0 1,
			    showmover=false,
			    printarea=b1-c3,
			    margintopwidth=0pt,
			    marginrightwidth=0pt,
			    marginleftwidth=8pt,
			    marginbottomwidth=10pt,
			    tinyboard
		    ]}}
		    $+ ic$
		    \raisebox{-0.3\height}{\fbox{\chessboard[
			    setfen=8/8/8/8/8/8/8/1N6 w - - 0 1,
			    showmover=false,
			    printarea=b1-c3,
			    margintopwidth=0pt,
			    marginrightwidth=0pt,
			    marginleftwidth=8pt,
			    marginbottomwidth=10pt,
			    tinyboard
		    ]}} 
	    }
	    \\
	    \begin{tabular}{c}
	    $\ket{c3,b1} = a\ket{01} + b\ket{11} + c\ket{10}$ \\
	    $\{v_t,v_s\} = \{N,N\}$ 
	    \end{tabular}
	    & & 
	    \begin{tabular}{c}
	    $\ket{c3,b1} = ia\ket{10} + b\ket{11} + ic\ket{01}$ \\
	    $\{v_t,v_s\} = \{N,N\}$
	    \end{tabular}
	\end{tabular}
	\end{centering}
	\caption{\small Example of Standard Jump $b1c3$ acting on a superposition. The classical piece information remains unchanged, but the amplitudes of the basis states have changed.}
	\label{fig:StandardJump}
\end{figure}
\begin{figure}[ht]
\captionsetup{width=.8\textwidth}
\centering
\mbox{
	\Qcircuit @C=1em @R=1em{
		& \lstick{\ket{s}} 			&\qw& \multigate{1}{U_{jump}} & \qw \\
		& \lstick{\ket{t}} 			&\qw& \ghost{U_{jump}} 	   	& \qw
	}
}
\caption{\small Quantum circuit diagram for applying the Standard Jump move to the $source(s)$ $target(t)$ qubits from the quantum state $\ket{\psi_\mathcal{B}}$.}
\label{fig:StandardJumpCircuit}
\end{figure}
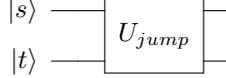

The Standard Jump is the Quantum Chess equivalent of the standard chess move for Knights and Kings (see figure \ref{fig:StandardJump}).\footnote{We may also use this option for any of the sliding pieces if they are moving to an adjacent square or if we know the path to be clear.} These pieces do not care about a path so we will make use of the Jump unitary (eq. \ref{eq:JumpUnitary}). This move attempts to fully swap a piece between the source(s) position and the target(t) position. We define the following possibility equation:
\begin{equation}\label{eq:StandardJumpPossibility}
	\mathcal{P}_{SJ}= (v_s \in \{N,K\}) \wedge valid(t,s,v_s) \wedge ((v_t = 0)\vee(v_t = v_s))
\end{equation}
This move will never result in Double Occupancy thus we can follow procedure \ref{alg:ExecuteMove} with $U_m = U_{jump}$. Figure \ref{fig:StandardJumpCircuit} shows a simple quantum circuit that applies the Standard Jump to the appropriate qubits for the occupancy superposition.

\subsection{Blocked Jump}\label{sec:BlockedJump}
\begin{figure}[htbp]
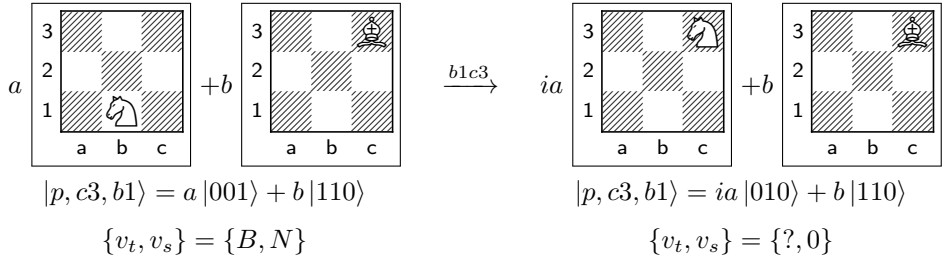

\captionsetup{width=.8\textwidth}
	\begin{centering}
	\def\arraystretch{1.5}
	\begin{tabular}{c c c}
		\hbox{
	    	$a$
		    \raisebox{-0.3\height}{\fbox{\chessboard[
			    setfen=8/8/8/8/8/8/8/1N6 w - - 0 1,
			    showmover=false,
			    printarea=a1-c3,
			    margintopwidth=0pt,
			    marginrightwidth=0pt,
			    marginleftwidth=8pt,
			    marginbottomwidth=10pt,
			    smallboard
		    ]}} 
		    $+ b$
		    \raisebox{-0.3\height}{\fbox{\chessboard[
			    setfen=8/8/8/8/8/2B5/8/8 w - - 0 1,
			    showmover=false,
			    printarea=a1-c3,
			    margintopwidth=0pt,
			    marginrightwidth=0pt,
			    marginleftwidth=8pt,
			    marginbottomwidth=10pt,
			    smallboard
		    ]}} 
	    }
	    & $\xrightarrow{\text{$b1c3$}}$ &
	    \hbox{
	    	$ia$
		    \raisebox{-0.3\height}{\fbox{\chessboard[
			    setfen=8/8/8/8/8/2N5/8/8 w - - 0 1,
			    showmover=false,
			    printarea=a1-c3,
			    margintopwidth=0pt,
			    marginrightwidth=0pt,
			    marginleftwidth=8pt,
			    marginbottomwidth=10pt,
			    smallboard
		    ]}} 
		    $+ b$
		    \raisebox{-0.3\height}{\fbox{\chessboard[
			    setfen=8/8/8/8/8/2B5/8/8 w - - 0 1,
			    showmover=false,
			    printarea=a1-c3,
			    margintopwidth=0pt,
			    marginrightwidth=0pt,
			    marginleftwidth=8pt,
			    marginbottomwidth=10pt,
			    smallboard
		    ]}}
	    }
	    \\
	    \begin{tabular}{c}
	    $\ket{p,c3,b1} = a\ket{001} + b\ket{110}$ \\
	    $\{v_t,v_s\} = \{B,N\}$ 
	    \end{tabular}
	    & & 
	    \begin{tabular}{c}
	    $\ket{p,c3,b1} = ia\ket{010} + b\ket{110}$ \\
	    $\{v_t,v_s\} = \{?,0\}$
	    \end{tabular}
	\end{tabular}
	\end{centering}
	\caption{\small Double Occupancy would occur if we attempted move b1c3 with the target (c3) occupancy encoded into the path (p) ancilla to act as a control for the swap. This must be prevented by applying a projective measurement before executing the move. }
	\label{fig:BlockedJump}
\end{figure}
\begin{figure}[ht]
\captionsetup{width=.8\textwidth}
\centering
\mbox{
	\Qcircuit @C=1em @R=1em{
		&							&	  &  			&\mbox{Measure}	&		 &&					&	\\
		& \lstick{\ket{s}} 			& \qw & \qw 		&\qw& \qw				 &\qw& \multigate{1}{U_{jump}} & \qw \\
		& \lstick{\ket{t}} 			& \qw & \ctrlo{1} 	&\qw& \qw				 &\qw& \ghost{U_{jump}} 	   	& \qw\\
		& \lstick{\ket{a}=\ket{0}} 	& \qw & \targ	  	&\qw& \measure{\mbox{a}} &\cw& \control \cw \cwx[-1] 
		\gategroup{3}{4}{4}{6}{0.7em}{--}
	}
}
\caption{\small Quantum circuit diagram for applying the Blocked Jump move to the source $(s)$ and target $(t)$ qubits from the quantum register $\underline{\psi_\mathcal{B}}$. First it encodes $M_1$ (eq. \ref{eq:BlockedJumpMeasurement}) into measurement ancilla ($a$) and then conditionally applies $U_{jump}$ (eq. \ref{eq:JumpUnitary}) if the measurement outcome is 1.}
\label{fig:BlockedJumpMeasureCircuit}
\end{figure}
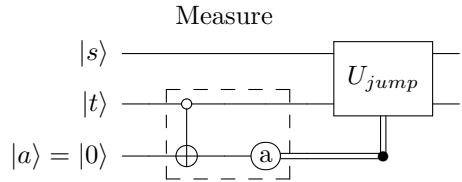

The Blocked variant of the Jump has the following possibility equation:
\begin{equation}\label{eq:BlockedJumpPossibility}
	\mathcal{P}_{BJ} = (v_s \in\{N,K\}) \wedge valid(t,s,v_s) \wedge (v_t \neq v_s) \wedge (v_t \in \{P,N,B,R,Q,K\})
\end{equation}
If this move is legal it will result in double occupancy (see figure \ref{fig:BlockedJump}). If a controlled operation were used to ensure the move acts only on the knight then the target square would be occupied by both a white knight and a white bishop after execution. Our state notation does not even support such a case. The basis states $\ket{01}$ and $\ket{10}$ are mutually exclusive for this move, they cannot both exist in the state if we want our move to never produce double occupancy. We find the following complete subsets of mutually exclusive basis states for the $\ket{t,s}$ Hilbert space: 
\begin{equation*}
\mathcal{M}_0=\{\ket{11},\ket{10}\} , \mathcal{M}_1=\{\ket{01}\}
\end{equation*}
From these two subsets we can construct two measurement operators, $M_0$ and $M_1$, with $\ket{00}$ included in either. Here we choose the following construction:
\begin{subequations}
\begin{IEEEeqnarray}{rCl}
M_0 &=& \ket{10}\bra{10} + \ket{11}\bra{11}\\
M_1 &=& \ket{00}\bra{00} + \ket{01}\bra{01} \label{eq:BlockedJumpMeasurement}
\end{IEEEeqnarray}
\end{subequations}
This can be reduced to a simple measurement of the target qubit in the computational basis. We can thus apply procedure \ref{alg:ExecuteMeasuringMove} with $U_m = U_{jump}$ (eq. \ref{eq:JumpUnitary}). Figure \ref{fig:BlockedJumpMeasureCircuit} shows a simple quantum circuit for applying the measurement and conditional move operator.

\subsection{Capture Jump}\label{sec:CaptureJump}
\begin{figure}[htbp]
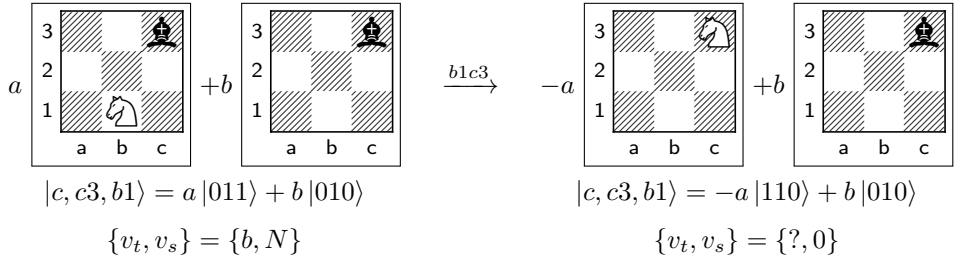

\captionsetup{width=.8\textwidth}
	\begin{centering}
	\def\arraystretch{1.5}
	\begin{tabular}{c c c}
		\hbox{
	    	$a$
		    \raisebox{-0.3\height}{\fbox{\chessboard[
			    setfen=8/8/8/8/8/2b5/8/1N6 w - - 0 1,
			    showmover=false,
			    printarea=a1-c3,
			    margintopwidth=0pt,
			    marginrightwidth=0pt,
			    marginleftwidth=8pt,
			    marginbottomwidth=10pt,
			    smallboard
		    ]}} 
		    $+ b$
		    \raisebox{-0.3\height}{\fbox{\chessboard[
			    setfen=8/8/8/8/8/2b5/8/8 w - - 0 1,
			    showmover=false,
			    printarea=a1-c3,
			    margintopwidth=0pt,
			    marginrightwidth=0pt,
			    marginleftwidth=8pt,
			    marginbottomwidth=10pt,
			    smallboard
		    ]}} 
	    }
	    & $\xrightarrow{\text{$b1c3$}}$ &
	    \hbox{
	    	$-a$
		    \raisebox{-0.3\height}{\fbox{\chessboard[
			    setfen=8/8/8/8/8/2N5/8/8 w - - 0 1,
			    showmover=false,
			    printarea=a1-c3,
			    margintopwidth=0pt,
			    marginrightwidth=0pt,
			    marginleftwidth=8pt,
			    marginbottomwidth=10pt,
			    smallboard
		    ]}} 
		    $+ b$
		    \raisebox{-0.3\height}{\fbox{\chessboard[
			    setfen=8/8/8/8/8/2b5/8/8 w - - 0 1,
			    showmover=false,
			    printarea=a1-c3,
			    margintopwidth=0pt,
			    marginrightwidth=0pt,
			    marginleftwidth=8pt,
			    marginbottomwidth=10pt,
			    smallboard
		    ]}}
	    }
	    \\
	    \begin{tabular}{c}
	    $\ket{c,c3,b1} = a\ket{011} + b\ket{010}$ \\
	    $\{v_t,v_s\} = \{b,N\}$ 
	    \end{tabular}
	    & & 
	    \begin{tabular}{c}
	    $\ket{c,c3,b1} = -a\ket{110} + b\ket{010}$ \\
	    $\{v_t,v_s\} = \{?,0\}$
	    \end{tabular}
	\end{tabular}
	\end{centering}
	\caption{\small Double Occupancy would occur if we attempted Capture Jump b1c3. This must be prevented by applying a projective measurement before executing the move. A "captured" ancilla (c) is added to hold the captured piece.}
	\label{fig:CaptureJump}
\end{figure}
\begin{figure}[ht]
\captionsetup{width=.8\textwidth}
\centering
\mbox{
	\Qcircuit @C=1em @R=1em{
	&							   &   &  		 &\mbox{Measure}&	&   &   &					 	&& 			 	&&\\
	&\lstick{\ket{s}} 	   &\qw&\ctrl{3}&\qw&\qw			&\qw&\qw&\qw 					&\qw&\multigate{1}{U_{jump}}&\qw&\qw\\
	&\lstick{\ket{t}} 	   &\qw&\qw  &\qw&\qw				&\qw&\qw&\multigate{1}{U_{jump}}&\qw&\ghost{U_{jump}} 		&\qw&\qw\\
	&\lstick{\ket{c}=\ket{0}}&\qw&\qw	 &\qw&\qw 				&\qw&\qw&\ghost{U_{jump}}		&\qw&\qw					&\qw&\qw\\
	&\lstick{\ket{a}=\ket{0}}&\qw&\targ 	&\qw&\measure{\mbox{a}}&\cw&\cw&\control \cw \cwx[-1]  &\cw&\control \cw \cwx[-2]
	\gategroup{2}{4}{5}{6}{0.7em}{--}
	}
}
\caption{\small Quantum circuit diagram for applying the Capture Jump move to the source ($s$) and target ($t$) qubits from the quantum register $\underline{\psi_\mathcal{B}}$. A captured ancilla ($c$) is added to hold the captured piece. $M_1$ (eq. \ref{eq:CaptureJumpMeasurement}) is encoded into measurement ancilla ($a$) and two $U_{jump}$ (eq. \ref{eq:JumpUnitary}) operations are conditionally applied based on the measurement outcome being 1.}
\label{fig:CaptureJumpMeasureCircuit}
\end{figure}
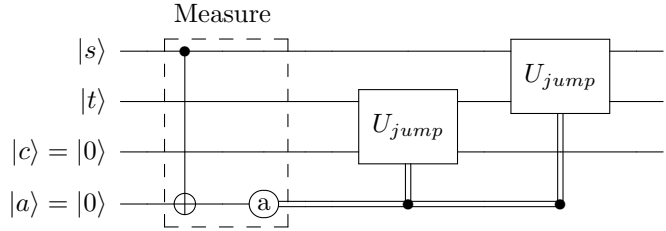

The Capture variant of the Jump has the following possibility equation:
\begin{equation}\label{eq:CaptureJumpPossibility}
	\mathcal{P}_{CJ} = (v_s \in\{N,K\}) \wedge valid(t,s,v_s) \wedge (v_t \in \{p,n,b,r,q,k\})
\end{equation}
This move could result in double occupancy (see figure \ref{fig:CaptureJump}). Given this example we can deduce that the basis states $\ket{11}$ and $\ket{10}$ are mutually exclusive. The complete subsets of mutually exclusive basis states for the $\ket{t,s}$ Hilbert space are: 
\begin{equation*}
\mathcal{M}_0=\{\ket{10}\} , \mathcal{M}_1=\{\ket{01},\ket{11}\}
\end{equation*}
We choose the following measurement operators:
\begin{subequations}
\begin{IEEEeqnarray}{rCl}
M_0 &=& \ket{10}\bra{10} + \ket{00}\bra{00}\\
M_1 &=& \ket{01}\bra{01} + \ket{11}\bra{11}\label{eq:CaptureJumpMeasurement}
\end{IEEEeqnarray}
\end{subequations}
This can be reduced to a simple measurement of the source qubit in the computational basis. In order to perform capture in a unitary way we must expand the Hilbert space to include a "captured" ancilla (\underline{c}) initialized in state $\ket{0}$. We can then apply procedure \ref{alg:ExecuteMeasuringMove} with $U_m = U_{jump}(s,t)U_{jump}(t,c)$. The circuit in figure \ref{fig:CaptureJumpMeasureCircuit} illustrates the move execution including the encoding of $M_1$ into an ancilla qubit $\underline{a}$ for the $Measure$ subroutine.

\subsection{Standard Slide}\label{sec:StandardSlide}
\begin{figure}[htbp]
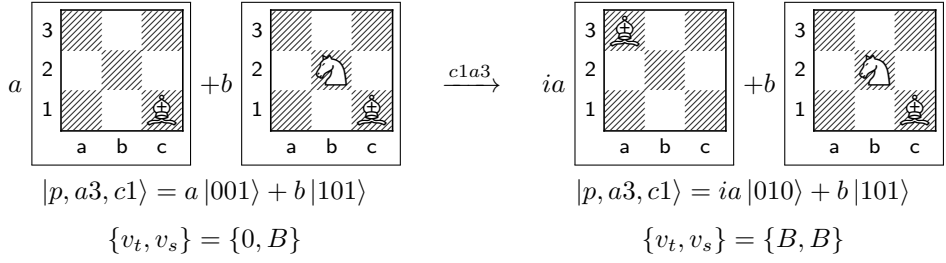

\captionsetup{width=.8\textwidth}
	\begin{centering}
	\def\arraystretch{1.5}
	\begin{tabular}{c c c}
		\hbox{
	    	$a$
		    \raisebox{-0.3\height}{\fbox{\chessboard[
			    setfen=8/8/8/8/8/8/8/2B5 w - - 0 1,
			    showmover=false,
			    printarea=a1-c3,
			    margintopwidth=0pt,
			    marginrightwidth=0pt,
			    marginleftwidth=8pt,
			    marginbottomwidth=10pt,
			    smallboard
		    ]}} 
		    $+ b$
		    \raisebox{-0.3\height}{\fbox{\chessboard[
			    setfen=8/8/8/8/8/8/1N6/2B5 w - - 0 1,
			    showmover=false,
			    printarea=a1-c3,
			    margintopwidth=0pt,
			    marginrightwidth=0pt,
			    marginleftwidth=8pt,
			    marginbottomwidth=10pt,
			    smallboard
		    ]}}
	    }
	    & $\xrightarrow{\text{$c1a3$}}$ &
	    \hbox{
	    	$ia$
		    \raisebox{-0.3\height}{\fbox{\chessboard[
			    setfen=8/8/8/8/8/B7/8/8 w - - 0 1,
			    showmover=false,
			    printarea=a1-c3,
			    margintopwidth=0pt,
			    marginrightwidth=0pt,
			    marginleftwidth=8pt,
			    marginbottomwidth=10pt,
			    smallboard
		    ]}} 
		    $+ b$
		    \raisebox{-0.3\height}{\fbox{\chessboard[
			    setfen=8/8/8/8/8/8/1N6/2B5 w - - 0 1,
			    showmover=false,
			    printarea=a1-c3,
			    margintopwidth=0pt,
			    marginrightwidth=0pt,
			    marginleftwidth=8pt,
			    marginbottomwidth=10pt,
			    smallboard
		    ]}}
	    }
	    \\
	    \begin{tabular}{c}
	    $\ket{p,a3,c1} = a\ket{001} + b\ket{101}$ \\
	    $\{v_t,v_s\} = \{0,B\}$ 
	    \end{tabular}
	    & & 
	    \begin{tabular}{c}
	    $\ket{p,a3,c1} = ia\ket{010} + b\ket{101}$ \\
	    $\{v_t,v_s\} = \{B,B\}$
	    \end{tabular}
	\end{tabular}
	\end{centering}
	\caption{\small Example of Standard Slide $c1a3$ acting on a superposition. The move results in the bishop being entangled with the knight.}
	\label{fig:StandardSlide}
\end{figure}
\begin{figure}[ht]
\captionsetup{width=.8\textwidth}
\centering
\subcaptionbox{\label{fig:FullStandardSlideCircuit}}
{
	\makebox[0.5\columnwidth]{
		\Qcircuit @C=1em @R=1em{
			& \lstick{\ket{s}} 			& \qw &\qw		 & \multigate{2}{U_{slide}} & \qw \\
			& \lstick{\ket{t}} 			& \qw &\qw		 & \ghost{U_{slide}} 	   	& \qw\\
			& \lstick{\ket{p} = \ket{1}}& \qw &\targ 	 & \ghost{U_{slide}}			& \qw\\
			& \lstick{\ket{p_1,...,p_n}}&/^n \qw &\ctrlo{-1}& \qw 	   	& \qw
		}
	}
}
\hfill
\subcaptionbox{\label{fig:SimpleStandardSlideCircuit}}
{
	\makebox[0.45\columnwidth]{
		\Qcircuit @C=1em @R=1em{
			& \lstick{\ket{s}} 			& \qw & \multigate{1}{U_{Jump}} & \qw \\
			& \lstick{\ket{t}} 			& \qw & \ghost{U_{Jump}} 	   	& \qw\\
			& \lstick{\ket{p_1,...,p_n}}&/^n \qw &\ctrlo{-1}& \qw 	   	& \qw
		}
	}
}
\caption{\small Quantum circuit diagrams for applying the Standard Slide move to the $source(s)$ $target(t)$ qubits with $n$ squares in the path $\{p_1,...,p_n\}$. In figure (a), a path ancilla will be in state $\ket{0}$ if no square in the path squares is occupied. This is in keeping with the definition of the Slide Unitary \ref{eq:SlideUnitary} acting on the $\ket{p,t,s}$ basis. Figure (b) shows a simplified circuit that has the same output with the slide operation converted to a zero-controlled $U_{jump}$.}
\label{fig:StandardSlideCircuits}
\end{figure}
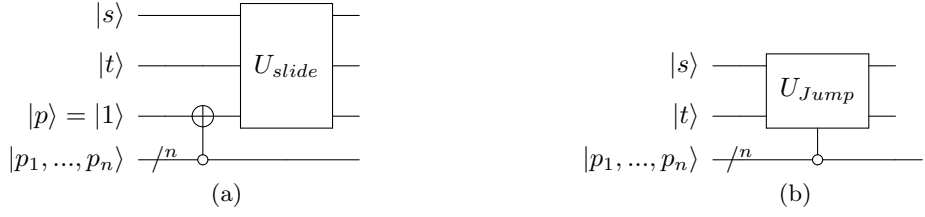

The Standard Slide is the Quantum Chess equivalent of the standard chess move for bishops, rooks, and queens (see figure \ref{fig:StandardSlide}). These pieces slide along a path so we must consider the occupancy of the squares between source and target. We introduce a "path" ancilla (p) and make use of the Slide unitary (eq. \ref{eq:SlideUnitary}). The Standard Slide possibility equation takes almost the same form as that of the Standard Jump:
\begin{equation}\label{eq:StandardSlidePossibility}
	\mathcal{P}_{SS}= (v_s \in \{B,R,Q\}) \wedge valid(t,s,v_s) \wedge ((v_t = 0)\vee(v_t = v_s))
\end{equation}
This move will never result in Double Occupancy thus we can follow procedure \ref{alg:ExecuteMove} with $U_m = U_{slide}$. Figure \ref{fig:FullStandardSlideCircuit} shows a quantum circuit that applies the Standard Slide where there are $n$ squares in the path.

\subsection{Blocked Slide}\label{sec:BlockedSlide}
\begin{figure}[htbp]
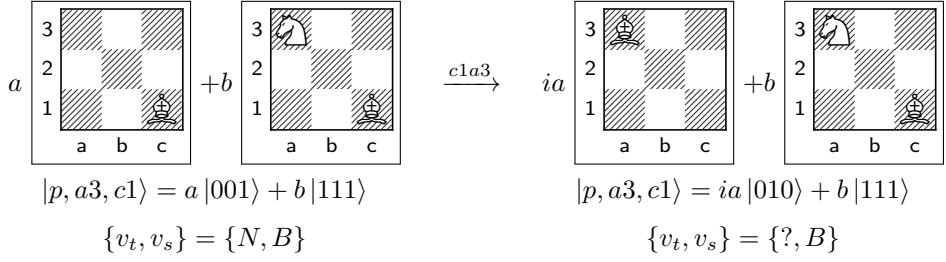

\captionsetup{width=.8\textwidth}
	\begin{centering}
	\def\arraystretch{1.5}
	\begin{tabular}{c c c}
		\hbox{
	    	$a$
		    \raisebox{-0.3\height}{\fbox{\chessboard[
			    setfen=8/8/8/8/8/8/8/2B5 w - - 0 1,
			    showmover=false,
			    printarea=a1-c3,
			    margintopwidth=0pt,
			    marginrightwidth=0pt,
			    marginleftwidth=8pt,
			    marginbottomwidth=10pt,
			    smallboard
		    ]}} 
		    $+ b$
		    \raisebox{-0.3\height}{\fbox{\chessboard[
			    setfen=8/8/8/8/8/N7/8/2B5 w - - 0 1,
			    showmover=false,
			    printarea=a1-c3,
			    margintopwidth=0pt,
			    marginrightwidth=0pt,
			    marginleftwidth=8pt,
			    marginbottomwidth=10pt,
			    smallboard
		    ]}}
	    }
	    & $\xrightarrow{\text{$c1a3$}}$ &
	    \hbox{
	    	$ia$
		    \raisebox{-0.3\height}{\fbox{\chessboard[
			    setfen=8/8/8/8/8/B7/8/8 w - - 0 1,
			    showmover=false,
			    printarea=a1-c3,
			    margintopwidth=0pt,
			    marginrightwidth=0pt,
			    marginleftwidth=8pt,
			    marginbottomwidth=10pt,
			    smallboard
		    ]}} 
		    $+ b$
		    \raisebox{-0.3\height}{\fbox{\chessboard[
			    setfen=8/8/8/8/8/N7/8/2B5 w - - 0 1,
			    showmover=false,
			    printarea=a1-c3,
			    margintopwidth=0pt,
			    marginrightwidth=0pt,
			    marginleftwidth=8pt,
			    marginbottomwidth=10pt,
			    smallboard
		    ]}}
	    }
	    \\
	    \begin{tabular}{c}
	    $\ket{p,a3,c1} = a\ket{001} + b\ket{111}$ \\
	    $\{v_t,v_s\} = \{N,B\}$ 
	    \end{tabular}
	    & & 
	    \begin{tabular}{c}
	    $\ket{p,a3,c1} = ia\ket{010} + b\ket{111}$ \\
	    $\{v_t,v_s\} = \{?,B\}$
	    \end{tabular}
	\end{tabular}
	\end{centering}
	\caption{\small Example of Blocked Slide $c1a3$ acting on a superposition. The occupancy of the target, a3, is encoded in the path control qubit. The move would result in Double Occupancy.}
	\label{fig:BlockedSlide}
\end{figure}
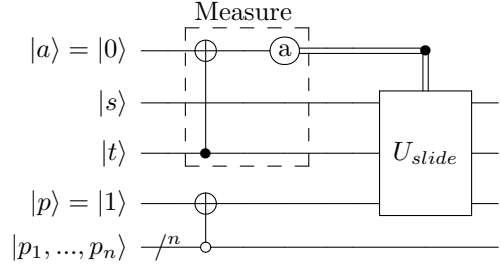
\begin{figure}[ht]
\captionsetup{width=.8\textwidth}
\centering
	\mbox{
		\Qcircuit @C=1em @R=1em{
			&						   &   &		  &	\mbox{Measure}				&	&&&&\\
			&\lstick{\ket{a}=\ket{0}}  &\qw&\targ 	  &\qw&\measure{\mbox{a}}&\cw&\cw&\control \cw \cwx[1]	  &\\
			&\lstick{\ket{s}} 		   &\qw&\qw		  &\qw&\qw               &\qw&\qw&\multigate{2}{U_{slide}}&\qw \\
			&\lstick{\ket{t}} 		   &\qw&\ctrl{-2} &\qw&\qw               &\qw&\qw&\ghost{U_{slide}} 	  &\qw\\
			&\lstick{\ket{p} = \ket{1}}&\qw&\targ 	  &\qw&\qw               &\qw&\qw&\ghost{U_{slide}}		  &\qw\\
			&\lstick{\ket{p_1,...,p_n}}&/^n\qw&\ctrlo{-1}&\qw&\qw               &\qw&\qw&\qw&\qw
			\gategroup{2}{4}{4}{6}{0.7em}{--}
		}
	}
\caption{\small Quantum circuit diagram for applying the Blocked Slide move to the $source(s)$ $target(t)$. The line $p_i$ represents the n qubits $\{p_1,...,p_n\}$  in the path. The state of $\ket{p}$ will remain $\ket{1}$ if \textit{any} square in the path is blocked.}
\label{fig:BlockedSlideCircuit}
\end{figure}

Here we outline the Blocked variant of a Slide move. The possibility equation for the Blocked Slide takes almost the same form as that of the Blocked Jump:
\begin{equation}\label{eq:BlockedSlidePossibility}
	\mathcal{P}_{BS} = (v_s \in\{B,R,Q\}) \wedge valid(t,s,v_s) \wedge (v_t \neq v_s) \wedge (v_t \in \{P,N,B,R,Q,K\})
\end{equation}
This move will result in double occupancy (see figure \ref{fig:BlockedSlide}). We find the following mutually exclusive subsets of basis states for the $\ket{p,t,s}$ Hilbert space: 
\begin{equation*}
\mathcal{M}_0=\{\ket{010},\ket{011},\ket{110},\ket{111}\} , \mathcal{M}_1=\{\ket{001}\}
\end{equation*}
Given the freedom of choice for the other three basis states, $\ket{000}$, $\ket{100}$, and $\ket{101}$ in either measurement operator we construct the following measurement operators.
\begin{subequations}
\begin{IEEEeqnarray}{rCl}
M_0 &=& \ket{010}\bra{010} + \ket{011}\bra{011} + \ket{110}\bra{110} + \ket{111}\bra{111}\\
M_1 &=& \ket{000}\bra{000} + \ket{001}\bra{001} + \ket{100}\bra{100} + \ket{101}\bra{101}\label{eq:BlockedSlideMeasurement}
\end{IEEEeqnarray}
\end{subequations}
Result $M_0$ means no move occurs because the target square is occupied. This can be reduced to a simple measurement of the target qubit in the computational basis. We can thus apply procedure \ref{alg:ExecuteMeasuringMove} with $U_m = U_{slide}$ (eq. \ref{eq:SlideUnitary}). Figure \ref{fig:BlockedSlideCircuit} shows a quantum circuit for applying the measurement and conditional move operator.

\subsection{Capture Slide}\label{sec:CaptureSlide}
\begin{figure}[htbp]
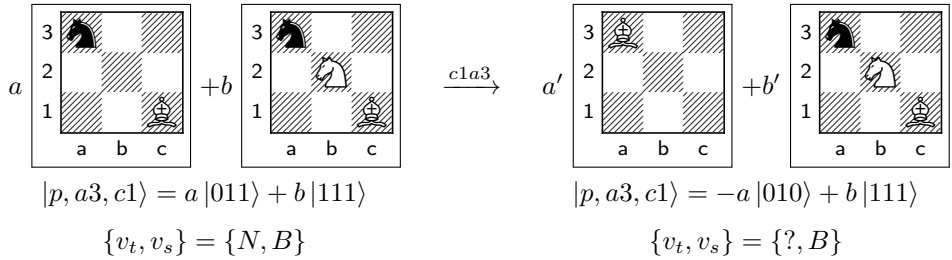

\captionsetup{width=.8\textwidth}
	\begin{centering}
	\def\arraystretch{1.5}
	\begin{tabular}{c c c}
		\hbox{
	    	$a$
		    \raisebox{-0.3\height}{\fbox{\chessboard[
			    setfen=8/8/8/8/8/n7/8/2B5 w - - 0 1,
			    showmover=false,
			    printarea=a1-c3,
			    margintopwidth=0pt,
			    marginrightwidth=0pt,
			    marginleftwidth=8pt,
			    marginbottomwidth=10pt,
			    smallboard
		    ]}} 
		    $+ b$
		    \raisebox{-0.3\height}{\fbox{\chessboard[
			    setfen=8/8/8/8/8/n7/1N6/2B5 w - - 0 1,
			    showmover=false,
			    printarea=a1-c3,
			    margintopwidth=0pt,
			    marginrightwidth=0pt,
			    marginleftwidth=8pt,
			    marginbottomwidth=10pt,
			    smallboard
		    ]}}
	    }
	    & $\xrightarrow{\text{$c1a3$}}$ &
	    \hbox{
	    	$a'$
		    \raisebox{-0.3\height}{\fbox{\chessboard[
			    setfen=8/8/8/8/8/B7/8/8 w - - 0 1,
			    showmover=false,
			    printarea=a1-c3,
			    margintopwidth=0pt,
			    marginrightwidth=0pt,
			    marginleftwidth=8pt,
			    marginbottomwidth=10pt,
			    smallboard
		    ]}} 
		    $+ b'$
		    \raisebox{-0.3\height}{\fbox{\chessboard[
			    setfen=8/8/8/8/8/n7/1N6/2B5 w - - 0 1,
			    showmover=false,
			    printarea=a1-c3,
			    margintopwidth=0pt,
			    marginrightwidth=0pt,
			    marginleftwidth=8pt,
			    marginbottomwidth=10pt,
			    smallboard
		    ]}}
	    }
	    \\
	    \begin{tabular}{c}
	    $\ket{p,a3,c1} = a\ket{011} + b\ket{111}$ \\
	    $\{v_t,v_s\} = \{N,B\}$ 
	    \end{tabular}
	    & & 
	    \begin{tabular}{c}
	    $\ket{p,a3,c1} = -a\ket{010} + b\ket{111}$ \\
	    $\{v_t,v_s\} = \{?,B\}$
	    \end{tabular}
	\end{tabular}
	\end{centering}
	\caption{\small Example of Capture Slide $c1a3$ acting on a superposition and resulting in Double Occupancy.}
	\label{fig:CaptureSlide}
\end{figure}
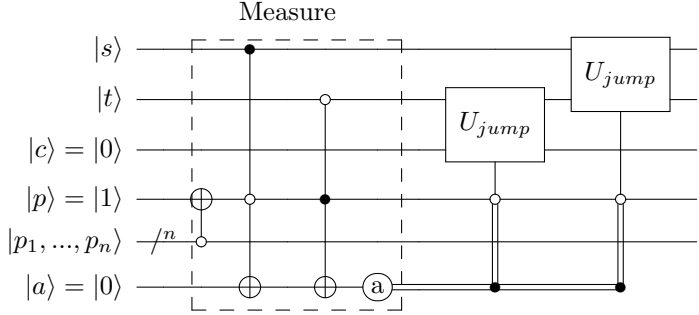
\begin{figure}[ht]
\captionsetup{width=.8\textwidth}
\centering
\mbox{
	\Qcircuit @C=1em @R=1em{
	&						   &   &		     &          &\mbox{Measure}&		     &	&&&&\\
	&\lstick{\ket{s}} 		   &\qw&\qw          &\ctrl{3}	&\qw&\qw	&\qw&\qw&\qw&\multigate{1}{U_{jump}}&\qw \\
	&\lstick{\ket{t}} 		   &\qw&\qw    	     &\qw       &\qw&\ctrlo{2} &\qw&\qw&\multigate{1}{U_{jump}}&\ghost{U_{jump}}&\qw\\
	&\lstick{\ket{c}=\ket{0}}  &\qw&\qw 	     &\qw       &\qw&\qw        &\qw&\qw&\ghost{U_{jump}}&\qw&\qw\\
	&\lstick{\ket{p} = \ket{1}}&\qw&\targ 	     &\ctrlo{2}&\qw&\ctrl{2}    &\qw&\qw&\ctrlo{-1}&\ctrlo{-2}		  &\qw\\
	&\lstick{\ket{p_1,...,p_n}}&/^n\qw&\ctrlo{-1}&\qw       &\qw&\qw                     &\qw             &\qw&\qw&\qw&\qw\\
	&\lstick{\ket{a}=\ket{0}}  &\qw&\qw          &\targ     &\qw&\targ  &\measure{\mbox{a}}&\cw&\control \cw \cwx[-2]&\control \cw \cwx[-2]
	\gategroup{2}{4}{7}{8}{0.7em}{--}
	}
}
\caption{\small Quantum circuit diagram for applying the Capture Slide move to the $source(s)$ $target(t)$. The line $p_i$ represents the n qubits $\{p_1,...,p_n\}$ in the path. The state of $\ket{p}$ will remain $\ket{1}$ if \textit{any} square in the path is blocked. Our measurement ancilla $\ket{a}$ will be in state $\ket{1}$ if $\ket{p}=\ket{0}$ and $\ket{s}=\ket{1}$, or if $\ket{p}=\ket{1}$ and $\ket{t}=\ket{0}$. Two zero-controlled $U_{jump}$ are applied if we see measurement outcome 1.}
\label{fig:CaptureSlideCircuit}
\end{figure}

Here we outline the Capture variant of a Slide move. The possibility equation for the Capture Slide takes almost the same form as that of the Capture Jump:
\begin{equation}\label{eq:CaptureSlidePossibility}
	\mathcal{P}_{CS} = (v_s \in\{B,R,Q\}) \wedge valid(t,s,v_s)\wedge (v_t \in \{p,n,b,r,q,k\})
\end{equation}
This move could result in double occupancy (see figure \ref{fig:CaptureSlide}). Given this example we can deduce that the basis states $\ket{011}$ and $\ket{111}$ are mutually exclusive. We find the following complete subsets of mutually exclusive basis states for the $\ket{p,t,s}$ Hilbert space: 
\begin{equation*}
\mathcal{M}_0=\{\ket{010},\ket{110},\ket{111}\} , \mathcal{M}_1=\{\ket{001},\ket{011}\}
\end{equation*}
With these two subsets we construct our measurement operators, $M_0$ and $M_1$.
\begin{subequations}
\begin{IEEEeqnarray}{rCl}
M_0 &=& \ket{000}\bra{000} + \ket{010}\bra{010} + \ket{110}\bra{110} + \ket{111}\bra{111}\\
M_1 &=& \ket{001}\bra{001} + \ket{011}\bra{011} + \ket{100}\bra{100} + \ket{101}\bra{101} \label{eq:CaptureSlideM1}
\end{IEEEeqnarray}
\end{subequations}
Result $M_1$ means either the path is not blocked and the source is occupied, or the path is blocked but the target is unoccupied. Adding a "captured" ancilla (\underline{c}) initialized in state $\ket{0}$ let's us maintain unitarity. We can then apply procedure \ref{alg:ExecuteMeasuringMove} with $U_m = U_{slide}(s,t)U_{slide}(t,c)$. The circuit in figure \ref{fig:CaptureSlideCircuit} illustrates the move execution including the encoding of $M_1$ into an ancilla qubit $\underline{a}$ for the $Measure$ subroutine. Note each $U_{slide}$ has been decomposed into a zero-controlled $U_{jump}$.

\subsection{Split Jump}\label{sec:SplitJump}
\begin{figure}[htbp]
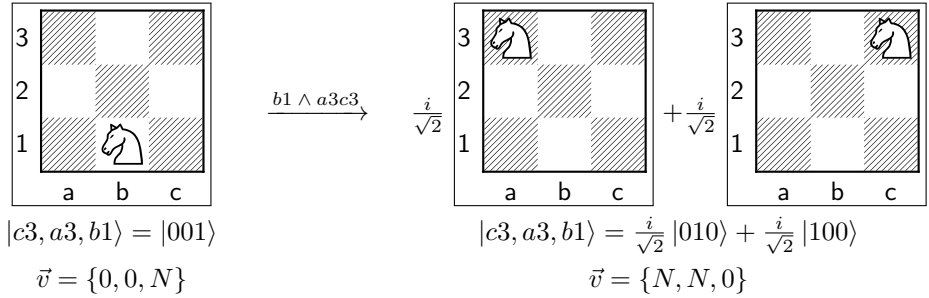

\captionsetup{width=.8\textwidth}
\def\arraystretch{1.5}
\begin{tabular}{c c c}
	\raisebox{-0.3\height}{\fbox{\chessboard[
	    setfen=8/8/8/8/8/8/8/1N6 w - - 0 1,
	    showmover=false,
	    printarea=a1-c3,
	    margintopwidth=0pt,
	    marginrightwidth=0pt,
	    marginleftwidth=8pt,
	    marginbottomwidth=10pt,
    ]}}
    & $\xrightarrow{\text{$b1\land a3c3$}}$ &
    \hbox{
    	$\frac{i}{\sqrt{2}}$
	    \raisebox{-0.3\height}{\fbox{\chessboard[
		    setfen=8/8/8/8/8/N7/8/8 w - - 0 1,
		    showmover=false,
		    printarea=a1-c3,
		    margintopwidth=0pt,
		    marginrightwidth=0pt,
		    marginleftwidth=8pt,
		    marginbottomwidth=10pt
	    ]}} 
	    $+ \frac{i}{\sqrt{2}}$
	    \raisebox{-0.3\height}{\fbox{\chessboard[
		    setfen=8/8/8/8/8/2N5/8/8 w - - 0 1,
		    showmover=false,
		    printarea=a1-c3,
		    margintopwidth=0pt,
		    marginrightwidth=0pt,
		    marginleftwidth=8pt,
		    marginbottomwidth=10pt
	    ]}}
    }
    \\
    \begin{tabular}{c}
    $\ket{c3,a3,b1} = \ket{001}$ \\
    $\vec{v} = \{0,0,N\}$ 
    \end{tabular}
    & & 
    \begin{tabular}{c}
    $\ket{c3,a3,b1} = \frac{i}{\sqrt{2}}\ket{010} + \frac{i}{\sqrt{2}}\ket{100}$ \\
    $\vec{v} = \{N,N,0\}$
    \end{tabular}
\end{tabular}
\caption{\small Split Jump $b1\land a3c3$ acting on a knight on a single board.}
\label{fig:SplitJump}
\end{figure}
\begin{figure}[ht]
\captionsetup{width=.8\textwidth}
\centering
\subcaptionbox{\label{fig:SplitJumpCircuit}}
{
	\makebox[0.45\columnwidth]{
		\Qcircuit @C=1em @R=1em{
			& \lstick{\ket{t1}} &\multigate{2}{U_{Split}} &\qw \\
			& \lstick{\ket{s}}&\ghost{U_{Split}} 	   	& \qw\\
			& \lstick{\ket{t2}}&\ghost{U_{Split}} 						& \qw 	   
		}
	}
}
\hfill
\subcaptionbox{\label{fig:SplitJumpDecomposedCircuit}}
{
	\makebox[0.45\columnwidth]{
		\Qcircuit @C=1em @R=1em{
			& \lstick{\ket{t1}} &\multigate{1}{U_{\sqrt{iSwap}}} & \qw &\qw \\
			& \lstick{\ket{s}}&\ghost{U_{\sqrt{iSwap}}} 	   	&\multigate{1}{U_{iSwap}}& \qw\\
			& \lstick{\ket{t2}}&\qw								&\ghost{U_{iSwap}}& \qw 	   
		}
	}
}
\caption{\small Quantum circuit diagrams for applying the Split Jump move to the source $(s)$, target $(t1)$, and target 2 $(t2)$ qubits. In figure (a) The Split unitary is applied as a three-qubit gate. Figure (b) shows its decomposition into two gates, the $\sqrt{iSwap}$ from eq. \ref{eq:iSwapRoot}, and the $iSwap$ from eq. \ref{eq:iSwap}.}
\label{fig:SplitJumpCircuits}
\end{figure}
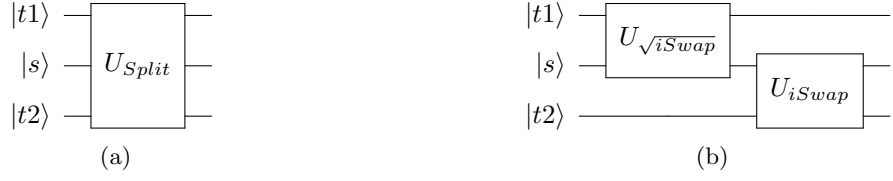

The Split Jump is the move by which the player creates superposition with Knights and Kings (see figure \ref{fig:SplitJump}). The move acts on three squares: source(s), target 1 (t1), and target 2 (t2). We define the following possibility equation:
\begin{equation}\label{eq:SplitJumpPossibility}
\begin{split}
	\mathcal{P}_{SPJ} =& (v_{s} \in\{N,K\}) \wedge valid(t_1,s,v_s) \wedge valid(t_2,s,v_s) \wedge (t_1 \neq t_2)\\ 
	&\wedge ( (v_{t_1} = 0) \vee (v_{t_1} = v_s)) \wedge (v_{t_2} = 0 \vee v_{t_2} = v_s)
\end{split}
\end{equation}
This move will never result in double occupancy, thus we can apply procedure \ref{alg:ExecuteMove} with $U_m=U_{split\_jump}$ (eq. \ref{eq:SplitJumpUnitary}). Figure \ref{fig:SplitJumpCircuit} shows a quantum circuit for the move.

\subsection{Merge Jump}\label{sec:MergeJump}
\begin{figure}[htbp]
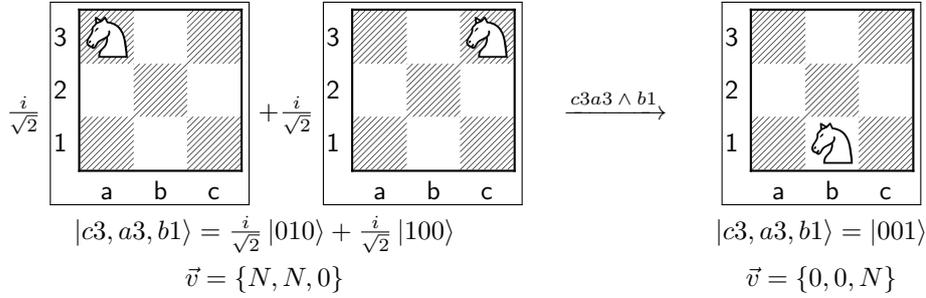

\captionsetup{width=.8\textwidth}
\def\arraystretch{1.5}
\begin{tabular}{c c c}
	\hbox{
    	$\frac{i}{\sqrt{2}}$
	    \raisebox{-0.3\height}{\fbox{\chessboard[
		    setfen=8/8/8/8/8/N7/8/8 w - - 0 1,
		    showmover=false,
		    printarea=a1-c3,
		    margintopwidth=0pt,
		    marginrightwidth=0pt,
		    marginleftwidth=8pt,
		    marginbottomwidth=10pt
	    ]}} 
	    $+ \frac{i}{\sqrt{2}}$
	    \raisebox{-0.3\height}{\fbox{\chessboard[
		    setfen=8/8/8/8/8/2N5/8/8 w - - 0 1,
		    showmover=false,
		    printarea=a1-c3,
		    margintopwidth=0pt,
		    marginrightwidth=0pt,
		    marginleftwidth=8pt,
		    marginbottomwidth=10pt
	    ]}}
    }
    & $\xrightarrow{\text{$c3a3\land b1$}}$ &
    \hbox{
	    \raisebox{-0.3\height}{\fbox{\chessboard[
		    setfen=8/8/8/8/8/8/8/1N6 w - - 0 1,
		    showmover=false,
		    printarea=a1-c3,
		    margintopwidth=0pt,
		    marginrightwidth=0pt,
		    marginleftwidth=8pt,
		    marginbottomwidth=10pt
	    ]}} 
    }
    \\
    \begin{tabular}{c}
    $\ket{c3,a3,b1} = \frac{i}{\sqrt{2}}\ket{010} + \frac{i}{\sqrt{2}}\ket{100}$ \\
    $\vec{v} = \{N,N,0\}$ 
    \end{tabular}
    & & 
    \begin{tabular}{c}
    $\ket{c3,a3,b1} = \ket{001}$ \\
    $\vec{v} = \{0,0,N\}$
    \end{tabular}
\end{tabular}
\caption{\small Merge Jump $c3a3\land b1$ is designed to \textit{undo} the SplitJump $b1\land a3c3$ on an otherwise empty board.}
\label{fig:MergeJump}
\end{figure}
\begin{figure}[ht]
\captionsetup{width=.8\textwidth}
\centering
\subcaptionbox{\label{fig:MergeJumpCircuit}}
{
	\makebox[0.45\columnwidth]{
		\Qcircuit @C=1em @R=1em{
			& \lstick{\ket{t1}} &\multigate{2}{U_{Merge}} &\qw \\
			& \lstick{\ket{s}}&\ghost{U_{Merge}} 	   	& \qw\\
			& \lstick{\ket{t2}}&\ghost{U_{Merge}} 						& \qw 	   
		}
	}
}
\hfill
\subcaptionbox{\label{fig:MergeJumpDecomposedCircuit}}
{
	\makebox[0.45\columnwidth]{
		\Qcircuit @C=1em @R=1em{
			& \lstick{\ket{t1}} & \qw &\multigate{1}{U_{\sqrt{iSwap}}} &\qw \\
			& \lstick{\ket{s}}&\multigate{1}{U_{iSwap}}&\ghost{U_{\sqrt{iSwap}}} 	   	& \qw\\
			& \lstick{\ket{t2}}&\ghost{U_{iSwap}}&\qw								& \qw 	   
		}
	}
}
\caption{\small Quantum circuit diagrams for applying the Merge Jump move to the source 1 $(s1)$, source 2 $(s2)$, and target $(t)$ qubits. In figure (a) The Merge unitary is applied as a three-qubit gate. Figure (b) shows its decomposition into two gates, the $\sqrt{iSwap}$ from eq. \ref{eq:iSwapRoot}, and the $iSwap$ from eq. \ref{eq:iSwap}.}
\label{fig:MergeJumpCircuits}
\end{figure}
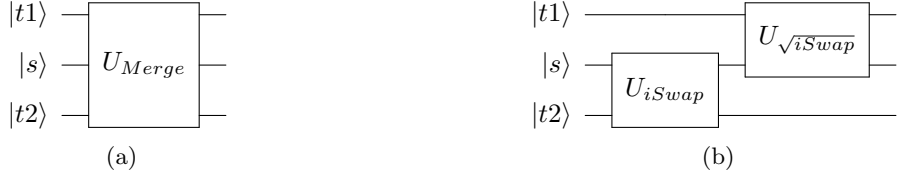

The Merge Jump is designed to undo the Split Jump (see figure \ref{fig:MergeJump}). The move acts on three squares: source 1 (s1), source 2 (s2), and target (t). We define the following possibility equation:
\begin{equation}\label{eq:MergeJumpPossibility}
\begin{split}
	\mathcal{P}_{MGJ} =& (v_{s_1} \in\{N,K\}) \wedge (v_{s_2} = v_{s_1}) \wedge valid(t,s_1,v_{s_1}) \wedge valid(t,s_2,v_{s_2})\\
	&\wedge (s_1 \neq s_2) \wedge (v_{t} = 0 \vee v_{t} = v_{s1})
\end{split}
\end{equation}
This move will never result in double occupancy, thus we can apply procedure \ref{alg:ExecuteMove} with $U_m=U_{merge\_jump}$ (eq. \ref{eq:MergeJumpUnitary}). Figure \ref{fig:MergeJumpCircuit} shows a quantum circuit for the move.

\subsection{Split Slide}\label{sec:SplitSlide}
\begin{figure}[htbp]
\captionsetup{width=.8\textwidth}
\def\arraystretch{1.5}
\begin{tabular}{c c c}
	\hbox{
    	$a$
	    \raisebox{-0.3\height}{\fbox{\chessboard[
		    setfen=8/8/8/8/8/8/8/R7 w - - 0 1,
		    showmover=false,
		    printarea=a1-b3,
		    margintopwidth=0pt,
		    marginrightwidth=0pt,
		    marginleftwidth=8pt,
		    marginbottomwidth=10pt,
		    tinyboard
	    ]}} 
	    $+ b$
	    \raisebox{-0.3\height}{\fbox{\chessboard[
		    setfen=8/8/8/8/8/8/N7/R7 w - - 0 1,
		    showmover=false,
		    printarea=a1-b3,
		    margintopwidth=0pt,
		    marginrightwidth=0pt,
		    marginleftwidth=8pt,
		    marginbottomwidth=10pt,
		    tinyboard
	    ]}}
	}
    & $\xrightarrow{\text{$a1\land a3b1$}}$ &
    \hbox{
    	$\frac{i}{\sqrt{2}}a$
	    \raisebox{-0.3\height}{\fbox{\chessboard[
		    setfen=8/8/8/8/8/R7/8/8 w - - 0 1,
		    showmover=false,
		    printarea=a1-b3,
		    margintopwidth=0pt,
		    marginrightwidth=0pt,
		    marginleftwidth=8pt,
		    marginbottomwidth=10pt,
		    tinyboard
	    ]}} 
	    $+ \frac{i}{\sqrt{2}}a$
	    \raisebox{-0.3\height}{\fbox{\chessboard[
		    setfen=8/8/8/8/8/8/8/1R6 w - - 0 1,
		    showmover=false,
		    printarea=a1-b3,
		    margintopwidth=0pt,
		    marginrightwidth=0pt,
		    marginleftwidth=8pt,
		    marginbottomwidth=10pt,
		    tinyboard
	    ]}}
	    $+ ib$
	    \raisebox{-0.3\height}{\fbox{\chessboard[
		    setfen=8/8/8/8/8/8/N7/1R6 w - - 0 1,
		    showmover=false,
		    printarea=a1-b3,
		    margintopwidth=0pt,
		    marginrightwidth=0pt,
		    marginleftwidth=8pt,
		    marginbottomwidth=10pt,
		    tinyboard
	    ]}}
    }
    \\
    \begin{tabular}{l}
    $\ket{p2,p1,b1,a3,a1}$\\
    $\qquad = a\ket{00001} + b\ket{01001}$ \\
    $\{b1,a3,a1\} = \{0,0,R\}$ 
    \end{tabular}
    & & 
    \begin{tabular}{l}
    $\ket{p2,p1,b1,a3,a1}$\\
    $\qquad = \frac{ia}{\sqrt{2}}(\ket{00010} + \ket{00100}) + ib\ket{01100}$ \\
    $\{b1,a3,a1\} = \{R,R,0\}$
    \end{tabular}
\end{tabular}
\caption{\small Split Slide $a1\land a3c1$ acting on a Rook in the presence of superposition. The move makes use of two path control qubits. Here we see the path ($p_1$) from s to t1 is partially blocked by a knight.}
\label{fig:SplitSlide}
\end{figure}
\begin{figure}[ht]
\captionsetup{width=.8\textwidth}
\centering
\mbox{
	\Qcircuit @C=1em @R=1em{
		& \lstick{\ket{t1}}         &\qw   &\qw&\qw       &\multigate{2}{U_{split}}&\multigate{1}{U_{jump}}&\qw                    &\qw\\
		& \lstick{\ket{s}}          &\qw   &\qw&\qw       &\ghost{U_{split}} 	   &\ghost{U_{jump}}       &\multigate{1}{U_{jump}}&\qw\\
		& \lstick{\ket{t2}}         &\qw   &\qw&\qw       &\ghost{U_{split}} 	   &\qw                    &\ghost{U_{jump}}      &\qw\\
		& \lstick{\ket{p1}=\ket{1}} & \qw  &\targ&\qw     &\ctrlo{-1}			   &\ctrlo{-2}             &\ctrl{-1}              &\qw\\
		& \lstick{\ket{p2}=\ket{1}} & \qw  &\qw&\targ     &\ctrlo{-1} 	           &\ctrl{-1}              &\ctrlo{-1}             &\qw\\
		& \lstick{\ket{x_1,...,x_n}}&/^n\qw&\ctrlo{-2}&\qw&\qw 	   	               &\qw                    &\qw 	   	           &\qw\\
		& \lstick{\ket{y_1,...,y_m}}&/^m\qw&\qw&\ctrlo{-2}&\qw 	   	               &\qw                    &\qw 	   	           &\qw
	}
}
\caption{\small Quantum circuit for applying the Split Slide move to the source $(s)$, target $(t1)$, and target 2 $(t2)$ qubits. The path from $s$ to $t1$ is composed of squares $\{x_1,...,x_n\}$ and the path from $s$ to $t2$ is composed of squares $\{y_1,...,y_m\}$. The controlled unitary operation reflects the blocked path effect expressed in eq. \ref{eq:SplitSlideUnitary}}
\label{fig:SplitSlideCircuit}
\end{figure}
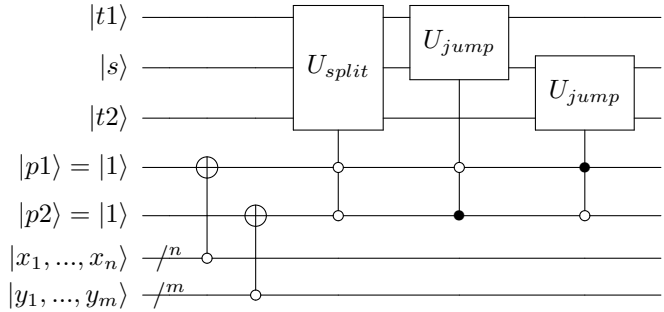

The Split Slide is the move by which the player creates superposition with bishops, rooks, and queens. The move acts on three squares: source(s), target 1 (t1), and target 2 (t2) (see figure \ref{fig:SplitSlide}). The possibility equation is similar to that of the Split Jump:
\begin{equation}\label{eq:SplitSlidePossibility}
\begin{split}
	\mathcal{P}_{SPS} =& (v_{s} \in\{B,R,Q\}) \wedge valid(t_1,s,v_s) \wedge valid(t_2,s,v_s) \wedge (t_1 \neq t_2)\\ 
	&\wedge ( (v_{t_1} = 0) \vee (v_{t_1} = v_s)) \wedge (v_{t_2} = 0 \vee v_{t_2} = v_s)
\end{split}
\end{equation}
This move will never result in double occupancy, thus we can apply procedure \ref{alg:ExecuteMove} with $U_m=U_{split\_slide}$ (eq. \ref{eq:SplitSlideUnitary}). Figure \ref{fig:SplitSlideCircuit} shows a quantum circuit for the move with $U_{split\_slide}$ decomposed to illustrate the path dependency.

\subsection{Merge Slide}\label{sec:MergeSlide}
\begin{figure}[htbp]
\captionsetup{width=.8\textwidth}
\def\arraystretch{1.5}
\begin{tabular}{c c c}
	\hbox{
    	$\frac{1}{\sqrt{2}}a$
	    \raisebox{-0.3\height}{\fbox{\chessboard[
		    setfen=8/8/8/8/8/R7/8/8 w - - 0 1,
		    showmover=false,
		    printarea=a1-b3,
		    margintopwidth=0pt,
		    marginrightwidth=0pt,
		    marginleftwidth=8pt,
		    marginbottomwidth=10pt,
		    tinyboard
	    ]}} 
	    $+ \frac{i}{\sqrt{2}}a$
	    \raisebox{-0.3\height}{\fbox{\chessboard[
		    setfen=8/8/8/8/8/8/8/1R6 w - - 0 1,
		    showmover=false,
		    printarea=a1-b3,
		    margintopwidth=0pt,
		    marginrightwidth=0pt,
		    marginleftwidth=8pt,
		    marginbottomwidth=10pt,
		    tinyboard
	    ]}}
	    $+ ib$
	    \raisebox{-0.3\height}{\fbox{\chessboard[
		    setfen=8/8/8/8/8/8/N7/1R6 w - - 0 1,
		    showmover=false,
		    printarea=a1-b3,
		    margintopwidth=0pt,
		    marginrightwidth=0pt,
		    marginleftwidth=8pt,
		    marginbottomwidth=10pt,
		    tinyboard
	    ]}}
	}
    & $\xrightarrow{\text{$b1a3 \land a1$}}$ &
    \hbox{
    	$a$
	    \raisebox{-0.3\height}{\fbox{\chessboard[
		    setfen=8/8/8/8/8/8/8/R7 w - - 0 1,
		    showmover=false,
		    printarea=a1-b3,
		    margintopwidth=0pt,
		    marginrightwidth=0pt,
		    marginleftwidth=8pt,
		    marginbottomwidth=10pt,
		    tinyboard
	    ]}} 
	    $+ b$
	    \raisebox{-0.3\height}{\fbox{\chessboard[
		    setfen=8/8/8/8/8/8/N7/R7 w - - 0 1,
		    showmover=false,
		    printarea=a1-b3,
		    margintopwidth=0pt,
		    marginrightwidth=0pt,
		    marginleftwidth=8pt,
		    marginbottomwidth=10pt,
		    tinyboard
	    ]}}
	}
    \\
    \begin{tabular}{l}
    $\ket{p2,p1,b1,a3,a1}$\\
    $\qquad = \frac{a}{\sqrt{2}}(\ket{00010} + i\ket{00100}) + ib\ket{01100}$ \\
    $\{b1,a3,a1\} = \{R,R,0\}$
    \end{tabular}
    & &
    \begin{tabular}{l}
    $\ket{p2,p1,b1,a3,a1}$\\
    $\qquad = a\ket{00001} + b\ket{01001}$ \\
    $\{b1,a3,a1\} = \{0,0,R\}$ 
    \end{tabular}
\end{tabular}
\caption{\small Merge Slide $b1a3\land a1$ acting on a Rook in the presence of superposition. The Merge undoes the Split that was illustrated in figure \ref{fig:SplitSlide}.}
\label{fig:MergeSlide}
\end{figure}
The Merge Slide is designed to undo the Split Slide (see figure \ref{fig:MergeSlide}). Its possibility equation takes a form similar to that of the Merge Jump:
\begin{equation}\label{eq:MergeSlidePossibility}
\begin{split}
	\mathcal{P}_{MGS} =& (v_{s_1} \in\{B,R,Q\}) \wedge (v_{s_2} = v_{s_1}) \wedge valid(t,s_1,v_{s_1}) \wedge valid(t,s_2,v_{s_2})\\
	& \wedge (s_1 \neq s_2) \wedge (v_{t} = 0 \vee v_{t} = v_{s_1})
\end{split}
\end{equation}
This move will never result in Double Occupancy, thus we can use procedure \ref{alg:ExecuteMove} with $U_m=U_{merge\_slide}$ (eq. \ref{eq:MergeSlideUnitary}). Figure \ref{fig:MergeSlideCircuit} shows a quantum circuit for the move with $U_{merge\_slide}$ decomposed to illustrate the path dependency.
\begin{figure}[ht]
\captionsetup{width=.8\textwidth}
\centering
\mbox{
	\Qcircuit @C=1em @R=1em{
		& \lstick{\ket{s1}}         &\qw   &\qw&\qw       &\multigate{2}{U_{Merge}}&\multigate{1}{U_{Jump}}&\qw                    &\qw\\
		& \lstick{\ket{t}}          &\qw   &\qw&\qw       &\ghost{U_{Merge}} 	   &\ghost{U_{Jump}}       &\multigate{1}{U_{Jump}}&\qw\\
		& \lstick{\ket{s2}}         &\qw   &\qw&\qw       &\ghost{U_{Merge}} 	   &\qw                    &\ghost{U_{Jump}}      &\qw\\
		& \lstick{\ket{p1}=\ket{1}} & \qw  &\targ&\qw     &\ctrlo{-1}			   &\ctrlo{-2}             &\ctrl{-1}              &\qw\\
		& \lstick{\ket{p2}=\ket{1}} & \qw  &\qw&\targ     &\ctrlo{-1} 	           &\ctrl{-1}              &\ctrlo{-1}             &\qw\\
		& \lstick{\ket{x_1,...,x_n}}&/^n\qw&\ctrlo{-2}&\qw&\qw 	   	               &\qw                    &\qw 	   	           &\qw\\
		& \lstick{\ket{y_1,...,y_m}}&/^m\qw&\qw&\ctrlo{-2}&\qw 	   	               &\qw                    &\qw 	   	           &\qw
	}
}
\caption{\small Quantum circuit for applying the Merge Slide move to the source 1 $(s1)$, source 2 $(s2)$, and target $(t)$ qubits. The path from $s1$ to $t$ is composed of squares $\{x_1,...,x_n\}$ and the path from $s2$ to $t$ is composed of squares $\{y_1,...,y_m\}$. The controlled unitary operation reflects the blocked path effect illustrated in eq. \ref{eq:MergeSlideUnitary}}
\label{fig:MergeSlideCircuit}
\end{figure}

\subsection{Pawn Movement}\label{sec:PawnMovement}
There are a number of special rules regarding pawn movement. We want pawns to behave similar to their counterpart in standard chess so here we detail the design of all of the ways in which they move, taking into account superposition and unitary movement restrictions. The following functions will be useful for defining the possibility of each type of pawn move.
\begin{itemize}
	\item $step(t,s)$ = True, if t is one step in the forward direction from s.
	\item $two\_step(t,s)$ = True, if t is two steps in the forward direction from s, \textbf{and} s is on the first pawn rank for the current player (rank 2 for white and rank 7 for black).
	\item $diagonal(t,s)$ = True, if t is one step in either forward diagonal direction from s.
	\item $rank(x)$ = The rank of square x.
	\item $file(x)$ = The file of square x.
	\item $ep(t,s,\mathcal{F})$ = True, if $file(t) = F_{ep}\in\{a,b,c,d,e,f,g,h\}$.
\end{itemize}
Pawn forward movement will be easy to handle with the tools we have already defined for Jump and Slide moves. However, Capture and En Passant will take some special care.

En passant (e.p.) is tricky in Quantum Chess. In standard Chess e.p. is possible if an opponent pawn begins its movement by advancing two steps and ends adjacent to one of our own pawns\cite{FIDE:online}. It has thus bypassed the square in which the player's pawn could have captured it. The player has have one turn in which we may chose to capture the opponent pawn by moving our pawn into the square it bypassed.

We will consider basis states of the form $\ket{e.p.\ target, move\ target, source}$, or $\ket{ep,t,s}$. We must determine what to do when the opponent Pawn being captured is in superposition. This can occur if it begins in superposition, and/or if it makes its two-step move through another piece that is in superposition. If there is another piece involved, and it belongs to the opponent, we would like to allow the capture of said piece. Thus pawn e.p. acts as a capturing move on both the target square (where attacking pawn ends up) as well as on the square occupied by the opponent's e.p. capturable pawn. There are three cases we must explore to fully determine how to implement e.p. in Quantum Chess:
\begin{enumerate}
\item Standard E.P.: The target square is either empty, or occupied in superposition by a pawn of the same color. 
\item Blocked E.P.: The target square is occupied in superposition by a non-pawn belonging to the current player, and there exists a subspace where e.p. is valid.
\item Capture E.P.: The target square is occupied in superposition by an opponent piece and there exists a subspace where e.p. is valid.
\end{enumerate}
In the following sections we fully detail all types of Pawn movement.

\subsubsection{Pawn Step}\label{sec:PawnSingleStep}
\begin{figure}[htbp]
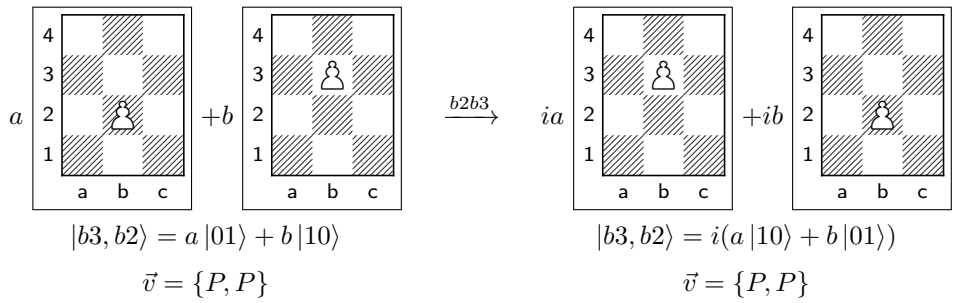

\captionsetup{width=.8\textwidth}
\begin{centering}
\def\arraystretch{1.5}
\begin{tabular}{c c c}
	\hbox{
    	$a$
	    \raisebox{-0.3\height}{\fbox{\chessboard[
		    setfen=8/8/8/8/8/8/1P6/8 w - - 0 1,
		    showmover=false,
		    printarea=a1-c4,
		    margintopwidth=0pt,
		    marginrightwidth=0pt,
		    marginleftwidth=8pt,
		    marginbottomwidth=10pt,
		    smallboard
	    ]}} 
	    $+ b$
	    \raisebox{-0.3\height}{\fbox{\chessboard[
		    setfen=8/8/8/8/8/1P6/8/8 w - - 0 1,
		    showmover=false,
		    printarea=a1-c4,
		    margintopwidth=0pt,
		    marginrightwidth=0pt,
		    marginleftwidth=8pt,
		    marginbottomwidth=10pt,
		    smallboard
	    ]}} 
    }
    & $\xrightarrow{\text{$b2b3$}}$ &
    \hbox{
    	$ia$
	    \raisebox{-0.3\height}{\fbox{\chessboard[
		    setfen=8/8/8/8/8/1P6/8/8 w - - 0 1,
		    showmover=false,
		    printarea=a1-c4,
		    margintopwidth=0pt,
		    marginrightwidth=0pt,
		    marginleftwidth=8pt,
		    marginbottomwidth=10pt,
		    smallboard
	    ]}} 
	    $+ ib$
	    \raisebox{-0.3\height}{\fbox{\chessboard[
		    setfen=8/8/8/8/8/8/1P6/8 w - - 0 1,
		    showmover=false,
		    printarea=a1-c4,
		    margintopwidth=0pt,
		    marginrightwidth=0pt,
		    marginleftwidth=8pt,
		    marginbottomwidth=10pt,
		    smallboard
	    ]}} 
    }
    \\
    \begin{tabular}{c}
	    $\ket{b3,b2} = a\ket{01} + b\ket{10}$ \\
	    $\vec{v} = \{P,P\}$ 
    \end{tabular}
    & & 
    \begin{tabular}{c}
	    $\ket{b3,b2} = i(a\ket{10} + b\ket{01})$ \\
	    $\vec{v} = \{P,P\}$
    \end{tabular}
\end{tabular}
\end{centering}
\caption{\small A pawn step is possible if the target is occupied by a pawn of the same color.}
\label{fig:PawnStepUnitarity}
\end{figure}
The Pawn step is the simplest of pawn moves. In standard chess a pawn may always move one square forward, so long as the target square is not occupied by any other piece. The possibility equation for a Pawn Step is:\begin{equation}\label{eq:PawnStepPossibility}
	\mathcal{P}_{PS} = (v_{s} = P)\wedge (v_{t} = 0 \vee v_{t} = P) \wedge step(t,s)
\end{equation}
This move will never result in Double Occupancy thus we can follow procedure \ref{alg:ExecuteMove} with $U_m = U_{jump}$. This is the same as what was defined for the Standard Jump. We can use the circuit shown in figure  \ref{fig:StandardJumpCircuit} to perform the move.

\subsubsection{Blocked Pawn Step}\label{sec:BlockedPawnStep}
\begin{figure}[htbp]
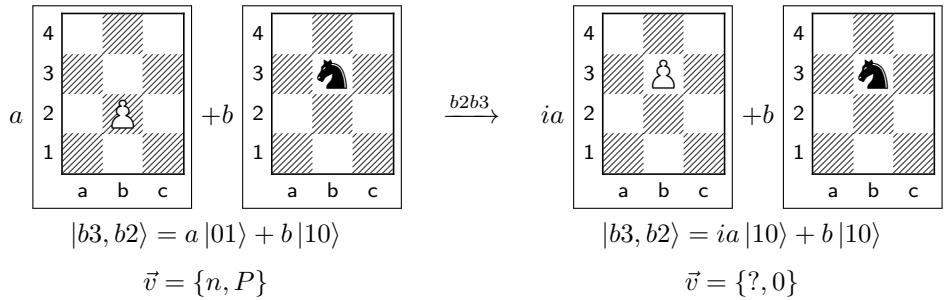

\captionsetup{width=.8\textwidth}
\begin{centering}
\def\arraystretch{1.5}
\begin{tabular}{c c c}
	\hbox{
    	$a$
	    \raisebox{-0.3\height}{\fbox{\chessboard[
		    setfen=8/8/8/8/8/8/1P6/8 w - - 0 1,
		    showmover=false,
		    printarea=a1-c4,
		    margintopwidth=0pt,
		    marginrightwidth=0pt,
		    marginleftwidth=8pt,
		    marginbottomwidth=10pt,
		    smallboard
	    ]}} 
	    $+ b$
	    \raisebox{-0.3\height}{\fbox{\chessboard[
		    setfen=8/8/8/8/8/1n6/8/8 w - - 0 1,
		    showmover=false,
		    printarea=a1-c4,
		    margintopwidth=0pt,
		    marginrightwidth=0pt,
		    marginleftwidth=8pt,
		    marginbottomwidth=10pt,
		    smallboard
	    ]}} 
    }
    & $\xrightarrow{\text{$b2b3$}}$ &
    \hbox{
    	$ia$
	    \raisebox{-0.3\height}{\fbox{\chessboard[
		    setfen=8/8/8/8/8/1P6/8/8 w - - 0 1,
		    showmover=false,
		    printarea=a1-c4,
		    margintopwidth=0pt,
		    marginrightwidth=0pt,
		    marginleftwidth=8pt,
		    marginbottomwidth=10pt,
		    smallboard
	    ]}} 
	    $+ b$
	    \raisebox{-0.3\height}{\fbox{\chessboard[
		    setfen=8/8/8/8/8/1n6/8/8 w - - 0 1,
		    showmover=false,
		    printarea=a1-c4,
		    margintopwidth=0pt,
		    marginrightwidth=0pt,
		    marginleftwidth=8pt,
		    marginbottomwidth=10pt,
		    smallboard
	    ]}} 
    }
    \\
    \begin{tabular}{c}
	    $\ket{b3,b2} = a\ket{01} + b\ket{10}$ \\
	    $\vec{v} = \{n,P\}$ 
    \end{tabular}
    & & 
    \begin{tabular}{c}
	    $\ket{b3,b2} = ia\ket{10} + b\ket{10}$ \\
	    $\vec{v} = \{?,0\}$
    \end{tabular}
\end{tabular}
\end{centering}
\caption{\small A blocked pawn step will result in double occupancy. Here square b3 is occupied by both a black knight and a white pawn after the move.}
\label{fig:BlockedPawnStep}
\end{figure}
The Blocked Pawn Step is similar to the Blocked Jump with a different condition on the occupancy of the target square. Any piece blocks the move because pawns may not capture with forward movement. The possibility equation is:
\begin{equation}\label{eq:BlockedPawnStepPossibility}
	\mathcal{P}_{BPS} = (v_s = P) \wedge (v_t \neq 0) \wedge (v_t \neq P) \wedge step(t,s)
\end{equation}
The move will result in double occupancy (see figure \ref{fig:BlockedPawnStep}). We use the same measurements defined for the Blocked Jump (eq. \ref{eq:BlockedJumpMeasurement}), as well as the same procedure and circuit to perform the move (fig. \ref{fig:BlockedJumpMeasureCircuit}).

\subsubsection{Pawn Two Step}\label{sec:PawnDoubleStep}
\begin{figure}[htbp]
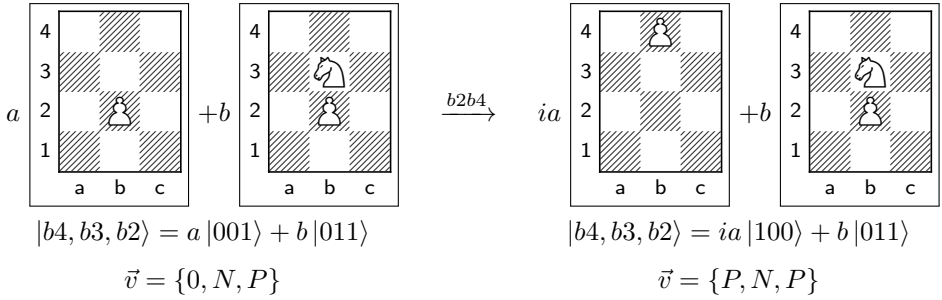

\captionsetup{width=.8\textwidth}
\begin{centering}
\def\arraystretch{1.5}
\begin{tabular}{c c c}
	\hbox{
    	$a$
	    \raisebox{-0.3\height}{\fbox{\chessboard[
		    setfen=8/8/8/8/8/8/1P6/8 w - - 0 1,
		    showmover=false,
		    printarea=a1-c4,
		    margintopwidth=0pt,
		    marginrightwidth=0pt,
		    marginleftwidth=8pt,
		    marginbottomwidth=10pt,
		    smallboard
	    ]}} 
	    $+ b$
	    \raisebox{-0.3\height}{\fbox{\chessboard[
		    setfen=8/8/8/8/8/1N6/1P6/8 w - - 0 1,
		    showmover=false,
		    printarea=a1-c4,
		    margintopwidth=0pt,
		    marginrightwidth=0pt,
		    marginleftwidth=8pt,
		    marginbottomwidth=10pt,
		    smallboard
	    ]}} 
    }
    & $\xrightarrow{\text{$b2b4$}}$ &
    \hbox{
    	$ia$
	    \raisebox{-0.3\height}{\fbox{\chessboard[
		    setfen=8/8/8/8/1P6/8/8/8 w - - 0 1,
		    showmover=false,
		    printarea=a1-c4,
		    margintopwidth=0pt,
		    marginrightwidth=0pt,
		    marginleftwidth=8pt,
		    marginbottomwidth=10pt,
		    smallboard
	    ]}} 
	    $+ b$
	    \raisebox{-0.3\height}{\fbox{\chessboard[
		    setfen=8/8/8/8/8/1N6/1P6/8 w - - 0 1,
		    showmover=false,
		    printarea=a1-c4,
		    margintopwidth=0pt,
		    marginrightwidth=0pt,
		    marginleftwidth=8pt,
		    marginbottomwidth=10pt,
		    smallboard
	    ]}} 
    }
    \\
    \begin{tabular}{c}
	    $\ket{b4,b3,b2} = a\ket{001} + b\ket{011}$ \\
	    $\vec{v} = \{0,N,P\}$ 
    \end{tabular}
    & & 
    \begin{tabular}{c}
	    $\ket{b4,b3,b2} = ia\ket{100} + b\ket{011}$ \\
	    $\vec{v} = \{P,N,P\}$
    \end{tabular}
\end{tabular}
\end{centering}
\caption{\small Pawn Two Step can entangle the pawn with another piece in superposition.}
\label{fig:PawnDoubleStepEntangle}
\end{figure}
In standard chess the pawn two step is allowed on any given pawn's first move. If the path is not blocked then the pawn may slide two squares forward. Here we define the equivalent move in the presence of superposition. The possibility equation is
\begin{equation}\label{eq:PawnTwoStepPossibility}
	\mathcal{P}_{PTS} = (v_s = P) \wedge (v_t \neq 0) \wedge (v_t \neq P) \wedge two\_step(t,s)
\end{equation}
This move acts similar to the Standard Slide (sec. \ref{fig:StandardSlide}). It wil never result in double occupancy. This move can entangle the pawn with any piece in superposition in the intermediate square (see figure \ref{fig:PawnDoubleStepEntangle}), making it one of the means by which a pawn may exist in superposition. We can use the same procedure as that used for the Standard Slide, as well as the same circuit seen in figure \ref{fig:StandardSlideCircuits}. At the end of the move procedure we will set the classical e.p. information in $\mathcal{F}$ if move outcome is determined to be legal by definition \ref{def:LegalMove}.

\subsubsection{Blocked Pawn Two Step}\label{sec:BlockedPawnTwoStep}
\begin{figure}[htbp]
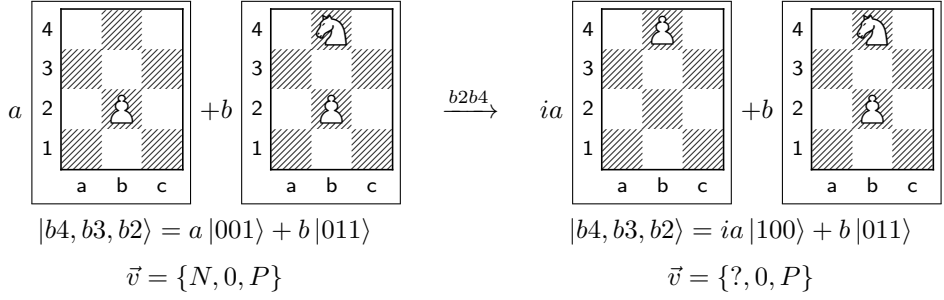

\captionsetup{width=.8\textwidth}
\begin{centering}
\def\arraystretch{1.5}
\begin{tabular}{c c c}
	\hbox{
    	$a$
	    \raisebox{-0.3\height}{\fbox{\chessboard[
		    setfen=8/8/8/8/8/8/1P6/8 w - - 0 1,
		    showmover=false,
		    printarea=a1-c4,
		    margintopwidth=0pt,
		    marginrightwidth=0pt,
		    marginleftwidth=8pt,
		    marginbottomwidth=10pt,
		    smallboard
	    ]}} 
	    $+ b$
	    \raisebox{-0.3\height}{\fbox{\chessboard[
		    setfen=8/8/8/8/1N6/8/1P6/8 w - - 0 1,
		    showmover=false,
		    printarea=a1-c4,
		    margintopwidth=0pt,
		    marginrightwidth=0pt,
		    marginleftwidth=8pt,
		    marginbottomwidth=10pt,
		    smallboard
	    ]}} 
    }
    & $\xrightarrow{\text{$b2b4$}}$ &
    \hbox{
    	$ia$
	    \raisebox{-0.3\height}{\fbox{\chessboard[
		    setfen=8/8/8/8/1P6/8/8/8 w - - 0 1,
		    showmover=false,
		    printarea=a1-c4,
		    margintopwidth=0pt,
		    marginrightwidth=0pt,
		    marginleftwidth=8pt,
		    marginbottomwidth=10pt,
		    smallboard
	    ]}} 
	    $+ b$
	    \raisebox{-0.3\height}{\fbox{\chessboard[
		    setfen=8/8/8/8/1N6/8/1P6/8 w - - 0 1,
		    showmover=false,
		    printarea=a1-c4,
		    margintopwidth=0pt,
		    marginrightwidth=0pt,
		    marginleftwidth=8pt,
		    marginbottomwidth=10pt,
		    smallboard
	    ]}} 
    }
    \\
    \begin{tabular}{c}
	    $\ket{b4,b3,b2} = a\ket{001} + b\ket{011}$ \\
	    $\vec{v} = \{N,0,P\}$ 
    \end{tabular}
    & & 
    \begin{tabular}{c}
	    $\ket{b4,b3,b2} = ia\ket{100} + b\ket{011}$ \\
	    $\vec{v} = \{?,0,P\}$
    \end{tabular}
\end{tabular}
\end{centering}
\caption{\small Blocked Pawn Two Step will result in double occupancy. Here square b4 would be occupied by both a white pawn and a white knight after the move.}
\label{fig:BlockedPawnTwoStep}
\end{figure}
The Blocked Pawn Two Step is similar to the Blocked Slide introduced in section \ref{sec:BlockedSlide}. The possibility equation for this move is:
\begin{equation}\label{eq:BlockedPawnTwoStepPossibility}
	\mathcal{P}_{BPTS} = (v_s = P) \wedge (v_t \neq 0) \wedge (v_t \neq P) \wedge two\_step(t,s)
\end{equation}
This move will result in double occupancy (see figure \ref{fig:BlockedPawnTwoStep}). We can use the same measurements defined for the Blocked Slide (eq. \ref{eq:BlockedSlideMeasurement}) as well as the same circuit (fig. \ref{fig:BlockedSlideCircuit}). At the end of the move procedure we will set the classical e.p. information in $\mathcal{F}$ if our measurement result was 1 \textit{and} the move outcome is determined to be legal by definition \ref{def:LegalMove}.

\subsubsection{Pawn Capture}\label{sec:PawnCapture}
\begin{figure}[htbp]
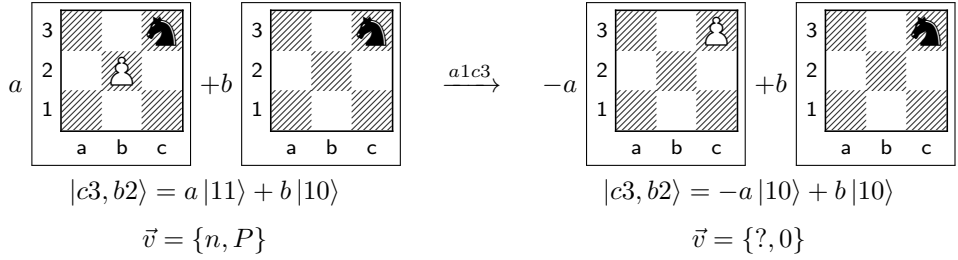

\captionsetup{width=.8\textwidth}
\begin{centering}
\def\arraystretch{1.5}
\begin{tabular}{c c c}
	\hbox{
    	$a$
	    \raisebox{-0.3\height}{\fbox{\chessboard[
		    setfen=8/8/8/8/8/2n5/1P6/8 w - - 0 1,
		    showmover=false,
		    printarea=a1-c3,
		    margintopwidth=0pt,
		    marginrightwidth=0pt,
		    marginleftwidth=8pt,
		    marginbottomwidth=10pt,
		    smallboard
	    ]}} 
	    $+ b$
	    \raisebox{-0.3\height}{\fbox{\chessboard[
		    setfen=8/8/8/8/8/2n5/8/8 w - - 0 1,
		    showmover=false,
		    printarea=a1-c3,
		    margintopwidth=0pt,
		    marginrightwidth=0pt,
		    marginleftwidth=8pt,
		    marginbottomwidth=10pt,
		    smallboard
	    ]}} 
    }
    & $\xrightarrow{\text{$a1c3$}}$ &
    \hbox{
    	$-a$
	    \raisebox{-0.3\height}{\fbox{\chessboard[
		    setfen=8/8/8/8/8/2P5/8/8 w - - 0 1,
		    showmover=false,
		    printarea=a1-c3,
		    margintopwidth=0pt,
		    marginrightwidth=0pt,
		    marginleftwidth=8pt,
		    marginbottomwidth=10pt,
		    smallboard
	    ]}} 
	    $+ b$
	    \raisebox{-0.3\height}{\fbox{\chessboard[
		    setfen=8/8/8/8/8/2n5/8/8 w - - 0 1,
		    showmover=false,
		    printarea=a1-c3,
		    margintopwidth=0pt,
		    marginrightwidth=0pt,
		    marginleftwidth=8pt,
		    marginbottomwidth=10pt,
		    smallboard
	    ]}}
    }
    \\
    \begin{tabular}{c}
    $\ket{c3,b2} = a\ket{11} + b\ket{10}$\\
    $\vec{v} = \{n,P\}$ 
    \end{tabular}
    & & 
    \begin{tabular}{c}
    $\ket{c3,b2} = -a\ket{10} + b\ket{10}$\\
    $\vec{v} = \{?,0\}$
    \end{tabular}
\end{tabular}
\end{centering}
\caption{\small Double Occupancy would occur if we attempted the pawn capturing move b2c3.}
\label{fig:PawnCaptureDoubleOccupancy}
\end{figure}
In standard chess pawn capture is possible if the target square is one step in either diagonal direction, and it is occupied by an opponent piece. We present the following possibility equation:
\begin{equation}\label{eq:PawnCapturePossibility}
	\mathcal{P}_{PC} = (v_s = P) \wedge (v_t \in \{p,n,b,r,q,k\}) \wedge diagonal(t,s)
\end{equation}
The move can result in double occupancy (see figure \ref{fig:PawnCaptureDoubleOccupancy}). We can use the measurements developed for the Capture Jump (eq. \ref{eq:CaptureJumpMeasurement}). However, we can't use the same procedure or circuit. The Jump Capture allows for the piece to move and occupy the target even if the target is not present in a particular basis state. Pawn diagonal movement is only allowed in the case where it is capturing a piece. We must define a different $U_m$ to use in procedure \ref{alg:ExecuteMeasuringMove}. 

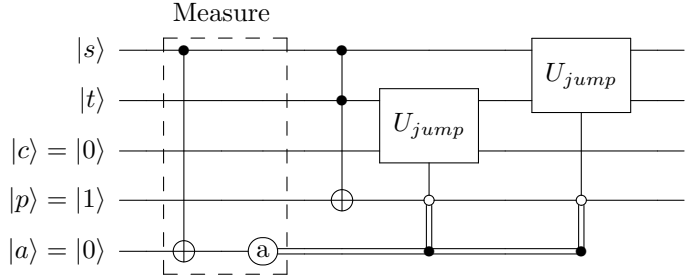
\begin{figure}[ht]
\captionsetup{width=.8\textwidth}
\centering
\mbox{
	\Qcircuit @C=1em @R=1em{
	&							   &   &  		 &\mbox{Measure}&	&        &   &					 && 			 	 &&\\
	&\lstick{\ket{s}} 	   &\qw&\ctrl{4}&\qw&\qw				&\qw&\ctrl{1}&\qw 					 &\qw&\multigate{1}{U_{jump}}&\qw&\qw\\
	&\lstick{\ket{t}} 	   &\qw&\qw  &\qw&\qw					&\qw&\ctrl{2}&\multigate{1}{U_{jump}}&\qw&\ghost{U_{jump}} 		 &\qw&\qw\\
	&\lstick{\ket{c}=\ket{0}}&\qw&\qw	 &\qw&\qw 				&\qw&\qw&\ghost{U_{jump}}		 &\qw&\qw					 &\qw&\qw\\
	&\lstick{\ket{p}=\ket{1}}&\qw&\qw	 &\qw&\qw 				&\qw&\targ&\ctrlo{-1}		         &\qw&\ctrlo{-2}		&\qw&\qw\\
	&\lstick{\ket{a}=\ket{0}}&\qw&\targ  &\qw&\measure{\mbox{a}}&\cw&\cw&\control\cw\cwx[-1]&\cw&\control\cw\cwx[-1]
	\gategroup{2}{4}{6}{6}{0.7em}{--}
	}
}
\caption{\small Quantum circuit diagram for applying the Pawn Capture move to the source ($s$) and target ($t$) qubits from the quantum register $\underline{\psi_\mathcal{B}}$. A captured ancilla ($c$) is added to hold the captured piece, and a "path" control ancilla ($p$) to encode the occupancy of $s$ and $t$. $M_1$ (eq. \ref{eq:CaptureJumpMeasurement}) is encoded into measurement ancilla ($a$) and two zero-controlled $U_{jump}$ (eq. \ref{eq:JumpUnitary}) operations are conditionally applied on with outcome 1 (source is occupied).}
\label{fig:PawnCaptureMeasureCircuit}
\end{figure}

The move should only act non-trivially if both the moving pawn and the target are present so we will need to use controlled operations. We want our control ancilla to have state $\ket{p} = \ket{0}$ if both the source and target occupancy qubits are in state $\ket{1}$. Our controlled unitaries are then equivalent to $U_{slide}$ (eq. \ref{eq:SlideUnitary}) acting on the $\ket{p,c,t}$ and $\ket{p,t,s}$ subspaces. Thus we can apply procedure \ref{alg:ExecuteMeasuringMove} with $U_m = U_{slide}(p,t,s)U_{slide}(p,c,t)$ as illustrated in the circuit in figure \ref{fig:PawnCaptureMeasureCircuit} where each $U_{slide}$ has been converted into a zero-controlled $U_{jump}$.

\subsubsection{Standard E.P.}\label{sec:epStandard}
\begin{figure}[htbp]
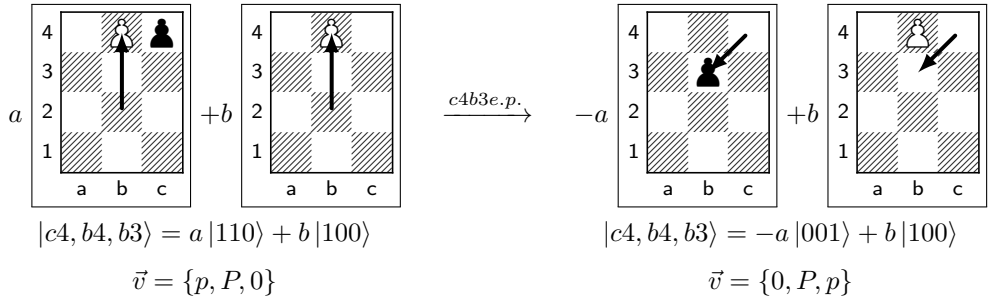

\captionsetup{width=.8\textwidth}
\begin{centering}
\def\arraystretch{1.5}
\begin{tabular}{c c c}
	\hbox{
    	$a$
	    \raisebox{-0.3\height}{\fbox{\chessboard[
		    setfen=8/8/8/8/1Pp5/8/8/8 w - - 0 1,
		    showmover=false,
		    printarea=a1-c4,
		    margintopwidth=0pt,
		    marginrightwidth=0pt,
		    marginleftwidth=8pt,
		    marginbottomwidth=10pt,
		    smallboard,
		    pgfstyle=straightmove,
		    markmoves={b2-b4}
	    ]}} 
	    $+ b$
	    \raisebox{-0.3\height}{\fbox{\chessboard[
		    setfen=8/8/8/8/1P6/8/8/8 w - - 0 1,
		    showmover=false,
		    printarea=a1-c4,
		    margintopwidth=0pt,
		    marginrightwidth=0pt,
		    marginleftwidth=8pt,
		    marginbottomwidth=10pt,
		    smallboard,
		    pgfstyle=straightmove,
		    markmoves={b2-b4}
	    ]}} 
    }
    & $\xrightarrow{\text{$c4b3 e.p.$}}$ &
    \hbox{
    	$-a$
	    \raisebox{-0.3\height}{\fbox{\chessboard[
		    setfen=8/8/8/8/8/1p6/8/8 w - - 0 1,
		    showmover=false,
		    printarea=a1-c4,
		    margintopwidth=0pt,
		    marginrightwidth=0pt,
		    marginleftwidth=8pt,
		    marginbottomwidth=10pt,
		    smallboard,
		    pgfstyle=straightmove,
		    markmoves={c4-b3}
	    ]}} 
	    $+ b$
	    \raisebox{-0.3\height}{\fbox{\chessboard[
		    setfen=8/8/8/8/1P6/8/8/8 w - - 0 1,
		    showmover=false,
		    printarea=a1-c4,
		    margintopwidth=0pt,
		    marginrightwidth=0pt,
		    marginleftwidth=8pt,
		    marginbottomwidth=10pt,
		    smallboard,
		    pgfstyle=straightmove,
		    markmoves={c4-b3}
	    ]}} 
    }
    \\
    \begin{tabular}{c}
	    $\ket{c4,b4,b3} = a\ket{110} + b\ket{100}$ \\
	    $\vec{v} = \{p,P,0\}$ 
    \end{tabular}
    & & 
    \begin{tabular}{c}
	    $\ket{c4,b4,b3} = -a\ket{001} + b\ket{100}$ \\
	    $\vec{v} = \{0,P,p\}$
    \end{tabular}
\end{tabular}
\end{centering}
\caption{\small Standard E.P. can result in entanglement in the presence of superposition. Here the black pawn in b3 is entangled with the white pawn in b4.}
\label{fig:EPStandard}
\end{figure}
\begin{figure}[ht]
\captionsetup{width=.8\textwidth}
\centering
\mbox{
	\Qcircuit @C=1em @R=1em{
	&\lstick{\ket{s}} 	     &\qw&\ctrl{2}&\qw 					 &\qw&\multigate{1}{U_{jump}}&\qw&\qw\\
	&\lstick{\ket{t}} 	     &\qw&\qw&\qw&\qw&\ghost{U_{jump}} 		 &\qw&\qw\\
	&\lstick{\ket{ep}} 	     &\qw&\ctrl{2}&\multigate{1}{U_{jump}}&\qw&\qw 		 &\qw&\qw\\
	&\lstick{\ket{c}=\ket{0}}&\qw&\qw  &\ghost{U_{jump}}		 &\qw&\qw					 &\qw&\qw\\
	&\lstick{\ket{p}=\ket{1}}&\qw&\targ&\ctrlo{-1}		         &\qw&\ctrlo{-3}			     &\qw&\qw\\
	}
}
\caption{\small Quantum circuit diagram for applying the Standard E.P. move to the source ($s$), target ($t$), and e.p. target (ep) qubits from the quantum register $\underline{\psi_\mathcal{B}}$. A captured ancilla ($c$) is added to hold the captured piece. Two zero-controlled $U_{jump}$ (eq. \ref{eq:JumpUnitary}) operations are applied with control $\ket{p}=\ket{0}$ if both the source and ep target are occupied.}
\label{fig:StandardEPCircuit}
\end{figure}
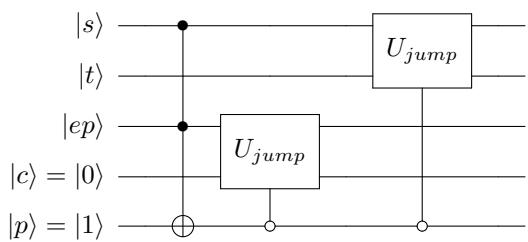

In standard Chess e.p. is possible if the source square is occupied by a pawn belonging to the current player, and the \textit{adjacent} square in the target direction is occupied by an opponent pawn that took two steps on the \textit{previous turn}. The possibility equation for Standard E.P. acting on a source (s), target (t) and e.p. target (ep) is:
\begin{equation}\label{eq:StandardEPPossibility}
	\mathcal{P}_{SEP} = (v_s = P) \wedge (v_t \in \{0,P\}) \wedge(v_{ep} = p) \wedge ep(t,s,\mathcal{F})\footnote{The constraint $(v_{ep} = p)$ is redundant since $ep(t,s,\mathcal{F})$ should only return true if ep is a valid target, meaning it must be occupied by an opponent pawn that has just jumped forward two squares. It is explicitly included for clarity.}
\end{equation}
This move will never result in Double Occupancy. As with other capturing moves we must make use of the captured ancilla space, as well as a control ancilla that encodes the occupancy of both the source and the ep target. We must also introduce a control qubit (p) to ensure the move only acts non-trivially if both the source and e.p. target are occupied. With a design similar to Pawn Capture we can apply procedure \ref{alg:ExecuteMove} where $U_m = U_{slide}(p,t,s)U_{slide}(p,c,ep)$ as illustrated with the circuit in figure \ref{fig:StandardEPCircuit}, where each $U_{slide}$ is converted to a zero-controlled $U_{jump}$.

\subsubsection{Blocked EP}\label{sec:epBlocked}
\begin{figure}[htbp]
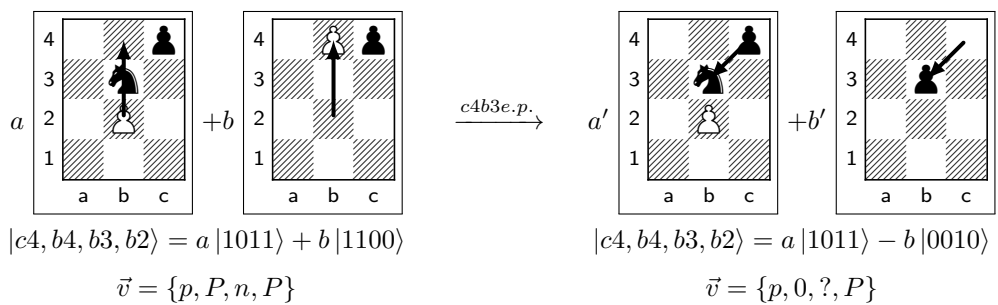

\captionsetup{width=.8\textwidth}
\begin{centering}
\def\arraystretch{1.5}
\begin{tabular}{c c c}
	\hbox{
    	$a$
	    \raisebox{-0.3\height}{\fbox{\chessboard[
		    setfen=8/8/8/8/2p5/1n6/1P6/8 w - - 0 1,
		    showmover=false,
		    printarea=a1-c4,
		    margintopwidth=0pt,
		    marginrightwidth=0pt,
		    marginleftwidth=8pt,
		    marginbottomwidth=10pt,
		    smallboard,
		    pgfstyle=straightmove,
		    markmoves={b2-b4}
	    ]}} 
	    $+ b$
	    \raisebox{-0.3\height}{\fbox{\chessboard[
		    setfen=8/8/8/8/1Pp5/8/8/8 w - - 0 1,
		    showmover=false,
		    printarea=a1-c4,
		    margintopwidth=0pt,
		    marginrightwidth=0pt,
		    marginleftwidth=8pt,
		    marginbottomwidth=10pt,
		    smallboard,
		    pgfstyle=straightmove,
		    markmoves={b2-b4}
	    ]}} 
    }
    & $\xrightarrow{\text{$c4b3 e.p.$}}$ &
    \hbox{
    	$a'$
	    \raisebox{-0.3\height}{\fbox{\chessboard[
		    setfen=8/8/8/8/2p5/1n6/1P6/8 w - - 0 1,
		    showmover=false,
		    printarea=a1-c4,
		    margintopwidth=0pt,
		    marginrightwidth=0pt,
		    marginleftwidth=8pt,
		    marginbottomwidth=10pt,
		    smallboard,
		    pgfstyle=straightmove,
		    markmoves={c4-b3}
	    ]}} 
	    $+ b'$
	    \raisebox{-0.3\height}{\fbox{\chessboard[
		    setfen=8/8/8/8/8/1p6/8/8 w - - 0 1,
		    showmover=false,
		    printarea=a1-c4,
		    margintopwidth=0pt,
		    marginrightwidth=0pt,
		    marginleftwidth=8pt,
		    marginbottomwidth=10pt,
		    smallboard,
		    pgfstyle=straightmove,
		    markmoves={c4-b3}
	    ]}} 
    }
    \\
    \begin{tabular}{c}
	    $\ket{c4,b4,b3,b2} = a\ket{1011} + b\ket{1100}$ \\
	    $\vec{v} = \{p,P,n,P\}$ 
    \end{tabular}
    & & 
    \begin{tabular}{c}
	    $\ket{c4,b4,b3,b2} = a\ket{1011} - b\ket{0010}$ \\
	    $\vec{v} = \{p,0,?,P\}$
    \end{tabular}
\end{tabular}
\end{centering}
\caption{\small Blocked E.P. is possible when the pawn being captured is entangled with another piece of the opposite color. Double occupancy occurs in this example.}
\label{fig:EPBlocked}
\end{figure}
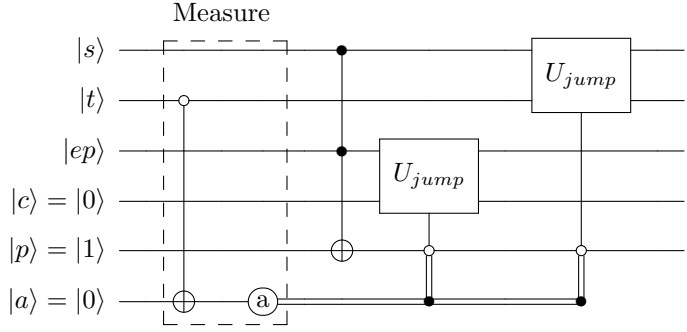
\begin{figure}[ht]
\captionsetup{width=.8\textwidth}
\centering
\mbox{
	\Qcircuit @C=1em @R=1em{
	&							   &   &  		 &\mbox{Measure}&	&        &   &					 && 			 	 &&\\
	&\lstick{\ket{s}} 	   &\qw&\qw&\qw&\qw				&\qw     &\ctrl{2}&\qw 					 &\qw&\multigate{1}{U_{jump}}&\qw&\qw\\
	&\lstick{\ket{t}} 	   &\qw&\ctrlo{4}&\qw&\qw					&\qw     &\qw&\qw&\qw&\ghost{U_{jump}} 		 &\qw&\qw\\
	&\lstick{\ket{ep}} 	   &\qw&\qw  &\qw&\qw					&\qw&\ctrl{2}&\multigate{1}{U_{jump}}&\qw&\qw 		 &\qw&\qw\\
	&\lstick{\ket{c}=\ket{0}}&\qw&\qw	 &\qw&\qw 				&\qw&\qw&\ghost{U_{jump}}		 &\qw&\qw					 &\qw&\qw\\
	&\lstick{\ket{p}=\ket{1}}&\qw&\qw	 &\qw&\qw 				&\qw&\targ&\ctrlo{-1}		         &\qw&\ctrlo{-3}			     &\qw&\qw\\
	&\lstick{\ket{a}=\ket{0}}&\qw&\targ  &\qw&\measure{\mbox{a}}&\cw&\cw&\control\cw\cwx[-1]&\cw&\control\cw\cwx[-1]
	\gategroup{2}{4}{7}{6}{0.7em}{--}
	}
}
\caption{\small Quantum circuit diagram for applying the Blocked E.P. move to the source ($s$), target ($t$), and e.p. target (ep) qubits from the quantum register $\underline{\psi_\mathcal{B}}$. A captured ancilla ($c$) is added to hold the captured piece, and a path ancilla ($p$) to encode the occupancy of $s$ and $t$. Two zero-controlled $U_{jump}$ (eq. \ref{eq:JumpUnitary}) operations are conditionally applied if the measurement outcome is $M_1$ (target is not occupied).}
\label{fig:BlockedEPCircuit}
\end{figure}

The existence of superposition means it is possible for en passant to be a legal move where the pawn being captured is entangled with another piece of the same color as the pawn performing the move. The possibility equation for Blocked E.P. is:
\begin{equation}\label{eq:BlockedEPPossibility}
	\mathcal{P}_{BEP} = (v_s = P) \wedge (v_t \in \{N,B,R,Q,K\}) \wedge(v_{ep} = p) \wedge ep(t,s,\mathcal{F})
\end{equation}
In this scenario the entangled piece will always be blocking the square the that the pawn is moving to. We can use the same measurements introduced for the Blocked Jump (eq. \ref{eq:BlockedJumpMeasurement}). Thus we can apply procedure \ref{alg:ExecuteMeasuringMove} with $U_m = U_{slide}(p,t,s)U_{slide}(p,c,ep)$ as illustrated with the circuit in figure \ref{fig:BlockedEPCircuit}, where each $U_{slide}$ is converted to a zero-controlled $U_{jump}$.

\subsubsection{Capture EP}\label{sec:epCapture}
\begin{figure}[htbp]
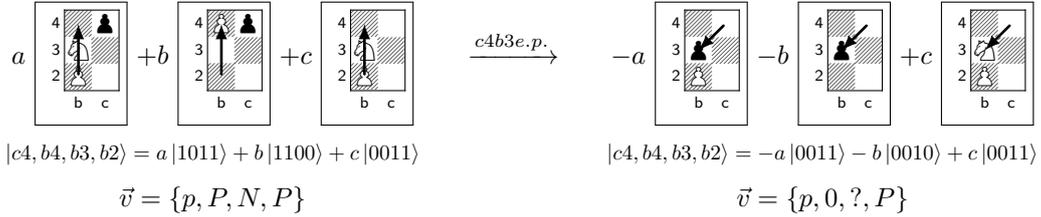

\captionsetup{width=.8\textwidth}
\begin{centering}
\def\arraystretch{1.5}
\begin{tabular}{c c c}
	\hbox{
    	$a$
	    \raisebox{-0.3\height}{\fbox{\chessboard[
		    setfen=8/8/8/8/2p5/1N6/1P6/8 w - - 0 1,
		    showmover=false,
		    printarea=b2-c4,
		    margintopwidth=0pt,
		    marginrightwidth=0pt,
		    marginleftwidth=8pt,
		    marginbottomwidth=10pt,
		    tinyboard,
		    pgfstyle=straightmove,
		    markmoves={b2-b4}
	    ]}} 
	    $+ b$
	    \raisebox{-0.3\height}{\fbox{\chessboard[
		    setfen=8/8/8/8/1Pp5/8/8/8 w - - 0 1,
		    showmover=false,
		    printarea=b2-c4,
		    margintopwidth=0pt,
		    marginrightwidth=0pt,
		    marginleftwidth=8pt,
		    marginbottomwidth=10pt,
		    tinyboard,
		    pgfstyle=straightmove,
		    markmoves={b2-b4}
	    ]}} 
	    $+ c$
	    \raisebox{-0.3\height}{\fbox{\chessboard[
		    setfen=8/8/8/8/8/1N6/1P6/8 w - - 0 1,
		    showmover=false,
		    printarea=b2-c4,
		    margintopwidth=0pt,
		    marginrightwidth=0pt,
		    marginleftwidth=8pt,
		    marginbottomwidth=10pt,
		    tinyboard,
		    pgfstyle=straightmove,
		    markmoves={b2-b4}
	    ]}} 
    }
    & $\xrightarrow{\text{$c4b3 e.p.$}}$ &
    \hbox{
    	$-a$
	    \raisebox{-0.3\height}{\fbox{\chessboard[
		    setfen=8/8/8/8/8/1p6/1P6/8 w - - 0 1,
		    showmover=false,
		    printarea=b2-c4,
		    margintopwidth=0pt,
		    marginrightwidth=0pt,
		    marginleftwidth=8pt,
		    marginbottomwidth=10pt,
		    tinyboard,
		    pgfstyle=straightmove,
		    markmoves={c4-b3}
	    ]}} 
	    $-b$
	    \raisebox{-0.3\height}{\fbox{\chessboard[
		    setfen=8/8/8/8/8/1p6/8/8 w - - 0 1,
		    showmover=false,
		    printarea=b2-c4,
		    margintopwidth=0pt,
		    marginrightwidth=0pt,
		    marginleftwidth=8pt,
		    marginbottomwidth=10pt,
		    tinyboard,
		    pgfstyle=straightmove,
		    markmoves={c4-b3}
	    ]}}
	    $+ c$
	    \raisebox{-0.3\height}{\fbox{\chessboard[
		    setfen=8/8/8/8/8/1N6/1P6/8 w - - 0 1,
		    showmover=false,
		    printarea=b2-c4,
		    margintopwidth=0pt,
		    marginrightwidth=0pt,
		    marginleftwidth=8pt,
		    marginbottomwidth=10pt,
		    tinyboard,
		    pgfstyle=straightmove,
		    markmoves={c4-b3}
	    ]}} 
    }
    \\
    \begin{tabular}{c}
	    \scalebox{0.8}{$\ket{c4,b4,b3,b2} = a\ket{1011} + b\ket{1100} + c\ket{0011}$} \\
	    $\vec{v} = \{p,P,N,P\}$ 
    \end{tabular}
    & & 
    \begin{tabular}{c}
	    \scalebox{0.8}{$\ket{c4,b4,b3,b2} = -a\ket{0011} -b\ket{0010} + c\ket{0011}$} \\
	    $\vec{v} = \{p,0,?,P\}$
    \end{tabular}
\end{tabular}
\end{centering}
\caption{\small Capture E.P. is possible when the pawn being captured is entangled with another piece of the same color. Capture is legal on both boards in the superposition. In one case it is through e.p. In the other it is just standard pawn capture. Double Occupancy can occur in the presence of superposition of the moving pawn.}
\label{fig:EPCaptureDoubleOccupancy}
\end{figure}

Capture E.P. is a special case of en passant where the target square is occupied by an opponent piece that is entangled with the pawn being captured by e.p. (see figure \ref{fig:EPCaptureDoubleOccupancy}). In this case the capture move is legal both under e.p. conditions, as well as conditions where normal pawn capture would be legal. This move can result in Double Occupancy in the presence of superposition. We can use the same measurement operator designed for the Capture Jump (eq. \ref{eq:CaptureJumpMeasurement}).

\begin{figure}[ht]
\captionsetup{width=.8\textwidth}
\centering
\mbox{
	\Qcircuit @C=1em @R=1em{
	&						   &   &  		&\mbox{Measure}&	   &   &        &	&   && 			  &&&\\
	&\lstick{\ket{s}} 	       &\qw&\ctrl{6}&\qw&\qw			   &\qw&\qw     &\qw&\qw&\qw 					     &\multigate{1}{U_{jump}}&\qw&\qw\\
	&\lstick{\ket{t}} 	       &\qw&\qw     &\qw&\qw			   &\qw&\qw     &\ctrl{4}&\qw&\multigate{1}{U_{jump}}&\ghost{U_{jump}} 		 &\qw&\qw\\
	&\lstick{\ket{c_2}=\ket{0}}&\qw&\qw	    &\qw&\qw 			   &\qw&\qw     &\qw &\qw&\ghost{U_{jump}}		  &\qw					 &\qw&\qw\\
	&\lstick{\ket{ep}} 	       &\qw&\qw     &\qw&\qw			   &\qw&\ctrl{2}&\qw&\multigate{1}{U_{jump}}&\qw&\qw 		 &\qw&\qw\\
	&\lstick{\ket{c_1}=\ket{0}}&\qw&\qw	    &\qw&\qw 			   &\qw&\qw      &\qw&\ghost{U_{jump}}		  &\qw&\qw					 &\qw&\qw\\
	&\lstick{\ket{p}=\ket{1}}  &\qw&\qw	    &\qw&\qw 			   &\qw&\targ   &\targ&\ctrlo{-1}   &\ctrlo{-3}&\ctrlo{-4}			     &\qw&\qw\\
	&\lstick{\ket{a}=\ket{0}}  &\qw&\targ   &\qw&\measure{\mbox{a}}&\cw&\cw&\cw&\control\cw\cwx[-1]&\control\cw\cwx[-1]&\control\cw\cwx[-1]
	\gategroup{2}{4}{8}{6}{0.7em}{--}
	}
}
\caption{\small Quantum circuit diagram for applying the Capture E.P. move to the source ($s$), target ($t$), and e.p. target (ep) qubits from the quantum register $\underline{\psi_\mathcal{B}}$. Two captured ancillas ($c_1$,$c_2$) are added to hold the captured pieces. Three zero-controlled $U_{jump}$ (eq. \ref{eq:JumpUnitary}) operations are conditionally applied if the measurement outcome is $M_1$ (source is occupied). Note: given the rules of the game and entanglement it is impossible for both the target and the e.p. target to be occupied in a single basis state if e.p. is valid.}
\label{fig:CaptureEPCircuit}
\end{figure}
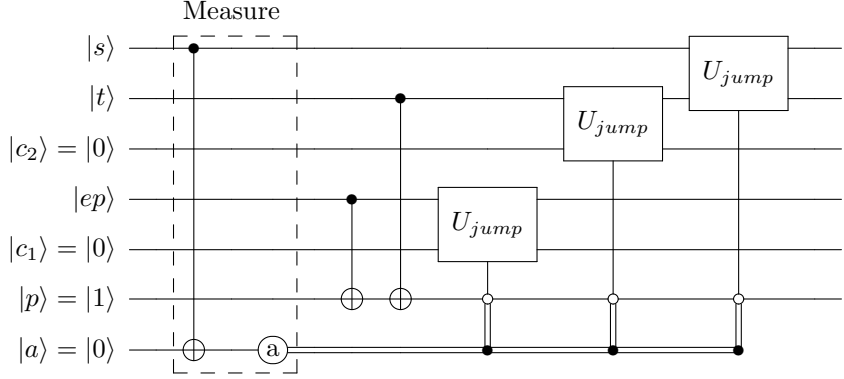

To perform the move we make use of two captured ancilla squares and a single control ($p$). The control should be set to $\ket{0}$ if the source is occupied, and either the target is occupied or the e.p. target is occupied. We can then follow procedure \ref{alg:ExecuteMeasuringMove} with $U_m = U_{slide}(p,t,s)U_{slide}(p,c_2,t)U_{slide}(p,c_1,ep)$ as illustrated with the circuit in figure \ref
{fig:CaptureEPCircuit}, where each $U_{slide}$ has been converted into zero-controlled $U_{jump}$.

\subsection{Castling}\label{sec:Castling}
We treat castling as a purely classical move, meaning we do not allow castling to be used in a Split or Merge. Similar to standard chess, the conditions under which castling is legal are
\begin{enumerate}
	\item The king has not been involved in any previous move.
	\item The rook which is participating in the castle has not been involved in any previous move.
\end{enumerate}
There are some subtle differences with the classical rules worth noting. Superposition means we could find the game in a state where either piece has been involved in a move and yet has not actually moved. We do not make any distinction for this ruleset. We say ANY move that involves with the king or rook in any way invalidates castling, even if the result is a projection into a subspace where the king or rook did not actually change.\footnote{It is possible to allow castling to be both legal and not legal, it would just require a corresponding superposition of "legality" flags for the occupancy boards in the superposition.} The other difference is that there is no concept of check, therefore any situation where check would invalidate castling in classical chess does not apply here.

\subsubsection{King Side Castle}\label{sec:ksMeasure}
\begin{figure}[htbp]
\captionsetup{width=.8\textwidth}
\begin{centering}
\def\arraystretch{1.5}
\begin{tabular}{c c c}
	\hbox{
    	$a$
	    \raisebox{-0.3\height}{\fbox{\chessboard[
		    setfen=8/8/8/8/8/8/8/4K2R w - - 0 1,
		    showmover=false,
		    printarea=e1-h2,
		    margintopwidth=0pt,
		    marginrightwidth=0pt,
		    marginleftwidth=8pt,
		    marginbottomwidth=10pt,
		    tinyboard
	    ]}} 
	    $+ b$
	    \raisebox{-0.3\height}{\fbox{\chessboard[
		    setfen=8/8/8/8/8/8/8/4K1BR w - - 0 1,
		    showmover=false,
		    printarea=e1-h2,
		    margintopwidth=0pt,
		    marginrightwidth=0pt,
		    marginleftwidth=8pt,
		    marginbottomwidth=10pt,
		    tinyboard
	    ]}} 
    }
    & $\xrightarrow{\text{$e1g1$}}$ &
    \hbox{
    	$-a$
	    \raisebox{-0.3\height}{\fbox{\chessboard[
		    setfen=8/8/8/8/8/8/8/5RK1 w - - 0 1,
		    showmover=false,
		    printarea=e1-h2,
		    margintopwidth=0pt,
		    marginrightwidth=0pt,
		    marginleftwidth=8pt,
		    marginbottomwidth=10pt,
		    tinyboard
	    ]}} 
	    $+ b$
	    \raisebox{-0.3\height}{\fbox{\chessboard[
		    setfen=8/8/8/8/8/8/8/4K1BR w - - 0 1,
		    showmover=false,
		    printarea=e1-h2,
		    margintopwidth=0pt,
		    marginrightwidth=0pt,
		    marginleftwidth=8pt,
		    marginbottomwidth=10pt,
		    tinyboard
	    ]}}
    }
    \\
    \begin{tabular}{c}
	    $\ket{e,f,g,h} = a\ket{1001} + b\ket{1011}$\\
	    $\vec{v} = \{K,0,B,R\}$ 
    \end{tabular}
    & & 
    \begin{tabular}{c}
	    $\ket{e,f,g,h} = -a\ket{0110} + b\ket{1011}$ \\
	    $\vec{v} = \{K,R,?,R\}$
    \end{tabular}
\end{tabular}
\end{centering}
\caption{\small Double Occupancy would occur if we attempted castling move e1g1. This must be prevented by applying a projective measurement before executing the move.}
\label{fig:KSDoubleOccupancy}
\end{figure}
\begin{figure}[ht]
\captionsetup{width=.8\textwidth}
\centering
\mbox{
	\Qcircuit @C=1em @R=1em{
	&							   &   &  		 &\mbox{Measure}&	&   &   &					 	&& 			 	&&\\
	&\lstick{\ket{e}} 	   &\qw&\qw&\qw&\qw			&\qw&\qw&\qw 					&\qw&\multigate{1}{U_{jump}}&\qw&\qw\\
	&\lstick{\ket{g}} 	   &\qw&\ctrlo{1}  &\qw&\qw				&\qw&\qw&\qw                    &\qw&\ghost{U_{jump}} 		&\qw&\qw\\
	&\lstick{\ket{f}}      &\qw&\ctrlo{2} &\qw&\qw 				&\qw&\qw&\multigate{1}{U_{jump}}		&\qw&\qw					&\qw&\qw\\
	&\lstick{\ket{h}}      &\qw&\qw	 &\qw&\qw 				&\qw&\qw&\ghost{U_{jump}}		&\qw&\qw					&\qw&\qw\\
	&\lstick{\ket{a}=\ket{0}}&\qw&\targ 	&\qw&\measure{\mbox{a}}&\cw&\cw&\control \cw \cwx[-1]  &\cw&\control \cw \cwx[-3]
	\gategroup{2}{4}{6}{6}{0.7em}{--}
	}
}
\caption{\small Quantum circuit diagram for applying the King Side Castle move to qubits representing the occupancy of squares $e,f,g,h$. $M_1$ (eq. \ref{eq:CastleMeasurement}) is encoded into measurement ancilla ($a$) and two $U_{jump}$ (eq. \ref{eq:JumpUnitary}) operations are conditionally applied based on the measurement outcome.}
\label{fig:KingSideCastleCircuit}
\end{figure}
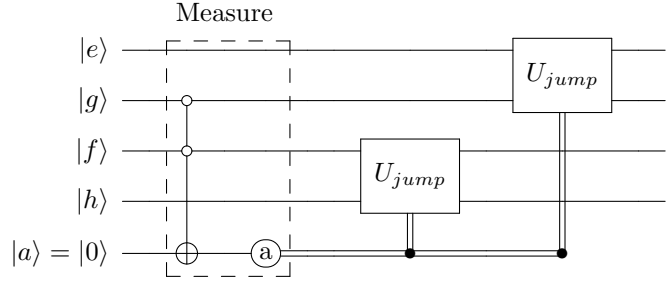

King side castling involves squares in columns $\{e,f,g,h\}$. The possibility equation for this move is:
\begin{equation}\label{eq:KingSideCastlePossibility}
	\mathcal{P}_{KS} = (v_e = K) \wedge (v_h = R) \wedge (F_K = True)
\end{equation}
This move can lead to double occupancy in cases of superposition of the target squares f and g (see figure \ref{fig:KSDoubleOccupancy}). We find the following mutually exclusive subsets of basis states in the $\ket{f,g}$ basis:
\begin{equation*}
\mathcal{M}_0=\{\ket{01},\ket{10},\ket{11} , \mathcal{M}_1=\{\ket{00}\}
\end{equation*}
From these two subsets we can begin constructing our measurement operators, $M_0$ and $M_1$ respectively.:
\begin{subequations}
\begin{IEEEeqnarray}{rCl}
M_0 &=& \ket{01}\bra{01} + \ket{10}\bra{10} + \ket{11}\bra{11}\\
M_1 &=& \ket{00}\bra{00} \label{eq:CastleMeasurement}
\end{IEEEeqnarray}
\end{subequations}
It is worth noting that if castling is determined to be possible, $\mathcal{F}_K = True$, it is not possible for squares e or h to be in superposition. We have defined castling to only be possible if neither the king or the rook has been involved in any previous move. If castling is possible, and we perform this measurement and find result 1, we can say with certainty that the state of squares e,f,g,h that we have projected into is $\ket{1001}$. We can thus apply procedure \ref{alg:ExecuteMeasuringMove} with $U_m = U_{jump}(e,g)U_{jump}(f,h)$, as illustrated in the circuit in figure \ref{fig:KingSideCastleCircuit}. These operators commute and can be performed simultaneously. 

\subsubsection{Queen Side Castle}\label{sec:qsMeasure}
\begin{figure}[htbp]
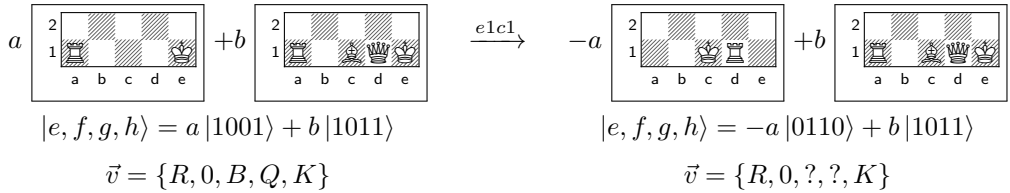

\captionsetup{width=.8\textwidth}
\begin{centering}
\def\arraystretch{1.5}
\begin{tabular}{c c c}
	\hbox{
    	$a$
	    \raisebox{-0.3\height}{\fbox{\chessboard[
		    setfen=8/8/8/8/8/8/8/R3K3 w - - 0 1,
		    showmover=false,
		    printarea=a1-e2,
		    margintopwidth=0pt,
		    marginrightwidth=0pt,
		    marginleftwidth=8pt,
		    marginbottomwidth=10pt,
		    tinyboard
	    ]}} 
	    $+ b$
	    \raisebox{-0.3\height}{\fbox{\chessboard[
		    setfen=8/8/8/8/8/8/8/R1BQK w - - 0 1,
		    showmover=false,
		    printarea=a1-e2,
		    margintopwidth=0pt,
		    marginrightwidth=0pt,
		    marginleftwidth=8pt,
		    marginbottomwidth=10pt,
		    tinyboard
	    ]}} 
    }
    & $\xrightarrow{\text{$e1c1$}}$ &
    \hbox{
    	$-a$
	    \raisebox{-0.3\height}{\fbox{\chessboard[
		    setfen=8/8/8/8/8/8/8/2KR4 w - - 0 1,
		    showmover=false,
		    printarea=a1-e2,
		    margintopwidth=0pt,
		    marginrightwidth=0pt,
		    marginleftwidth=8pt,
		    marginbottomwidth=10pt,
		    tinyboard
	    ]}} 
	    $+ b$
	    \raisebox{-0.3\height}{\fbox{\chessboard[
		    setfen=8/8/8/8/8/8/8/R1BQK w - - 0 1,
		    showmover=false,
		    printarea=a1-e2,
		    margintopwidth=0pt,
		    marginrightwidth=0pt,
		    marginleftwidth=8pt,
		    marginbottomwidth=10pt,
		    tinyboard
	    ]}}
    }
    \\
    \begin{tabular}{c}
	    $\ket{e,f,g,h} = a\ket{1001} + b\ket{1011}$\\
	    $\vec{v} = \{R,0,B,Q,K\}$ 
    \end{tabular}
    & & 
    \begin{tabular}{c}
	    $\ket{e,f,g,h} = -a\ket{0110} + b\ket{1011}$ \\
	    $\vec{v} = \{R,0,?,?,K\}$
    \end{tabular}
\end{tabular}
\end{centering}
\caption{\small Double Occupancy would occur if we attempted castling move e1c1. This must be prevented by applying a projective measurement before executing the move.}
\label{fig:QSDoubleOccupancy}
\end{figure}
\begin{figure}[ht]
\captionsetup{width=.8\textwidth}
\centering
\mbox{
	\Qcircuit @C=1em @R=1em{
	&					   &   &  		 &\mbox{Measure}&	    &   &   &   &&                     &   &\\
	&\lstick{\ket{a}} 	   &\qw&\qw      &\qw&\qw			    &\qw&\qw 	      &\qw&\multigate{1}{U_{jump}}&\qw\\
	&\lstick{\ket{d}} 	   &\qw&\ctrlo{1}&\qw&\qw				&\qw&\qw          &\qw&\ghost{U_{jump}} 	  &\qw\\
	&\lstick{\ket{c}}      &\qw&\ctrlo{3}&\qw&\qw 				&\qw&\qw          &\multigate{1}{U_{jump}}&\qw&\qw\\
	&\lstick{\ket{e}}      &\qw&\qw	     &\qw&\qw 				&\qw&\qw          &\ghost{U_{jump}}		  &\qw&\qw\\
	&\lstick{\ket{b}}      &\qw&\qw      &\qw&\qw 				&\qw&\qw&\ctrlo{-1}&\ctrlo{-3}&\qw\\
	&\lstick{\ket{a}=\ket{0}}&\qw&\targ  &\qw&\measure{\mbox{a}}&\cw&\cw&\control \cw \cwx[-1]&\control \cw \cwx[-1]
	\gategroup{2}{4}{7}{6}{0.7em}{--}
	}
}
\caption{\small Quantum circuit diagram for applying the Queen Side Castle move to the qubits representing squares in files $a,b,c,d,e$. $M_1$ (eq. \ref{eq:CaptureJumpMeasurement}) is encoded into measurement ancilla ($a$) and two zero-controlled $U_{jump}$ (eq. \ref{eq:JumpUnitary}) operations are conditionally applied if the measurement outcome is 1.}
\label{fig:QueenSideCastleCircuit}
\end{figure}
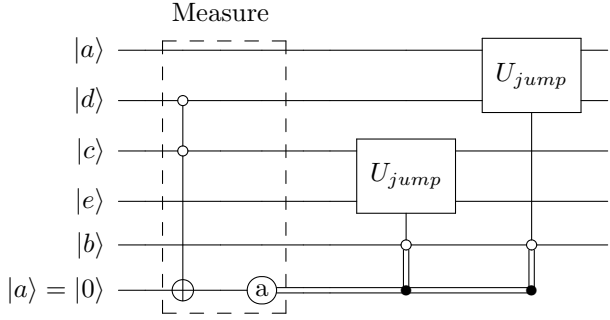

Queen side castling involves squares in columns $\{a,b,c,d,e\}$. The possibility equation for this move is:
\begin{equation}\label{eq:QueenSideCastlePossibility}
	\mathcal{P}_{QS} = (v_e = K) \wedge (v_a = R) \wedge (F_Q = True)
\end{equation}
This move can lead to double occupancy in cases of superposition of the target squares c and d (see figure \ref{fig:QSDoubleOccupancy}). We can use the same measurement designed for King Side Castling (eq. \ref{eq:CastleMeasurement}). As was the case with King Side Castling, if this move is determined to be possible, $F_Q = True$, it is not possible for squares a or e to be in superposition. If castling is possible, and we perform this measurement and find result 1, we can say with certainty that squares a and e are in state $\ket{1}$. The move must still be a controlled operation because square b may be in superposition. We can thus apply procedure \ref{alg:ExecuteMeasuringMove} with $U_m = U_{slide}(b,a,d)U_{slide}(b,c,e)$, as illustrated in the circuit in figure \ref{fig:QueenSideCastleCircuit} with the $U_{slide}$ operators converted to zero-controlled $U_{jump}$ operators.

\section{Classical Simulation}
\begin{figure}[h]
\begin{centering}
\captionsetup{width=.8\textwidth}
\includegraphics[width=0.5\textwidth]{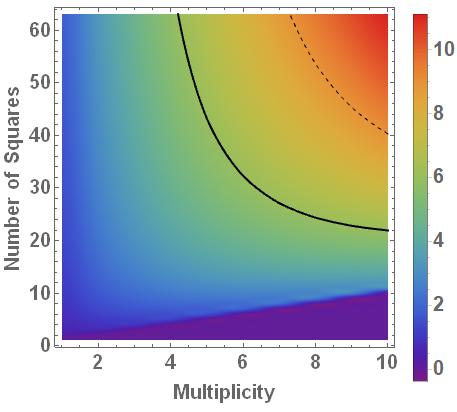}
\caption{\small Heat map of the $Log_{10}$ of the upper bound on the size of the superposition as we vary the multiplicity of a single piece in superposition and the number of squares that piece occupies. The solid black contour line shows the $10^6$ cutoff, where the maximum size of the superposition is about 1 million basis states. The dashed line is the $10^9$ contour.}
\label{fig:SizeMultiplicity}
\end{centering}
\end{figure}

When we consider the classical simulability of the game we must worry about the size ($\mathcal{S}$) of the superposition. As our state grows to include more possible board configurations we must contend with limits in both memory and execution time. If we wish for the game to be playable by the general public we need to keep the use of these resources at a reasonable level. 

Modern CPUs have chips that run in the GHz range. If we imagine an algorithm for calculating the outcome of a move to be O(N) in the size of the superposition $\mathcal{S}$ then our run time will be on the order of $\mathcal{S}\times 10^{-9}$ seconds. A reasonable maximum state size should likely not be much larger than $10^{9}$ for single move calculation times on the order of one second. Considering our state representation this also gives us a memory size on the order of a gigabyte which is on the high end of reasonable assumption for modern personal machines. To account for the actual execution time of a move algorithm (more than one clock cycle), and in an attempt to decrease the worst case memory consumption we will shoot for a more modest maximum size of $10^6$.

A basic approach to calculating the upper-bound on the size of the superposition would be to consider that we are simply storing the occupancy in superposition. For a given number of pieces (k), the maximum size of the superposition would simply be the number of ways of arranging k 1's on the board. Superposition allows for varying numbers of pieces to exist when some pieces have been partially captured. A naive upper bound would then be the sum of the maximum size for a given k over the possible values of k. The maximum number of pieces on a board is 32 so we get:
\begin{equation}
	\label{eq:NaiveSize} \mathcal{S} = \sum_{k=0}^{32}\binom{64}{k} \approx 1.0\times 10^{19}
\end{equation}
This is still quite large. We can improve on this by considering the effect of the No Double Occupancy rule. Each piece of a different value must be confined to a separate subset of squares. Any overlap will result in a collapse. Thus our formula becomes:
\begin{subequations}
\begin{IEEEeqnarray}{rCl}
\label{eq:SubspaceSize} S_v &=& \sum_{m=0}^{m_v} \binom{s_v}{m} \\
\label{eq:SuperpositionSize} \mathcal{S} &=& \prod_{v_w\in V_w} S_{v_w} \prod_{v_b\in V_b} S_{v_b}
\end{IEEEeqnarray}
\end{subequations}
where $V_w = \{P,N,B,R,Q,K\}$ and $V_b = \{p,n,b,r,q,k\}$ are the sets of distinct piece values for the white and black players respectively, $m_v$ is the maximum multiplicity of the pieces of value v, and $s_v$ is the number of squares occupied by the pieces of value v with the following restrictions:
\begin{subequations}
\begin{IEEEeqnarray}{rCl}
\sum_{v\in V_w \cup V_b} s_v &\leq& 64\\
m_k &=& m_K = 1\\
m_{v_w} &\leq& (8+m_{v_w,0}) - \sum_{u\in V_w\backslash {P,K}} (m_{u_w}-m_{u_w,0}) \\
m_{v_b} &\leq& (8+m_{v_b,0}) - \sum_{u\in V_b\backslash {p,k}} (m_{u_b}-m_{u_b,0})
\end{IEEEeqnarray}
\end{subequations}
where $m_{v,0}$ is the number of pieces of value v that start on the board. The restrictions on $m_{v_{w(b)}}$ come from the promotion rules of chess. Pieces that begin the game with multiplicity 2 can possibly achieve multiplicity 10 by promoting all 8 pawns. Let $\zeta_v=\{s_v,m_v\}$ denote the number of occupied squares and multiplicity for a piece value v. Solving equation \ref{eq:SuperpositionSize} numerically we find that the new upper bound is about $7.9 \times 10^{17}$ and occurs when $\zeta_p = \zeta_P = \{0,0\}$, $\zeta_k = \zeta_K = \zeta_q = \zeta_Q = \{1,1\}$, $\zeta_n = \zeta_N = \zeta_b = \zeta_B = \{3,2\}$, and $\zeta_r = \zeta_R = \{24,10\}$\footnote{Swapping rooks with knights will give the same upper bound}. This upper bound is dominated by the single piece with maximum multiplicity subset. We can see how the upper bound behaves as a player spreads a single piece in superposition over the board. Figure \ref{fig:SizeMultiplicity} shows that the simulation should remain tractable until we begin seeing a piece with $m_v > 5$ and $s_v > 40$.

\section{Rules}
Now that we have completed the design of the tools we will need to make Quantum Chess we can lay out the rules.
\begin{enumerate}
	\item A 64x64 board is set up exactly as in standard Chess.
	\item No square can ever be in a state of being occupied by more than one piece. A square is considered to be occupied by a piece if there is a non-zero probability of finding the piece in that square.
	\item Players alternate turns with \textit{white} moving first.
	\item Pieces follow the same movement patterns as in standard Chess.
	\item There is no concept of check or checkmate. Kings are captured like any other piece.
	\item All pieces have access to a "standard equivalent" move.
	\item Non-pawns may access Split and Merge moves.
	\item A player may choose from any of the moves outlined in section \ref{sec:QuantumChessMoves}. The player indicates the move to be done as follows:
	\begin{itemize}
		\item Two Square Move: For any move involving a single source and single target for the moving piece. The player specifies a single source square occupied by a piece belonging to the player, and a target square that defines a valid possible move as determined by the possibility equations. This applies to all standard Chess equivalent moves, including e.p. and castling for which the other squares involved can be extrpolated by the indicated source and target.
		\item Split Move: The player specifies a single source occupied by a piece belonging to the player, and two target squares that define a valid possible move as determined by the Split Jump (\ref{sec:SplitJump}) and Split Slide (\ref{sec:SplitSlide}) possibility equations. 
		\item Merge Move: The player specifies two source squares each occupied by the same piece belonging to the player, and a single target square that defines a valid possible move as determined by the Merge Jump (\ref{sec:MergeJump}) and Merge Slide (\ref{sec:MergeSlide}) possibility equations. 
	\end{itemize}
	\item If the move has no effect the player repeats their turn with a different move. See \ref{def:LegalMove}
	\item If a move results in a state where one player has zero probability of having a king on the board then the opposing player wins. 
	\item If a move results in a state where both players have zero probability of having a king on the board then the game ends in a draw.
\end{enumerate}

\section{Quantum Effects In Game}
Now that we have finished the construction of Quantum Chess it would be helpful to determine whether it can actually result in non-trivial quantum effects. More specifically, does the construction achieve the goal of giving players access to superposition, entanglement, and interference? Superposition is trivially accessible through simple use of Split moves. Here we will show how to generate both entanglement and interference in game.

\subsection{Entanglement: Quantum Chess Bell States}
\begin{figure}[htbp]
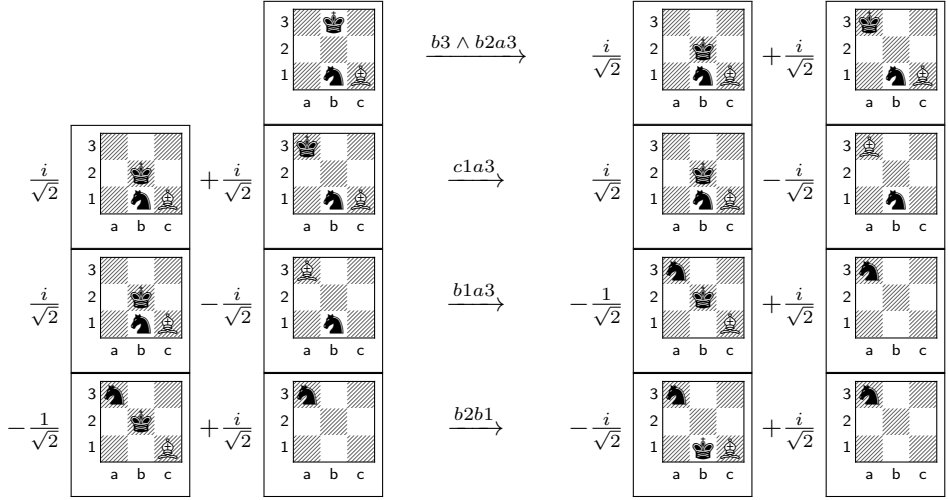

\captionsetup{width=.8\textwidth}
\begin{centering}
\def\arraystretch{1.5}
\begin{tabular}{r c r}
	\hbox{
	    \raisebox{-0.3\height}{\fbox{\chessboard[
		    setfen=8/8/8/8/8/1k6/8/1nB5 w - - 0 1,
		    showmover=false,
		    printarea=a1-c3,
		    margintopwidth=0pt,
		    marginrightwidth=0pt,
		    marginleftwidth=8pt,
		    marginbottomwidth=10pt,
		    tinyboard
	    ]}} 
    }
    & $\xrightarrow{\text{$b3\wedge b2a3$}}$ &
    \hbox{
    	$\frac{i}{\sqrt2}$
	    \raisebox{-0.3\height}{\fbox{\chessboard[
		    setfen=8/8/8/8/8/8/1k6/1nB5 w - - 0 1,
		    showmover=false,
		    printarea=a1-c3,
		    margintopwidth=0pt,
		    marginrightwidth=0pt,
		    marginleftwidth=8pt,
		    marginbottomwidth=10pt,
		    tinyboard
	    ]}} 
	    $+ \frac{i}{\sqrt2}$
	    \raisebox{-0.3\height}{\fbox{\chessboard[
		    setfen=8/8/8/8/8/k7/8/1nB5 w - - 0 1,
		    showmover=false,
		    printarea=a1-c3,
		    margintopwidth=0pt,
		    marginrightwidth=0pt,
		    marginleftwidth=8pt,
		    marginbottomwidth=10pt,
		    tinyboard
	    ]}}
    }
    \\
    \hbox{
    	$\frac{i}{\sqrt2}$
	    \raisebox{-0.3\height}{\fbox{\chessboard[
		    setfen=8/8/8/8/8/8/1k6/1nB5 w - - 0 1,
		    showmover=false,
		    printarea=a1-c3,
		    margintopwidth=0pt,
		    marginrightwidth=0pt,
		    marginleftwidth=8pt,
		    marginbottomwidth=10pt,
		    tinyboard
	    ]}} 
	    $+ \frac{i}{\sqrt2}$
	    \raisebox{-0.3\height}{\fbox{\chessboard[
		    setfen=8/8/8/8/8/k7/8/1nB5 w - - 0 1,
		    showmover=false,
		    printarea=a1-c3,
		    margintopwidth=0pt,
		    marginrightwidth=0pt,
		    marginleftwidth=8pt,
		    marginbottomwidth=10pt,
		    tinyboard
	    ]}}
    }
    & $\xrightarrow{\text{$c1a3$}}$ &
    \hbox{
    	$\frac{i}{\sqrt2}$
	    \raisebox{-0.3\height}{\fbox{\chessboard[
		    setfen=8/8/8/8/8/8/1k6/1nB5 w - - 0 1,
		    showmover=false,
		    printarea=a1-c3,
		    margintopwidth=0pt,
		    marginrightwidth=0pt,
		    marginleftwidth=8pt,
		    marginbottomwidth=10pt,
		    tinyboard
	    ]}} 
	    $- \frac{i}{\sqrt2}$
	    \raisebox{-0.3\height}{\fbox{\chessboard[
		    setfen=8/8/8/8/8/B7/8/1n6 w - - 0 1,
		    showmover=false,
		    printarea=a1-c3,
		    margintopwidth=0pt,
		    marginrightwidth=0pt,
		    marginleftwidth=8pt,
		    marginbottomwidth=10pt,
		    tinyboard
	    ]}}
    }
    \\
    \hbox{
    	$\frac{i}{\sqrt2}$
	    \raisebox{-0.3\height}{\fbox{\chessboard[
		    setfen=8/8/8/8/8/8/1k6/1nB5 w - - 0 1,
		    showmover=false,
		    printarea=a1-c3,
		    margintopwidth=0pt,
		    marginrightwidth=0pt,
		    marginleftwidth=8pt,
		    marginbottomwidth=10pt,
		    tinyboard
	    ]}} 
	    $- \frac{i}{\sqrt2}$
	    \raisebox{-0.3\height}{\fbox{\chessboard[
		    setfen=8/8/8/8/8/B7/8/1n6 w - - 0 1,
		    showmover=false,
		    printarea=a1-c3,
		    margintopwidth=0pt,
		    marginrightwidth=0pt,
		    marginleftwidth=8pt,
		    marginbottomwidth=10pt,
		    tinyboard
	    ]}}
    }
    & $\xrightarrow{\text{$b1a3$}}$ &
    \hbox{
    	$-\frac{1}{\sqrt2}$
	    \raisebox{-0.3\height}{\fbox{\chessboard[
		    setfen=8/8/8/8/8/n7/1k6/2B5 w - - 0 1,
		    showmover=false,
		    printarea=a1-c3,
		    margintopwidth=0pt,
		    marginrightwidth=0pt,
		    marginleftwidth=8pt,
		    marginbottomwidth=10pt,
		    tinyboard
	    ]}} 
	    $+ \frac{i}{\sqrt2}$
	    \raisebox{-0.3\height}{\fbox{\chessboard[
		    setfen=8/8/8/8/8/n7/8/8 w - - 0 1,
		    showmover=false,
		    printarea=a1-c3,
		    margintopwidth=0pt,
		    marginrightwidth=0pt,
		    marginleftwidth=8pt,
		    marginbottomwidth=10pt,
		    tinyboard
	    ]}}
    }
    \\
    \hbox{
    	$-\frac{1}{\sqrt2}$
	    \raisebox{-0.3\height}{\fbox{\chessboard[
		    setfen=8/8/8/8/8/n7/1k6/2B5 w - - 0 1,
		    showmover=false,
		    printarea=a1-c3,
		    margintopwidth=0pt,
		    marginrightwidth=0pt,
		    marginleftwidth=8pt,
		    marginbottomwidth=10pt,
		    tinyboard
	    ]}} 
	    $+ \frac{i}{\sqrt2}$
	    \raisebox{-0.3\height}{\fbox{\chessboard[
		    setfen=8/8/8/8/8/n7/8/8 w - - 0 1,
		    showmover=false,
		    printarea=a1-c3,
		    margintopwidth=0pt,
		    marginrightwidth=0pt,
		    marginleftwidth=8pt,
		    marginbottomwidth=10pt,
		    tinyboard
	    ]}}
    }
    & $\xrightarrow{\text{$b2b1$}}$ &
    \hbox{
    	$-\frac{i}{\sqrt2}$
	    \raisebox{-0.3\height}{\fbox{\chessboard[
		    setfen=8/8/8/8/8/n7/8/1kB5 w - - 0 1,
		    showmover=false,
		    printarea=a1-c3,
		    margintopwidth=0pt,
		    marginrightwidth=0pt,
		    marginleftwidth=8pt,
		    marginbottomwidth=10pt,
		    tinyboard
	    ]}} 
	    $+ \frac{i}{\sqrt2}$
	    \raisebox{-0.3\height}{\fbox{\chessboard[
		    setfen=8/8/8/8/8/n7/8/8 w - - 0 1,
		    showmover=false,
		    printarea=a1-c3,
		    margintopwidth=0pt,
		    marginrightwidth=0pt,
		    marginleftwidth=8pt,
		    marginbottomwidth=10pt,
		    tinyboard
	    ]}}
    }
\end{tabular}
\end{centering}
\caption{\small A Split Jump ($b3\wedge b2a3$) followed by a Capture Slide ($c1a3$), a Capture Jump ($b1a3$), and a Standard Jump ($b2b1$) help create one of the Bell States. The occupancy of the final state of squares c1, b1, and a3 is $\ket{a3,b1,c1}=\frac{i}{\sqrt{2}}(-\ket{111} + \ket{100})$}
\label{fig:CaptureBellState}
\end{figure}
The prototypical entangled states are the Bell States \cite{Nielsen:2011}:
\begin{IEEEeqnarray}{rCl}
\ket{\Phi^+}&=&\frac{1}{\sqrt{2}}(\ket{00}+\ket{11}) \label{eq:BellPhi+}\\
\ket{\Phi^-}&=&\frac{1}{\sqrt{2}}(\ket{00}-\ket{11}) \label{eq:BellPhi-}\\
\ket{\Psi^+}&=&\frac{1}{\sqrt{2}}(\ket{01}+\ket{10}) \label{eq:BellPsi+}\\
\ket{\Psi^-}&=&\frac{1}{\sqrt{2}}(\ket{01}-\ket{10}) \label{eq:BellPsi-}
\end{IEEEeqnarray}
These states are useful in a number of areas of quantum information including superdense coding \cite{BennettSuperDense:1992}\cite{Nielsen:2011}, and are used to rule out local hidden variable theories of quantum mechanics via the Bell-CHSH inequality\cite{CHSH:1969}. Here we show how to generate the Bell States in Quantum Chess by applying various sequences of moves to a simple starting state.

Section \ref{sec:SplitJump} shows us that the Split Jump, which applies a simple Split unitary to a three-qubit state, gets us close to our first Bell state $\ket{\Psi^-}$. Consider applying the moves $a1\wedge a2b1,\ b1a1$, and $b1a2$ to a board with a single king in position a1. Since there is only a single piece we will ignore the classical piece type information. The game begins in state $\ket{a2,b1,a1}=\ket{001}$. Applying the move unitaries we have
\begin{subequations}
\begin{IEEEeqnarray}{rCl}
U_{split}(a1,a2,b1)\ket{001} &=& \frac{i}{\sqrt{2}}(\ket{100}+\ket{010}) \\
U_{jump}(b1,a1)\frac{i}{\sqrt{2}}(\ket{100}+\ket{010}) &=& \frac{1}{\sqrt{2}}(i\ket{100}-\ket{001}) \\
U_{jump}(b1,a2)\frac{1}{\sqrt{2}}(i\ket{100}-\ket{001}) &=& \frac{1}{\sqrt{2}}(-\ket{010}-\ket{001})\label{eq:FinalToBellPsi+}
\end{IEEEeqnarray}
\end{subequations}
We can rewrite the final state in eq. \ref{eq:FinalToBellPsi+} to get $\ket{\Psi^+}$:
\begin{equation}
\ket{a2}\otimes \ket{b1,a1} = -\ket{0}\otimes [\frac{1}{\sqrt{2}}(\ket{10}+\ket{01})] = \ket{0}\otimes\ket{\Psi^+}
\end{equation}
We can phase rotate the $\ket{01}$ state by applying two more jump moves, $a1a2, a2a1$. In the basis $\ket{a2,b1,a1}$ we have:
\begin{subequations}
\begin{IEEEeqnarray}{rCl}
U_{jump}(a1,a2)\frac{1}{\sqrt{2}}(-\ket{010}-\ket{001}) &=& \frac{1}{\sqrt{2}}(-\ket{010}-i\ket{100}) \label{eq:JumpRotate1}\\
U_{jump}(a2,a1)\frac{1}{\sqrt{2}}(-\ket{010}-i\ket{100}) &=& \frac{1}{\sqrt{2}}(-\ket{010}+\ket{001}) \label{eq:JumpRotate2}
\end{IEEEeqnarray}
\end{subequations}
which can be rewritten to get $\ket{\Psi^-}$:
\begin{equation}
\ket{a2}\otimes \ket{b1,a1} = -\ket{0}\otimes [\frac{1}{\sqrt{2}}(\ket{10}-\ket{01})] = \ket{0}\otimes\ket{\Psi^-}
\end{equation}
To get $\ket{\Phi^+}$ and $\ket{\Phi^-}$ we will need to add more pieces. Consider the state and sequence of moves shown in figure \ref{fig:CaptureBellState}. These moves act on squares c1, b1, b2, b3, and a3, and two captured ancillas ($x_1$,$x_2$) must be introduced which will be thrown away at the end. The evolution of the occupancy state in the basis $\ket{x_2,x_1}\otimes\ket{a3,b3,b2,b1,c1}$ is as follows:
\begin{subequations}
\begin{IEEEeqnarray}{l}
U_{split}(a3,b2,b3)\ket{00}\ket{01011}=\frac{1}{\sqrt{2}}(i\ket{00}\ket{00111} + i\ket{00}\ket{10011})
\\
\begin{aligned}
U_{slide}(b2,a3,c1)U_{slide}(b2,x_1,a3) & \frac{1}{\sqrt{2}}(i\ket{00}\ket{00111} + i\ket{00}\ket{10011})\\
	& = \frac{1}{\sqrt{2}}(i\ket{00}\ket{00111} - i\ket{01}\ket{10010})
\end{aligned}
\\
\begin{aligned}
U_{jump}(b1,a3)U_{jump}(x_2,a3) & \frac{1}{\sqrt{2}}(i\ket{00}\ket{00111} - i\ket{01}\ket{10010})\\
	& = \frac{1}{\sqrt{2}}(-\ket{00}\ket{10101} + i\ket{11}\ket{10000})
\end{aligned}
\\
\begin{aligned}
U_{jump}(b2,b1) & \frac{1}{\sqrt{2}}(-\ket{00}\ket{10101} + i\ket{11}\ket{10000})\\
	& \qquad = \frac{1}{\sqrt{2}}(-i\ket{00}\ket{10011} + i\ket{11}\ket{10000})
\end{aligned}
\end{IEEEeqnarray}
\end{subequations}
If we trace out the ancillas and the qubits for empty squares we can rewrite the state in the basis $\ket{a3,b1,c1}$ to get $\ket{\Phi^-}$:
\begin{equation}
\ket{a3}\otimes \ket{b1,c1} = \ket{1}\otimes [\frac{i}{\sqrt{2}}(\ket{00}-\ket{11})] = i\ket{1}\otimes\ket{\Phi^-}
\end{equation}
As we did in equations \ref{eq:JumpRotate1} and \ref{eq:JumpRotate2}, we can pick up a phase rotation on the $\ket{11}$ state by applying two more king moves, $b1b2$ and $b2b1$, to get $\ket{\Phi^+}$:
\begin{equation}
\ket{a3}\otimes \ket{b1,c1} = \ket{1}\otimes [\frac{i}{\sqrt{2}}(\ket{00} + \ket{11})] = i\ket{1}\otimes\ket{\Phi^+}
\end{equation}

\subsection{Interference}
\begin{figure}[htbp]
\captionsetup{width=.8\textwidth}
\begin{centering}
\def\arraystretch{1.5}
\begin{tabular}{c}
	\begin{tabular}{r c r c r}
		& \text{$a1\wedge a2b1$} & & \text{$a2\wedge a1b2$} &\\
		\hbox{
		    \raisebox{-0.3\height}{\fbox{\chessboard[
			    setfen=8/8/8/8/8/8/8/K7 w - - 0 1,
			    showmover=false,
			    printarea=a1-b2,
			    margintopwidth=0pt,
			    marginrightwidth=0pt,
			    marginleftwidth=8pt,
			    marginbottomwidth=10pt,
			    smallboard
		    ]}} 
	    }\tikzmark{a}
		&&\begin{tabular}{r}
			\tikzmark{b}\hbox{
		    	$\frac{i}{\sqrt2}$
			    \raisebox{-0.3\height}{\fbox{\chessboard[
				    setfen=8/8/8/8/8/8/K7/8 w - - 0 1,
				    showmover=false,
				    printarea=a1-b2,
				    margintopwidth=0pt,
				    marginrightwidth=0pt,
				    marginleftwidth=8pt,
				    marginbottomwidth=10pt,
				    smallboard
			    ]}}
			}\tikzmark{d}
		    \\
		    $+\qquad$
		    \\
			\tikzmark{c}\hbox{
				$\frac{i}{\sqrt2}$
			    \raisebox{-0.3\height}{\fbox{\chessboard[
				    setfen=8/8/8/8/8/8/8/1K6 w - - 0 1,
				    showmover=false,
				    printarea=a1-b2,
				    margintopwidth=0pt,
				    marginrightwidth=0pt,
				    marginleftwidth=8pt,
				    marginbottomwidth=10pt,
				    smallboard
			    ]}}
		    }\tikzmark{e}
	    \end{tabular}
	    && 
	    \begin{tabular}{r}
			\tikzmark{f}\hbox{
		   		$-\frac{1}{2}$
			    \raisebox{-0.3\height}{\fbox{\chessboard[
				    setfen=8/8/8/8/8/8/8/K7 w - - 0 1,
				    showmover=false,
				    printarea=a1-b2,
				    margintopwidth=0pt,
				    marginrightwidth=0pt,
				    marginleftwidth=8pt,
				    marginbottomwidth=10pt,
				    smallboard
			    ]}} 
			} 
		    \\
		    $+\qquad$
		    \\
		    \tikzmark{g}\hbox{
				$- \frac{1}{2}$
			    \raisebox{-0.3\height}{\fbox{\chessboard[
				    setfen=8/8/8/8/8/8/1K6/8 w - - 0 1,
				    showmover=false,
				    printarea=a1-b2,
				    margintopwidth=0pt,
				    marginrightwidth=0pt,
				    marginleftwidth=8pt,
				    marginbottomwidth=10pt,
				    smallboard
			    ]}}
		    }
		    \\
		    $+\qquad$
		    \\
		    \tikzmark{h}\hbox{
				$\frac{i}{\sqrt2}$
			    \raisebox{-0.3\height}{\fbox{\chessboard[
				    setfen=8/8/8/8/8/8/8/1K6 w - - 0 1,
				    showmover=false,
				    printarea=a1-b2,
				    margintopwidth=0pt,
				    marginrightwidth=0pt,
				    marginleftwidth=8pt,
				    marginbottomwidth=10pt,
				    smallboard
			    ]}}
		    }
	    \end{tabular}
	\end{tabular}
	\\
	\begin{tabular}{r c r c r}
		& \text{$b1\wedge a1b2$} & & \text{interference} &\\
		\begin{tabular}{r}
			\hbox{
		   		$-\frac{1}{2}$
			    \raisebox{-0.3\height}{\fbox{\chessboard[
				    setfen=8/8/8/8/8/8/8/K7 w - - 0 1,
				    showmover=false,
				    printarea=a1-b2,
				    margintopwidth=0pt,
				    marginrightwidth=0pt,
				    marginleftwidth=8pt,
				    marginbottomwidth=10pt,
				    smallboard
			    ]}} 
			}\tikzmark{i}
		    \\
		    $+\qquad$
		    \\
		    \hbox{
				$- \frac{1}{2}$
			    \raisebox{-0.3\height}{\fbox{\chessboard[
				    setfen=8/8/8/8/8/8/1K6/8 w - - 0 1,
				    showmover=false,
				    printarea=a1-b2,
				    margintopwidth=0pt,
				    marginrightwidth=0pt,
				    marginleftwidth=8pt,
				    marginbottomwidth=10pt,
				    smallboard
			    ]}}
		    }\tikzmark{j}
		    \\
		    $+\qquad$
		    \\
		    \hbox{
				$\frac{i}{\sqrt2}$
			    \raisebox{-0.3\height}{\fbox{\chessboard[
				    setfen=8/8/8/8/8/8/8/1K6 w - - 0 1,
				    showmover=false,
				    printarea=a1-b2,
				    margintopwidth=0pt,
				    marginrightwidth=0pt,
				    marginleftwidth=8pt,
				    marginbottomwidth=10pt,
				    smallboard
			    ]}}
		    }\tikzmark{k}
	    \end{tabular}
		&&
		\begin{tabular}{r}
			\tikzmark{i1}\hbox{
		   		$-\frac{1}{2\sqrt{2}}$
			    \raisebox{-0.3\height}{\fbox{\chessboard[
				    setfen=8/8/8/8/8/8/8/K7 w - - 0 1,
				    showmover=false,
				    printarea=a1-b2,
				    margintopwidth=0pt,
				    marginrightwidth=0pt,
				    marginleftwidth=8pt,
				    marginbottomwidth=6pt,
				    tinyboard
			    ]}} 
			}\tikzmark{l}
		    \\
		    $+\qquad$
		    \\
		    \tikzmark{i2}\hbox{
				$\frac{1}{2\sqrt{2}}$
			    \raisebox{-0.3\height}{\fbox{\chessboard[
				    setfen=8/8/8/8/8/8/1K6/8 w - - 0 1,
				    showmover=false,
				    printarea=a1-b2,
				    margintopwidth=0pt,
				    marginrightwidth=0pt,
				    marginleftwidth=8pt,
				    marginbottomwidth=6pt,
				    tinyboard
			    ]}}
		    }\tikzmark{m}
		    \\
		    $+\qquad$
		    \\
		    \tikzmark{j1}\hbox{
				$- \frac{i}{2}$
			    \raisebox{-0.3\height}{\fbox{\chessboard[
				    setfen=8/8/8/8/8/8/8/1K6 w - - 0 1,
				    showmover=false,
				    printarea=a1-b2,
				    margintopwidth=0pt,
				    marginrightwidth=0pt,
				    marginleftwidth=8pt,
				    marginbottomwidth=6pt,
				    tinyboard
			    ]}}
		    }\tikzmark{n}
		    \\
		    $+\qquad$
		    \\
		    \tikzmark{k1}\hbox{
		   		$-\frac{1}{2}$
			    \raisebox{-0.3\height}{\fbox{\chessboard[
				    setfen=8/8/8/8/8/8/8/K7 w - - 0 1,
				    showmover=false,
				    printarea=a1-b2,
				    margintopwidth=0pt,
				    marginrightwidth=0pt,
				    marginleftwidth=8pt,
				    marginbottomwidth=6pt,
				    tinyboard
			    ]}} 
			}\tikzmark{o}
		    \\
		    $+\qquad$
		    \\
		    \tikzmark{k2}\hbox{
				$- \frac{1}{2}$
			    \raisebox{-0.3\height}{\fbox{\chessboard[
				    setfen=8/8/8/8/8/8/1K6/8 w - - 0 1,
				    showmover=false,
				    printarea=a1-b2,
				    margintopwidth=0pt,
				    marginrightwidth=0pt,
				    marginleftwidth=8pt,
				    marginbottomwidth=6pt,
				    tinyboard
			    ]}}
		    }\tikzmark{p}
		\end{tabular}
    	&& 
	    \begin{tabular}{r}
			\tikzmark{q}\hbox{
		   		$-\underset{\text{constructive}}{(\frac{1}{2\sqrt{2}}+\frac{1}{2})}$
			    \raisebox{-0.3\height}{\fbox{\chessboard[
				    setfen=8/8/8/8/8/8/8/K7 w - - 0 1,
				    showmover=false,
				    printarea=a1-b2,
				    margintopwidth=0pt,
				    marginrightwidth=0pt,
				    marginleftwidth=8pt,
				    marginbottomwidth=10pt,
				    smallboard
			    ]}} 
			} 
			\\
		    $+\qquad$
		    \\
		    \tikzmark{r}\hbox{
				$- \frac{i}{2}$
			    \raisebox{-0.3\height}{\fbox{\chessboard[
				    setfen=8/8/8/8/8/8/8/1K6 w - - 0 1,
				    showmover=false,
				    printarea=a1-b2,
				    margintopwidth=0pt,
				    marginrightwidth=0pt,
				    marginleftwidth=8pt,
				    marginbottomwidth=10pt,
				    smallboard
			    ]}}
		    }
		    \\
		    $+\qquad$
		    \\
		    \tikzmark{s}\hbox{
				$\underset{\text{destructive}}{(\frac{1}{2\sqrt{2}}-\frac{1}{2})}$
			    \raisebox{-0.3\height}{\fbox{\chessboard[
				    setfen=8/8/8/8/8/8/1K6/8 w - - 0 1,
				    showmover=false,
				    printarea=a1-b2,
				    margintopwidth=0pt,
				    marginrightwidth=0pt,
				    marginleftwidth=8pt,
				    marginbottomwidth=10pt,
				    smallboard
			    ]}}
		    }
		\end{tabular}
	\end{tabular}
\end{tabular}
\begin{tikzpicture}[overlay, remember picture, yshift=.25\baselineskip, shorten >=.5pt, shorten <=.5pt, 
					decoration = {snake, pre length=3pt,post length=7pt,}]
    \draw [->] ([yshift=.75pt]{pic cs:a}) -- ({pic cs:b});
    \draw [->] ([yshift=-.75pt]{pic cs:a}) -- ({pic cs:c});
    \draw [->] ([yshift=.75pt]{pic cs:d}) -- ({pic cs:f});
    \draw [->] ([yshift=-.75pt]{pic cs:d}) -- ({pic cs:g});
    \draw [->] ({pic cs:e}) -- ({pic cs:h});
    \draw [->] ([yshift=.75pt]{pic cs:i}) -- ({pic cs:i1});
    \draw [->] ([yshift=-.75pt]{pic cs:i}) -- ({pic cs:i2});
    \draw [->] ({pic cs:j}) -- ({pic cs:j1});
    \draw [->] ([yshift=.75pt]{pic cs:k}) -- ({pic cs:k1});
    \draw [->] ([yshift=-.75pt]{pic cs:k}) -- ({pic cs:k2});
    \draw [->,decorate,blue] ({pic cs:l}) -- ([yshift=.75pt]{pic cs:q});
    \draw [->,decorate,blue] ({pic cs:o}) -- ([yshift=-.75pt]{pic cs:q});
    \draw [->] ({pic cs:n}) -- ({pic cs:r});
    \draw [->,decorate,red] ({pic cs:m}) -- ([yshift=.75pt]{pic cs:s});
    \draw [->,decorate,red] ({pic cs:p}) -- ([yshift=-.75pt]{pic cs:s});
\end{tikzpicture}
\end{centering}
\caption{\small A series of three Split Jumps results in constructive and destructive interference.}
\label{fig:KingInterference}
\end{figure}
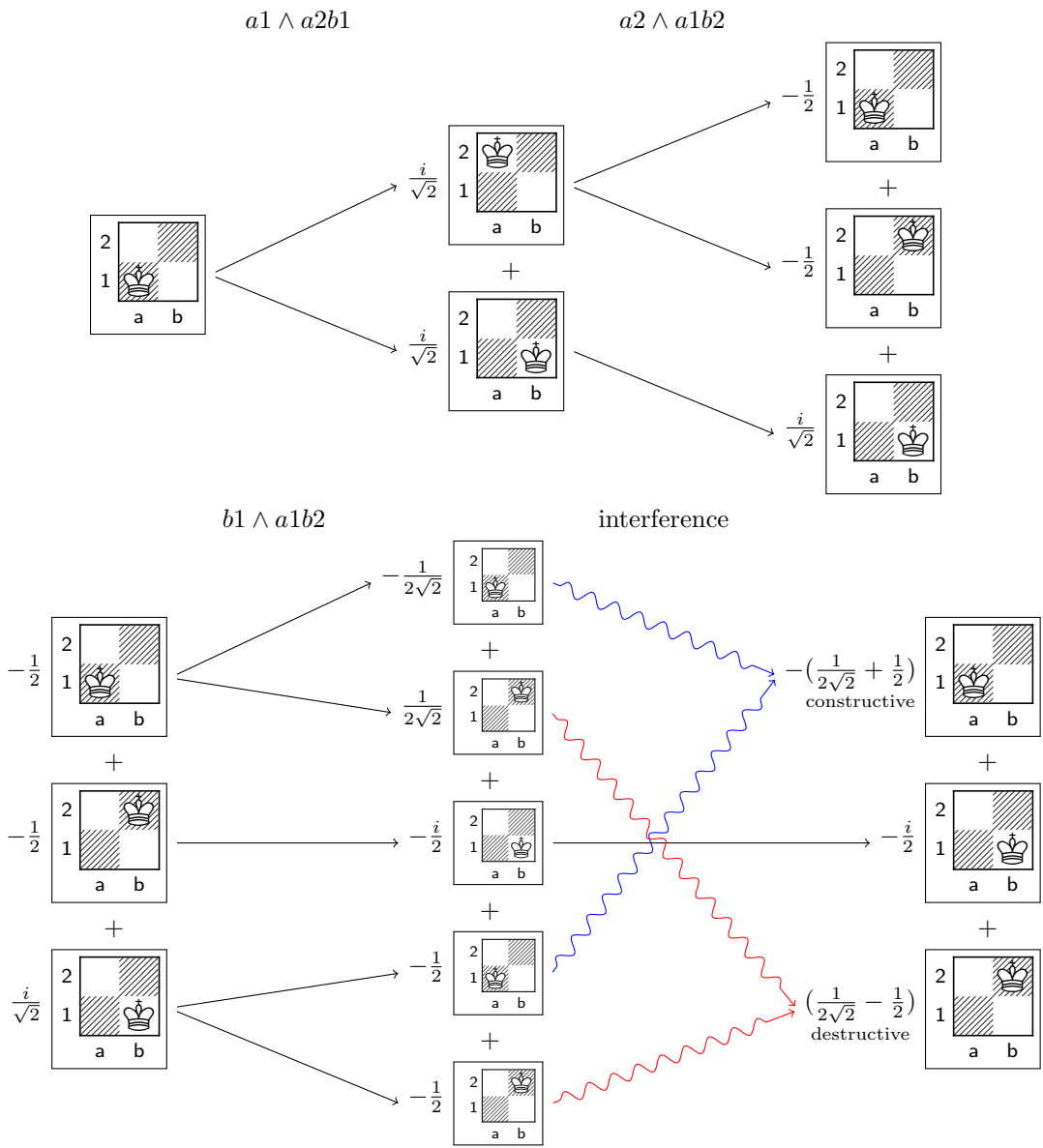
Interference can be seen when like pieces interact. Consider a single king in square a1, and the sequence of moves $a1\wedge a2b1$, $a2\wedge a1b2$, and $b1\wedge a1b2$. By applying the $U_{split}$ unitary \ref{eq:SplitJumpUnitary} to the appropriately ordered subspace\footnote{Correct ordering of the qubits is imperative when applying the unitary operations. For example, to apply $U_{split}(b1,a1,b2$) to a state in the basis $\ket{b2,b1,a2,a1}$ one must first map the occupancy values to the basis $\ket{b2,a1,b1}$, apply the unitary, then map back.} of the occupancy state in the basis $\ket{b2,b1,a2,a1}$ we have the following evolution:
\begin{IEEEeqnarray*}{r l}
\ket{0001} &\xrightarrow{\text{$a1\wedge a2b1$}} \frac{i}{\sqrt{2}}(\ket{0010} + \ket{0100})\\
\frac{i}{\sqrt{2}}\ket{0010} + \frac{i}{\sqrt{2}}\ket{0100} &\xrightarrow{\text{$a2\wedge a1b2$}} -\frac{1}{2}(\ket{0001} + \ket{1000}) +\frac{i}{\sqrt{2}}\ket{0100}\\
-\frac{1}{2}\ket{0001} - \frac{1}{2}\ket{1000} +\frac{i}{\sqrt{2}}\ket{0100} &\xrightarrow{\text{$b1\wedge a1b2$}}
	 -\frac{1}{2\sqrt{2}}(\ket{0001}-\ket{1000}) - \frac{i}{2}\ket{0100}\\ 
	&\qquad \qquad -\frac{1}{2}(\ket{0001}+\ket{1000})\\
	&=-\underset{\text{constructive}}{(\frac{1}{2\sqrt{2}}+\frac{1}{2})}\ket{0001} + \underset{\text{destructive}}{(\frac{1}{2\sqrt{2}}-\frac{1}{2})}\ket{1000} - \frac{i}{2}\ket{0100}
\end{IEEEeqnarray*}
If we square the amplitudes we get a probability distribution of $p(a1=1) \approx 0.73$, $p(a2=1) \approx 0.02$, and $p(b1=1) = 0.25$. Figure \ref{fig:KingInterference} shows how the splits act on each board in the superposition resulting in the final state. Note that the outcome depends on the order of the squares chosen during the move. If we follow the same procedure as above, but for the last split we instead do $b1\wedge b2a1$ the probability distribution changes to $p(a1=1) \approx 0.02$ , $p(a2=1) \approx 0.73$, $p(b1=1) = 0.25$. This difference may allow players to come up with interesting strategies involving interference.

\section{Conclusion}
Here we have detailed the process of designing Quantum Chess and discussed some of the considerations that went into certain design decisions. We see that it is possible to design a game using fully unitary dynamics and non-deterministic measurements while allowing the player to use quantum effects as a resource in game play. Through careful crafting of the rules of play we can see that it is possible to limit the size of the superposition so the game remains simulable and at the same time allow for significant quantum effects to be present (i.e. not just random). There is much future work that can be done with both Quantum Chess, and its future variants. There are many other classic games that we would also like to see receive a similar quantum treatment, such as Checkers and Reversi. We can also investigate implementation of these games using quantum resources. Recent work in the area of implementing games for quantum computers can be seen in \textit{The History of Games for Quantum Computers} \cite{Wooten:online}. AI for these types of games is another area of research to be explored. Standard Chess has served as a test bed for AI development for decades, perhaps Quantum Chess can do the same for quantum AI development. The future is quantum, and quantum games can be fun.

\section{Acknowledgements}
This project had very humble beginnings. Initially it was not even related to Chess. It was a game where each player had a single piece type on a 5x5 board and the pieces moved using the square root of a swap. It accomplished the goals of superposition, entanglement, and interference, but it was decidedly not fun. The story likely would have ended there if not for my advisor, Todd Brun, who recognized its potential and put me in contact with someone he knew who "would be interested in this sort of thing". The introduction to Spiros Michalakis led to an amazing friendship and a project that has been more fulfilling than anything else I have worked on to date.

It is safe to say Quantum Chess would not be what is today without Spiros' enthusiasm and support. Our frequent meetings at Caltech's Institute for Quantum Information and Matter to discuss gameplay mechanics and how to create a fun experience helped shape what it would become. And then IQIM's video \textit{Anyone Can Quantum} \cite{AnyoneCanQuantum}, written by Spiros and Jose Gonzalez, and subsequent outreach funding from the Institute gave Quantum Chess a life of its own which led to a number of incredible opportunities that ultimately pushed me to finish this long overdue paper.

A.I. for the game has been a challenge, and work on that front is still far from complete. To that end Spiros introduced me to Evert van Nieuwenburg who has been pivotal in its progress. We've had a number of discussions about A.I. for Quantum Chess and for his new version of Quantum Tic Tac Toe \cite{QuantumTicTacToe:2019}, as well as other quantum games, and I look forward to working on future projects with him.

I must also mention the Quantum Chess community, and particularly Elmar van Kordenoordt and M. Sol\'{e}. Frequent discussions in the Quantum Chess Discord channel led to the development of Quantum Chess Algebraic Notation, and community input on gameplay mechanics and knowledge of Chess strategy has helped shape the current ruleset. 

Finally I want to acknowledge my wife Laurie who pushed me to step outside my comfort zone, and the rest of my family and friends who have been incredibly supportive and enthusiastic about the entire project. 

To all of you, and the many I have not mentioned by name but who have acted as sounding boards for my ideas and as moral support, I thank you.

\clearpage
\bibliography{QuantumChessBib}{}
\bibliographystyle{hplain}

\end{document}